\documentclass[epj]{svjour}
\usepackage{amsmath, amssymb, amstext}
\usepackage{bbm}
\usepackage{booktabs}
\usepackage{array}
\usepackage{overpic}
\usepackage{pdflscape}
\usepackage{longtable}
\usepackage{caption}

\usepackage{graphicx} \graphicspath{ {./figures/} {./graphs/} }
\usepackage[varg]{txfonts} 
\usepackage[latin1]{inputenc}
\newcommand{\Kb}{{\bar K}}
\newcommand{\La}{{\Lambda}}
\newcommand{\Si}{{\Sigma}}
\newcommand{\SiS}{{\Sigma^*}}
\newcommand{\XiS}{{\Xi^*}}
\newcommand{\D}{{\Delta}}
\newcommand{\Om}{{\Omega}}
 \newcommand{\ONE}{\mathbbm1} 
\newcommand\numberthis{\addtocounter{equation}{1}\tag{\theequation}}
\newcommand{\ir}[1]{\mathbf{#1}}

\newcommand{\sharedhead}{}
\newcolumntype{t}{>{$}c<{$}} 
\newcommand{\psd}{\ensuremath{P_{1/2}^{(\sigma)}}}
\newcommand{\psq}{\ensuremath{P_{3/2}^{(\sigma)}}}
\newcommand{\psdq}{\ensuremath{P_{1/2,3/2}^{(\sigma)}}}
\newcommand{\psqd}{\ensuremath{P_{1/2,3/2}^{(\sigma)\,\dagger}}}
\newcommand{\Czz}{\ensuremath{c_{00}}}
\newcommand{\Czo}{\ensuremath{c_{01}}}
\newcommand{\Czt}{\ensuremath{c_{02}}}
\newcommand{\Coo}{\ensuremath{c_{11}}}
\newcommand{\Cot}{\ensuremath{c_{12}}}
\newcommand{\Ctt}{\ensuremath{c_{22}}}
\newcommand{\Gzz}{\ensuremath{C_{00}}}
\newcommand{\Gzo}{\ensuremath{C_{01}}}
\newcommand{\Gzt}{\ensuremath{C_{02}}}
\newcommand{\Goo}{\ensuremath{C_{11}}}
\newcommand{\Got}{\ensuremath{C_{12}}}
\newcommand{\Gtt}{\ensuremath{C_{22}}}
\newcommand{\GTzz}{\ensuremath{C}}
\newcommand{\GTzo}{\ensuremath{C}}
\newcommand{\GTzt}{\ensuremath{C}}
\newcommand{\GToo}{\ensuremath{C}}
\newcommand{\GTot}{\ensuremath{C}}
\newcommand{\GTtt}{\ensuremath{C}}
\newcommand{\sumCzza}{\ensuremath{\sum_{f=1,3,5} \Czz^f}}
\newcommand{\sumCzzb}{\ensuremath{\sum_{f=2,4,6} \Czz^f}}
\newcommand{\sumCzoa}{\ensuremath{\sum_{f=1,2} \Czo^f}}
\newcommand{\sumCooa}{\ensuremath{\sum_{f=1,3,5,7} \Coo^f}}
\newcommand{\sumCoob}{\ensuremath{\sum_{f=2,4,6,8} \Coo^f}}
\newcommand{\sumCzta}{\ensuremath{\Czt^1}}
\newcommand{\sumCztb}{\ensuremath{\Czt^2}}
\newcommand{\sumCota}{\ensuremath{\Cot^1}}
\newcommand{\sumCotb}{\ensuremath{\Cot^2}}
\newcommand{\sumCtta}{\ensuremath{\sum_{f=1,5} \Ctt^f}}
\newcommand{\sumCttb}{\ensuremath{\sum_{f=2,6} \Ctt^f}}
\newcommand{\sumCttc}{\ensuremath{\sum_{f=3,7} \Ctt^f}}
\newcommand{\sumCttd}{\ensuremath{\sum_{f=4,8} \Ctt^f}}
\newcommand{\Fpic}[5]{
   \parbox[c][2.2cm][c]{1.05cm}{
   \begin{overpic}[scale=.4]{#5}
   \put(-3,105){$#3$}\put(51,105){$#4$}
   \put(-3,-17){$#1$}\put(51,-17){$#2$}
   \end{overpic}}
}

\begin{document}
\title{%
Scattering of decuplet baryons in chiral effective field theory 
}%
\author{
J.~Haidenbauer\inst{1} \and
S.~Petschauer\inst{2} \and
N.~Kaiser\inst{2} \and
Ulf-G. Mei\ss{}ner \inst{3,1} \and
W.~Weise\inst{2}
}


\institute{
Institute for Advanced Simulation and J\"ulich Center for Hadron Physics,
Institut f\"ur Kernphysik, Forschungszentrum J\"ulich, D-52425 J\"ulich, Germany
\and
Physik Department, Technische Universit\"at M\"unchen, D-85747 Garching, Germany
\and
Helmholtz-Institut f\"ur Strahlen- und Kernphysik and Bethe Center
for Theoretical Physics, Universit\"at Bonn, D-53115 Bonn, Germany
}
\abstract{A formalism for treating the scattering of decuplet baryons in chiral 
effective field theory is developed. The minimal Lagrangian and potentials in 
leading-order SU(3) chiral effective field theory for the
interactions of octet baryons (\(B\)) and decuplet baryons (\(D\)) for the 
transitions \(BB\to BB\), \(BB\leftrightarrow DB\), \(DB\to DB\), \(BB\leftrightarrow DD\),
\(DB\leftrightarrow DD\), and \(DD\to DD\) are provided. 
As an application of the formalism we compare with results from
lattice QCD simulations for $\Omega\Omega$ and $N\Omega$ scattering. 
Implications of our results pertinent to  the quest for dibaryons are discussed. 
\PACS{
        {12.39.Fe}{} \and 
        {13.75.Ev}{} \and 
        {12.38.Gc}{} \and 
        {14.20.Pt}{} 
      }
} 
\maketitle
\section{Introduction} 
\label{sec:Intro} 

{An important} and basically timeless topic in hadron physics is the possible existence 
of dibaryons \cite{Dyson:1964xwa,Clement:2016}. 
Historically, the so-called dibaryons have {primarily} been conceived as tightly bound 
six-quark objects. However, the {terminology is also frequently adopted for} 
shallow (or not so shallow)
bound states of two baryons, as in practice it might be difficult to distinguish 
between these {six-quark and two-baryon} configurations. 
We note that methods to disentangle compact multi-quark states from loosely 
bound molecular systems have recently been reviewed in ref.~\cite{Guo:2017jvc}.

{A famous dibaryon} proposed to be of the first category is the
$H$-particle, predicted by Jaffe within the bag model as a compact $|uuddss\rangle$
state~\cite{Jaffe:1976yi}. The $H$-dibaryon has the quantum numbers of the $\Lambda\Lambda$ 
system {in an $S$-wave state}, namely strangeness ${S}=-2$, isospin $I=0$, and $J^P=0^+$. 
After decades of failed efforts to establish its existence experimentally,
it regained popularity a few years ago due to lattice simulations that provided
evidence for its presence~\cite{Beane:2010hg,Inoue:2010es}, 
though only for quark masses that are larger than the physical ones.

Definitely of the second category is the deuteron, a bound state in the coupled $^3S_1$-$^3D_1$ 
partial waves of {the} neutron-proton ($np$) {system}. Also of that category is a possible
(unstable) {quasibound} state in the $^3S_1$-$^3D_1$ partial waves of the $\Sigma N$ channel 
with isospin $I=1/2$ which appears as a pronounced cusp-like enhancement in the 
$\Lambda p$ invariant-mass spectrum of reactions like 
$K^-d \to \pi^-\Lambda p$ and $pp\to K^+\Lambda p$ very close to the $\Sigma N$ threshold,
see the review \cite{Machner:2013}.  

In refs.~\cite{Polinder:2006zh,Haidenbauer:2013} it has been shown that chiral 
effective field theory (EFT), an approach initially suggested for 
deriving and describing the forces between 
nucleons~\cite{Wei90,Wei91,Ordonez:1993,Epelbaum:2008,Machleidt:2011}, 
can be straightforwardly extended to the interaction of baryons with strange\-ness. 
{In this} framework the symmetries of QCD together with the appropriate low-energy 
degrees of freedom are exploited to construct the baryon-baryon interactions. Moreover,
there is an underlying power counting {that} allows one to improve the results
systematically by going {progressively} to higher orders in a perturbative expansion.
This approach is well {prepared} to shed light on the $H$-dibaryon, should it indeed exist, 
and on other possible dibaryons in the strange\-ness $S=-3$ and $-4$ sectors, as
demonstrated in refs.~\cite{Haidenbauer:2011ah,Haidenbauer:2011za}.
In particular, one can study implications of the imposed (approximate)
SU(3) symmetry and further explore the dependence of the properties of such dibaryons 
on the pion\footnote{{The Gell-Mann-Oakes-Renner relation states that the squared} pion 
mass is proportional to the average light quark mass. Therefore, the notions
``quark mass dependence'' and ``pion mass dependence'' {are} used synonymously.}
and baryon masses. The latter aspect is important since, as mentioned, the pertinent lattice QCD 
(LQCD) calculations were not performed for physical quark masses. 

In the present paper we extend the {investigations} of 
refs.~\cite{Polinder:2006zh,Haidenbauer:2011ah,Haidenbauer:2011za}
to decuplet baryons. 
The main goal is to provide the basic elements for treating the scattering of
octet baryons with decuplet baryons ($BD$) and of two decuplet baryons ($DD$) 
within SU(3) chiral EFT. {Here} we restrict ourselves, {as a first step,} to leading order (LO) 
in the Weinberg counting \cite{Epelbaum:2008}.  
In this case {the interaction potential
is given} by single pseudoscalar-meson ($\pi$, $\eta$, $K$) exchange {plus}
four-baryon contact terms without derivatives \cite{Polinder:2006zh}. The latter
represent the short-ranged part of the baryon-baryon force and involve
low-energy constants (LECs),  parameters to be fixed by fits to data.
The resulting potential is then used {in} a regularized Lippmann-Schwin\-ger (LS) 
equation to generate possible bound states and to evaluate two-body
scattering amplitudes. 
As in refs.~\cite{Polinder:2006zh,Haidenbauer:2011ah,Haidenbauer:2011za}
we assume the (approximate) validity of SU(3) flavor symmetry for the coupling strengths involved\footnote{SU(3) 
breaking effects are nonetheless incorporated using physical baryon and meson masses.}. 
This assumption allows one to establish relations between the forces in 
channels with different isospin and strange\-ness, and we recapitulate the
implications of SU(3) symmetry for the interactions in the $BD$ and $DD$ systems.
In addition we investigate the consequences of SU(3) symmetry for transitions from 
these systems to states composed of two octet baryons ($BB$). 
Note that a work in similar spirit but aiming at the charm and beauty sector,
where instead heavy quark spin symmetry plays a fundamental role, has been 
presented recently \cite{Lu:2017}. 
 
{A primary} motivation for our study comes from recent LQCD calculations
for $\Om\Om$ and $N\Om$ scattering
\cite{Buchoff:2012,Yamada:2015,Etminan:2014,Sasaki:2016}. 
In some of those {computations} signals for possible dibaryons were found 
\cite{Yamada:2015,Etminan:2014,Sasaki:2016}.
Thus, as a first application of our formalism we analyze the results for 
$\Om\Om$ and $N\Om$ scattering presented in those works. {We also} discuss 
implications on possible other dibaryon states, notably in the $\SiS\D$, 
$\Om\XiS$, and $\D\D$ systems.  
{Apart from} indications for dibaryons in the $\Om\Om$ and $N\Om$
systems from LQCD calculations, 
there are {further} longstanding claims for dibaryons in those channels from 
studies {based} on the constituent quark model, see 
refs.~\cite{Huang:2015,Zhang:2006,Dai:2006} for more recent efforts.
Possible bound states in $\Xi \Om$ and $\Xi^* \Om$ \cite{Li:1999}, 
and $\D\Om$ \cite{Li:1999bc,Dai:2007} have been discussed, too. 
And finally, there has been a noticeable revival of dibaryons in the $\D\D$ sector,
in the context of measurements of the reactions $pn\to d \pi^0\pi^0$ and 
$pn\to d \pi^+\pi^-$ by the WASA-at-COSY collaboration in J\"ulich  
\cite{Adlarson:2011}. A resonance in the $^3D_3$-$^3G_3$ $NN$ partial 
wave required to describe the data \cite{Adlarson:2014}, termed 
$d^*(2380)$ dibaryon, 
could be a reflection of an $S$-wave {quasibound} state in the $\D\D$ system 
\cite{Huang:2014,Gal:2014,Chen:2014}. 

In the analysis of the LQCD simulations we follow closely the strategy 
of our previous works \cite{Haidenbauer:2011ah,Haidenbauer:2011za}:
(i) the LECs, i.e. the only free parameters in the potential, are determined by 
a fit to LQCD results (phase shifts, scattering lengths) employing the inherent 
baryon and meson masses of the lattice simulation;
(ii) Results at the physical point are obtained via a calculation in which
the pertinent physical masses of the mesons are substituted in the evaluation of 
the potential and those of the baryons in the baryon-baryon propagators appearing
in the LS equation. 
Such an ``extrapolation'' {can be} expected to work only qualitatively, given the still 
{unphysically large} hadron masses in the LQCD simulations for $\Om\Om$, 
$N\Om$, and $\D\D$. Nonetheless  recent (and still preliminary) lattice
results for the $H$-dibaryon, and for $\La\La$ and $N\Xi$ phase shifts, respectively, 
close to the physical point \cite{Sasaki:2016,Sasaki:20162} reveal that the 
prediction/extrapolation in refs.~\cite{Haidenbauer:2011ah,Haidenbauer:2011za}, 
performed in the way described above, may be fairly realistic. In this context let
us mention that also for other baryon-baryon channels LQCD simulations have been 
performed for almost physical quark masses~\cite{Doi:2017,Ishii:2017,Nemura:2017} 
and preliminary results for phase shifts have become available \footnote{Notably, 
the method employed in these works is yet under discussion in the LQCD community.}
for some channels. 

{This} paper is structured {as follows}:
A basic outline of our formalism is {given} in sec.~\ref{sec:Formalism}. 
Specifically, we establish the underlying Lagrangians {for constructing} the
interactions in the $BB$, $BD$, and $DD$ sectors at leading order in chiral EFT. 
Furthermore, the essentials of the imposed SU(3) flavor symmetry are described. 
Finally, explicit expressions for the potentials resulting from meson-exchange and
from the four-baryon contact terms are given. 
In sec.~\ref{sec:Results}, as exemplary applications of our formalism,
results for selected $DD$ and $BD$ reactions are presented, where we focus
on {such reaction channels for which} predictions from lattice QCD simulations 
for phase shifts and/or scattering lengths are available: 
$\Om\Om$ scattering in the $^1S_0$ partial wave 
and $N\Om$ in the $^5S_2$ partial wave. Furthermore, we explore the 
$\D\D$ system with total angular momentum $J=3$ and isospin $I=0$, 
and the possible emergence of a bound state that could be associated 
with the $d^*(2380)$ dibaryon.
A brief summary and an outlook is presented in sec.~\ref{sec:Summary}. 
Technical details of our calculation are summarized in three appendices.


\section{Formalism}
\label{sec:Formalism}

Our calculation is based on a non-relativistic approach in leading-order SU(3) chiral EFT.
Possible (isospin violating) $\pi^0$-$\eta$ or  $\Sigma^0$-$\Lambda$  mixing is neglected.
We use the conventional building blocks as laid down, e.g., in 
refs.~\cite{Polinder:2006zh,Haidenbauer:2013}, {with} the pseudoscalar meson 
octet represented by the \(3\times3\) matrix in flavor space:
\begin{equation} \label{eq:mesonmat}
 \phi=
 \begin{pmatrix}
  \pi^0 + \frac{\eta}{\sqrt 3} & \sqrt 2 \pi^+ & \sqrt 2 K^+ \\
  \sqrt 2 \pi^- & -\pi^0 + \frac{\eta}{\sqrt 3} & \sqrt 2 K^0 \\
  \sqrt 2 K^- & \sqrt 2 \bar K^0 & -\frac{2\eta}{\sqrt 3}
 \end{pmatrix}\,.
\end{equation}
The octet baryons are represented by {the} \(3\times3\) matrix
\begin{equation} \label{eq:baryonmat}
 B=
 \begin{pmatrix}
  \frac{\Sigma^0}{\sqrt 2} + \frac{\Lambda}{\sqrt 6} & \Sigma^+ & p \\
  \Sigma^- & -\frac{\Sigma^0}{\sqrt 2} + \frac{\Lambda}{\sqrt 6} & n \\
  \Xi^- & \Xi^0 & -\frac{2\Lambda}{\sqrt 6}
 \end{pmatrix}\, \ .
\end{equation}
{The} decuplet baryons are represented by the totally symmetric three-index tensor \(T\):
\begin{align*} 
T_{111}&=\Delta^{++}\,, \ \
T_{112}=\tfrac{1}{\sqrt{3}}\Delta^{+}\,, \ \
T_{122}=\tfrac{1}{\sqrt{3}}\Delta^{0}\,, \ \
T_{222}=\Delta^{-}\,,\\
T_{113}&=\tfrac{1}{\sqrt{3}}\Sigma^{*+}\,, \ \
T_{123}=\tfrac{1}{\sqrt{6}}\Sigma^{*0}\,, \ \
T_{223}=\tfrac{1}{\sqrt{3}}\Sigma^{*-}\,,\\
T_{133}&=\tfrac{1}{\sqrt{3}}\Xi^{*0}\,, \ \
T_{233}=\tfrac{1}{\sqrt{3}}\Xi^{*-}\,,\\
T_{333}&=\Omega^-\, \ . \numberthis
\label{eq:baryonDecTensor}
\end{align*}

{The} construction of the chiral Lagrangian {requires} a complete set of spin operators, {covering all combinations of spin 1/2 and spin 3/2 states}.
These have been established employing the Wigner-Eckart theorem.
One obtains for spin 1/2 the identity matrix \(\ONE\) and \(\sigma^i\), i.e. the usual Pauli matrices, corresponding to scalar and vector operators.
For the spin 1/2 to spin 3/2 transition operators, the resulting \(4\times2\) matrices connect the two-component spinors of octet baryons with the four-component spinors of decuplet baryons \cite{Ericson:1988gk}. 
{The corresponding vector and rank-2 tensor operators are \(S^{i\dagger}\) and \(S^{ij\dagger}\), respectively.}
{The reverse} transition, 3/2 to 1/2, {is described by} the {Hermitian-conju\-gate} spin transition operators \(S^i\) and \(S^{ij}\).
Finally, {the \(4\times4\) 
matrices \(\ONE\), \(\Sigma^i\), \(\Sigma^{ij}\), \(\Sigma^{ijk}\), corresponding in this order to scalar, vector, rank-2 and rank-3 tensor operators, are introduced for the description of the spin 3/2 sector.}
These sets of matrices for the different types of transitions form a basis.
More details on the construction of the spin operators and their explicit expressions 
can be found in Appendix \ref{app:Spin}.

For the two-body interactions {considered} in the following, tensor products of two of 
these spin operators are involved.

\subsection{Lagrangians {generating} meson exchange} \label{subsec:LagMes}

In this subsection we consider the {meson-baryon Lagrangians for the construction of the 
leading-order meson exchange} interactions of octet and decuplet baryons.

The SU(3) invariant Lagrangian for octet baryons {coupled to the pseudoscalar meson octet} is given by
\cite{Polinder:2006zh,Petschauer:2013}
\begin{align*}
 \mathcal{L}_\mathrm{BB\phi} &= -\frac D {2f_0} {\rm tr}( \bar B \vec\sigma\cdot \lbrace \vec\nabla\phi,B\rbrace) - \frac F {2f_0} {\rm tr}(\bar B \vec\sigma\cdot \left[\vec\nabla\phi,B\right]) \, \ , \numberthis 
\label{eq:BBM}
\end{align*}
where $f_0$ is the weak pseudoscalar-meson decay constant $f_0 \approx f_\pi$.
The coupling constants \(F\) and \(D\) satisfy \(F + D = g_A \),
with \(g_A\) the axial-vector strength measured in neutron \(\beta\)-decay\footnote{Empirical 
values of these constants are $f_\pi \simeq 92.2$ MeV and $g_A = 1.272\pm 0.002$ \cite{PDG}.}.
{Introducing standard isospin operators}, the Lagrangian can be written in its well known
form~\cite{deSwart:1963gc}:
\begin{eqnarray}
{\mathcal L}_\mathrm{BB\phi}&=&-f_{NN\pi}\bar{N}\mbox{\boldmath $\tau$}N\cdot\mbox{\boldmath $\pi$} 
+if_{\Sigma\Sigma\pi}\bar{\mbox{\boldmath $ \Sigma$}}\times{\mbox{\boldmath $ \Sigma$}}\cdot\mbox{\boldmath $\pi$} \nonumber \\
&&-f_{\Lambda\Sigma\pi}\left[\bar{\Lambda}{\mbox{\boldmath $ \Sigma$}}+\bar{\mbox{\boldmath $\Sigma$}}
\Lambda\right]\cdot\mbox{\boldmath $\pi$}-f_{\Xi\Xi\pi}\bar{\Xi}\mbox{\boldmath $\tau$}\Xi\cdot\mbox{\boldmath $\pi$} \nonumber \\
&&-f_{\Lambda NK}\left[\bar{N}\Lambda K+\bar{\Lambda} N K^\dagger\right]
\nonumber \\&&
-f_{\Xi\Lambda K}\left[\bar{\Xi}\Lambda K_c+\bar{\Lambda}\Xi K_c^\dagger\right]
\nonumber \\&&
-f_{\Sigma NK}\left[\bar{\mbox{\boldmath $ \Sigma$}}\cdot K^\dagger\mbox{\boldmath $\tau$}N+\bar{N}\mbox{\boldmath $\tau$} K\cdot{\mbox{\boldmath $ \Sigma$}}\right]
\nonumber \\&&
-f_{\Xi \Sigma K}\left[\bar{\mbox{\boldmath $ \Sigma$}}\cdot K_c^\dagger\mbox{\boldmath $\tau$}\Xi+\bar{\Xi}
\mbox{\boldmath $\tau$} K_c\cdot{\mbox{\boldmath $ \Sigma$}}\right]
\nonumber \\&&
-f_{NN\eta_8}\bar{N} N\eta
\nonumber \\&&
-f_{\Lambda\Lambda\eta_8}\bar{\Lambda}\Lambda \eta-f_{\Sigma\Sigma\eta_8}\bar{\mbox{\boldmath $ \Sigma$}}\cdot{\mbox{\boldmath $ \Sigma$}}\eta
\nonumber \\&& 
-f_{\Xi\Xi\eta_8}\bar{\Xi}\Xi\eta \ .
\label{eq:BBMI}
\end{eqnarray}
Here, we have introduced the isospin doublets
\begin{equation}
N=\left(\begin{array}{r}p\\n\end{array}\right)\ ,\ \ \Xi=\left(\begin{array}{r}\Xi^0\\\Xi^-\end{array}\right)\ ,\ \ 
K=\left(\begin{array}{r}K^+\\K^0\end{array}\right)\ ,\ \  K_c=\left(\begin{array}{r}\Kb^0\\-K^-\end{array}\right)\ .
\label{eq:3.8}
\end{equation}
The signs have been chosen according to the conventions of ref.~\cite{deSwart:1963gc}.
With regard to isovectors the assignment $\Si^+ = -|1,1\rangle$, etc., is used so 
that the inner product of $\mbox{\boldmath $\Sigma$}$ (or $\mbox{\boldmath $\pi$}$) defined 
in spherical components reads
\begin{equation}
\mbox{\boldmath $\Sigma$}\cdot\mbox{\boldmath $\Sigma$}=\sum_m (-1)^m\Sigma_m\Sigma_{-m}=
\Sigma^+\Sigma^-+\Sigma^0\Sigma^0+\Sigma^-\Sigma^+\ .
\label{eq:A2.8}
\end{equation}
{Spin- and momentum operators (\(\sigma  \cdot  \nabla\)) have been omitted in eq.~(\ref{eq:BBMI}) to 
simplify the presentation.} That structure is the same for all $B_iB_j\phi_k$ vertices. 
The coupling constants introduced in eq.~(\ref{eq:BBMI}) are given by 
\begin{equation}
\begin{array}{rlrl}
f_{NN\pi}  = & f, & f_{NN\eta_8}  = & \frac{1}{\sqrt{3}}(4\alpha -1)f, \\
f_{\Lambda NK} = & -\frac{1}{\sqrt{3}}(1+2\alpha)f, & f_{\Xi\Xi\pi}  = & -(1-2\alpha)f, \\
f_{\Xi\Xi\eta_8}  = & -\frac{1}{\sqrt{3}}(1+2\alpha )f, & f_{\Xi\Lambda K} = & \frac{1}{\sqrt{3}}(4\alpha-1)f, \\
f_{\Lambda\Sigma\pi}  = & \frac{2}{\sqrt{3}}(1-\alpha)f, & f_{\Sigma\Sigma\eta_8}  = & \frac{2}{\sqrt{3}}(1-\alpha )f,\\
f_{\Sigma NK} = & (1-2\alpha)f, & f_{\Sigma\Sigma\pi}  = & 2\alpha f, \\
f_{\Lambda\Lambda\eta_8}  = & -\frac{2}{\sqrt{3}}(1-\alpha )f, & f_{\Xi\Sigma K} = & -f.
\end{array}
\label{eq:BBC}
\end{equation}
with $f\equiv g_A/(2f_0)$ and $\alpha\equiv F/(F+D)$. 
In the present {work} we use the same values as in our {previous} LO study of the $YN$ system \cite{Polinder:2006zh},
namely $f_0=93$~MeV, $g_A = 1.26$, $\alpha=0.4$. 

The Lagrangian involving an octet and a decuplet baryon {is} written in the form 
\cite{Sasaki,Petschauer:2016pbn}
\begin{align*}
\mathcal{L}_\mathrm{DB\phi} &= \frac C{f_0} \sum_{a,b,c,d,e=1}^3 \epsilon_{abc} \Big( \bar T_{ade} \vec S^{\,\dagger} \cdot \left(\vec\nabla \phi_{db}\right)B_{ec} \\
&\qquad\qquad\qquad\qquad+
\bar B_{ce} \vec S \cdot \left(\vec\nabla\phi_{bd}\right)T_{ade} \Big)  \,. \numberthis
\label{eq:BD} 
\end{align*}
The coupling constant \(C\) can be fixed from the decay \(\Delta\to\pi N\). 
{One finds} \(C=3g_A/4\) \cite{Gersten}. The vector spin operator 
\mbox{\boldmath $S^{\,\dagger}$} consists of $4\times 2$ matrices {representing} the transition
from spin $1/2$ to $3/2$. {Their explicit forms are} given in Appendix \ref{app:Spin}. 

Omitting again the spin-momentum dependent part of the Lagrangian, eq.~(\ref{eq:BD}),
one can cast the remainder into a simple form using isospin operators
\begin{eqnarray}
{\mathcal L}_\mathrm{DB\phi}&=&-f_{N\Delta\pi}\bar{\Delta}\mbox{\boldmath $T^\dagger$}N\cdot\mbox{\boldmath $\pi$} 
+if_{\Sigma\SiS\pi}\bar{\mbox{\boldmath $ \SiS$}}\times{\mbox{\boldmath $ \Sigma$}}\cdot\mbox{\boldmath $\pi$} \nonumber \\
&&-f_{\Lambda\SiS\pi}\bar{\mbox{\boldmath $\SiS$}}
\Lambda\cdot\mbox{\boldmath $\pi$}-f_{\Xi\XiS\pi}\bar{\XiS}\mbox{\boldmath $\tau$}\Xi\cdot\mbox{\boldmath $\pi$} 
\nonumber \\&&
-f_{N\SiS K}\bar{\mbox{\boldmath $ \SiS$}}\cdot K^\dagger\mbox{\boldmath $\tau$}N
-f_{\Sigma \Delta K}\bar{\Delta}\mbox{\boldmath $T^\dagger$} K\cdot{\mbox{\boldmath $ \Sigma$}}
\nonumber \\&&
-f_{\Lambda \XiS K}\bar{\XiS}\Lambda K_c
-f_{\Sigma \XiS K}\bar{\XiS}\mbox{\boldmath $\tau$} K_c\cdot{\mbox{\boldmath $ \Sigma$}}
\nonumber \\&&
-f_{\Xi \SiS K}\bar{\mbox{\boldmath $ \SiS$}}\cdot K_c^\dagger\mbox{\boldmath $\tau$}\Xi
-f_{\Xi \Om K}\bar{\Om}\Xi K^\dagger
\nonumber \\&&
-f_{\Sigma\SiS\eta_8}\bar{\mbox{\boldmath $ \SiS$}}\cdot{\mbox{\boldmath $ \Sigma$}}\eta
-f_{\Xi\XiS\eta_8}\bar{\XiS}\Xi\eta 
\nonumber \\&&
+ \ {\rm h.c.} \ , 
\label{eq:BDI}
\end{eqnarray}
where the $\Delta$ isobar is represented by an isospin quartet, i.e. \mbox{$(\D^{++},\D^{+},\D^{0},\D^{-})^T$}
and $\XiS$ and $\SiS$ by isospin states analogous to those in eqs.~(\ref{eq:3.8})  
and (\ref{eq:A2.8}).
The SU(3) relations for the coupling constants {of octet to decuplet 
baryons are} expressed in terms of the coupling constant $f_{BD}$ {and} read
\begin{eqnarray}
f_{N\Delta\pi} = 
f_{\Xi \Om K} = 
-f_{\Sigma \Delta K} &=& f_{BD}, \nonumber \\
f_{\Sigma\SiS\pi} =
f_{N\SiS K} =
f_{\Sigma \XiS K} =
f_{\Xi \SiS K} &=& -\frac{1}{\sqrt{6}} f_{BD}, \nonumber \\
f_{\Lambda\SiS\pi} =
-f_{\Lambda \XiS K} =
-f_{\Sigma\SiS\eta_8} =
f_{\Xi\XiS\eta_8} &=& \frac{1}{\sqrt{2}} f_{BD}, \nonumber \\
f_{\Xi\XiS\pi} &=& \frac{1}{\sqrt{6}} f_{BD} \ .
\label{eq:BDC}
\end{eqnarray}
The relation between the coupling constants in eqs.~(\ref{eq:BD})
and (\ref{eq:BDI}) is $f_{BD} = \sqrt{2}\,C/f_0$ and results from the different normalization
of the isospin {transition} operator \mbox{\boldmath $T^\dagger$}, cf. eq.~(\ref{eq:Spin1}) in 
Appendix \ref{app:Spin}, and the tensor representation for the decuplet baryons
in eq.~(\ref{eq:baryonDecTensor}). 
The $N\D\pi$ coupling constant $h_A$, commonly used in studies of $\pi N$ scattering
within chiral perturbation theory \cite{Hemmert:1997,Fettes:2000,Yao:2016}, coincides with $f_{BD}$. 
Accordingly, the choice of $C$ mentioned above is in line with the large-$N_c$ value 
$h_{A} = 3g_A/(2\sqrt{2})$.

The Lagrangian involving two decuplet baryons and a meson expressed in the conventional building 
blocks is given by
\begin{align*}
\mathcal{L}_\mathrm{DD\phi} &= \frac H{f_0} \sum_{a,b,c,d=1}^3 (\bar T_{abc}\vec\Sigma T_{abd}) \cdot (\vec\nabla\phi_{cd}) \, ,  \numberthis
\label{eq:DD}
\end{align*}
with a new coupling constant \(H\) and the spin-3/2
operator ${\mathbf \Sigma}$ (see appendix A).  Although the same
symbol is used for the isotriplet of $\Sigma^{+,0,-}$ hyperons, the
respective meaning of  ${\mathbf \Sigma}$ should be clear from the context.  
Casting the Lagrangian eq.~(\ref{eq:DD}) into a simple form utilizing isospin operators leads to 
\begin{eqnarray}
{\mathcal L}_\mathrm{DD\phi}&=&-f_{\D\D\pi}\bar{\D}\mbox{\boldmath $\theta$}\D\cdot\mbox{\boldmath $\pi$} 
+if_{\SiS\SiS\pi}\bar{\mbox{\boldmath $ \SiS$}}\times{\mbox{\boldmath $ \SiS$}}\cdot\mbox{\boldmath $\pi$} \nonumber \\
&&-f_{\XiS\XiS\pi}\bar{\XiS}\mbox{\boldmath $\tau$}\XiS\cdot\mbox{\boldmath $\pi$} 
\nonumber \\&&
-f_{\SiS \D K}\left[\bar{\mbox{\boldmath $\SiS$}}\cdot K^\dagger\mbox{\boldmath $T$}\D
  +\bar{\D}\mbox{\boldmath $T^\dagger$} K\cdot{\mbox{\boldmath $ \SiS$}}\right]
\nonumber \\&&
-f_{\XiS \SiS K}\left[\bar{\mbox{\boldmath $\SiS$}}\cdot K_c^\dagger\mbox{\boldmath $\tau$}\XiS+\bar{\XiS}
\mbox{\boldmath $\tau$} K_c\cdot{\mbox{\boldmath $ \SiS$}}\right]
\nonumber \\&&
-f_{\Om\XiS K}\left[\bar{\Om}\XiS K^\dagger + \bar{\XiS}\Om K\right]
\nonumber \\&&
-f_{\D\D \eta_8}\bar{\D} \D\eta
-f_{\SiS\SiS\eta_8}\bar{\mbox{\boldmath $ \SiS$}}\cdot{\mbox{\boldmath $ \SiS$}}\eta
\nonumber \\&& 
-f_{\XiS\XiS\eta_8}\bar{\XiS}\XiS\eta  
-f_{\Om \Om \eta_8}\bar{\Om }\Om \eta \ ,
\label{eq:DDI}
\end{eqnarray}
omitting once again the spin-momentum part. 
{Here the} SU(3) relations for the coupling constants are:
\begin{eqnarray} f_{\D\D\pi} = f_{\XiS\XiS\pi} &=& f_{DD}, \nonumber \\
-f_{\SiS \D K} = f_{\Om\XiS K} &=& \sqrt{6} f_{DD}, \nonumber \\
f_{\SiS\SiS\pi} = f_{\XiS \SiS K} &=& 2f_{DD}, \nonumber \\
f_{\D\D \eta_8} = -f_{\XiS\XiS\eta_8} =-f_{\Om \Om \eta_8}/2 &=& \sqrt{3} f_{DD}, \nonumber \\
f_{\SiS\SiS\eta_8} &=& 0 \ ,
\label{eq:DDC}
\end{eqnarray}
{where the} normalization of the decuplet-baryon
tensor and the definition of the isospin 3/2 operator \mbox{\boldmath $\theta$} {as given} in
Appendix \ref{app:Spin} imply $f_{DD} = H / (3 f_0)$. We note that a remarkable correlation between the 
leading $\pi \Delta N$ and $\pi \Delta\Delta$ couplings beyond the large-$N_C$ expansion has
recently been found \cite{Gegelia:2016pjm}.

The general form of the LO meson-exchange potential is analogous to that for $BB$ systems 
\cite{Polinder:2006zh,Haidenbauer:2013}, namely 
\begin{eqnarray}
V_{P}&=&-\frac{\left(\mbox{$\mathcal{O}$}_1\cdot{\bf q}\right)\left(\mbox{$\mathcal{O}$}_2\cdot{\bf q}\right)}
{{\bf q}^2+m^2_P}\ ,
\label{OPE}
\end{eqnarray}
where $m_P$ is the mass of the exchanged pseudoscalar meson $P$. 
The symbol $\mathcal{O}_i$ stands generically for the spin operators
at the two vertices. 
For spin $1/2 \to 1/2$ it {is} given by \mbox{\boldmath $ \sigma_i$},
for $3/2 \to 3/2$ by \mbox{\boldmath $ \Sigma_i$},
for $1/2 \to 3/2$ by \mbox{\boldmath $S_i^\dagger$}, and 
for $3/2 \to 1/2$ by \mbox{\boldmath $S_i$}.
The transferred momentum ${\bf q}$ is defined in terms of the final and initial 
center-of-mass (c.m.) momenta of the baryons, ${\bf p}'$ and ${\bf p}$, as 
${\bf q}={\bf p}'-{\bf p}$. 
As already mentioned in the Introduction, we use either the physical masses of the exchanged 
pseudoscalar mesons or masses corresponding to the lattice QCD simulations. 
Thus, the explicit ${\rm SU(3)}$ breaking reflected in the mass splitting between 
the pseudoscalar mesons is taken into account. 

Following our previous works \cite{Polinder:2006zh,Haidenbauer:2013} 
the $\eta$ meson is identified with the octet state $\eta_8$ and its 
physical mass is used. A possible coupling of two octet baryons to a
{singlet meson} ($\eta_0$) which would introduce a further coupling constant 
\cite{deSwart:1963gc}, {is} ignored {in the present context}. As far as the vertex of
two decuplet baryons and a meson is concerned, since 
according to the decomposition 
$\mathbf{10}\otimes \mathbf{\overline{10}} = \mathbf{64} \oplus \mathbf{27} \oplus \mathbf{8} \oplus \mathbf{1}$ \cite{deSwart:1963gc},
one can {also} couple the two decuplet baryons to a singlet and then with a singlet
meson to obtain an SU(3) invariant. Again this would introduce a further coupling 
constant which we likewise ignore.

The actual LO meson-exchange potential is obtained {by} multiplying the 
spin-momentum part of the potential, eq.~(\ref{OPE}) with the  
coupling constants as given in eqs.~(\ref{eq:BBC}), (\ref{eq:BDC}), or (\ref{eq:DDC})
and with the pertinent isospin factors, i.e. 
\begin{eqnarray}
V^{\mathcal{B}_1\mathcal{B}_2\to \mathcal{B}_1'\mathcal{B}_2'}_{OBE}&=& \sum_{\pi,\eta,K} 
f_{\mathcal{B}_1\mathcal{B}_1'P}f_{\mathcal{B}_2\mathcal{B}_2'P} \, 
\mathcal{I}_P^{\mathcal{B}_1\mathcal{B}_2\to \mathcal{B}_1'\mathcal{B}_2'} \, V_{P} \ , 
\label{eq:15}
\end{eqnarray}
where $\mathcal{B}$ stands here for octet or decuplet baryons. 

A complete list of isospin factors for the $DD$ case is provided in 
sec.~\ref{sec:Results} where $DD$ scattering results 
are presented for various channels and partial waves. 
{Here}, we refrain from listing those for $BD$ {scattering} and for the $BB\to BD$, $BB\to DD$, 
and $BD\to DD$ transitions. {The} ones for diagrams 
involving $\pi$ and/or $\eta$ exchange can be straightforwardly calculated by using 
eq.~(27) of ref.~\cite{Wiringa}. Furthermore, many coefficients can be also inferred from the tables in 
refs.~\cite{Polinder:2006zh,Polinder:2007mp} since for transitions of baryons with the
same isospins in the initial- and final states, the isospin coefficients are also the
same. Note, however that symmetrization factors ($\sqrt{2}$ and/or a $(-1)^{I-I_1-I_2}$ from
recoupling) {may} differ and that has to be taken into account appropriately. 
Some isospin coefficients for $BB\to BD$ can be found in refs.~\cite{Holzenkamp:1989,Holzenkamp}. 

An explicit representation of the {potentials} in the particle basis can be found in {Appendix}~\ref{app:BBpart}. This form of the potential has been used, {in particular}, to obtain the SU(3) relations in 
Appendix~\ref{app:SU3}, starting from the contact term Lagrangians.

\subsection{Partial wave decomposition of the meson-exchange contribution} 

The partial wave projection of the meson-exchange contribution to the potential is {performed using} 
the helicity basis, see refs.~\cite{Holzenkamp,Erkelenz:1971} or Appendix B of \cite{Holzenkamp:1989}. The partial-wave projected potential is given by 
\begin{eqnarray} 
&&\langle \lambda_1'\lambda_2'| V^J(p',p) |\lambda_1\lambda_2  \rangle \nonumber \\
&&= 2\pi \int_{-1}^{+1} d(\cos\theta) d^J_{\lambda,\lambda'}(\theta ) 
\langle \lambda_1'\lambda_2'| V(\vec p',\vec p) |\lambda_1\lambda_2  \rangle \ , 
\end{eqnarray}
with $\lambda=\lambda_1-\lambda_2$, $\lambda'=\lambda_1'-\lambda_2'$. 
Here $|\lambda_i\rangle$ and $|\lambda_i'\rangle$ represent the helicity states of 
the incoming and outgoing baryons. The total angular momentum is denoted by $J$. 
The quantities $d^J_{\lambda,\lambda'}(\theta )$ are the reduced rotation
matrices. We calculate those numerically based on the definition of the $d$-functions 
in terms of Jacobi polynomials \cite{Edmonds}. {Analytic} expressions for the $d$-functions needed for $J$ up to $2$ 
can be found, e.g. in ref.~\cite{vanFaassen:1983}. An incomplete set of the 
$d^J_{\lambda,\lambda'}$'s with $\lambda = 3$ {is listed} in ref.~\cite{vanFaassen:1984}. 

The amplitudes in the standard $JMLS$ basis are then {constructed} via a
unitary transformation, also described in Appendix B of \cite{Holzenkamp:1989}: 
\begin{eqnarray} 
&&\langle JML'S'| V^J(p',p) |JMLS  \rangle =  \sum_{\lambda_1\lambda_2,\lambda_1'\lambda_2'}\\
&&\langle JML'S'| JM\lambda_1'\lambda_2' \rangle 
  \langle \lambda_1'\lambda_2'| V^J(p',p) |\lambda_1\lambda_2  \rangle 
  \langle JM\lambda_1\lambda_2 |JMLS  \rangle,  \nonumber 
\end{eqnarray}
with 
\begin{equation} 
\langle JMLS|JM\lambda_1\lambda_2\rangle = \sqrt{\frac{2L+1}{2J+1}}
\langle L0S\lambda|J\lambda\rangle
\langle s_1\lambda_1 s_2 -\lambda_2|S\lambda\rangle,
\end{equation} 
where $\lambda = \lambda_1 -\lambda_2$
and $\langle s_1\lambda_1 s_2\lambda_2 | S\lambda\rangle$ denotes a
 Clebsch-Gordan coefficient.
 
The wave function of the decuplet baryons is constructed from 
the standard Rarita-Schwinger spinor  
\begin{equation}
\psi^\mu (\vec p, \Lambda) = \sum_{\lambda_1,\lambda_2}
\langle 1\lambda_1 \frac{1}{2} \lambda_2 | \frac{3}{2} \Lambda \rangle 
\epsilon^\mu (\vec p, \lambda_1) u (\vec p, \lambda_2) \ . 
\end{equation}

In the non-relativistic (static) approach that we follow here, the 
polarization vectors $\epsilon^\mu(\vec p, \lambda_1)$ reduce to \cite{Hemmert:1997,Brown} 
\begin{equation} 
\epsilon^i (\vec p,\pm 1) = \frac{1}{\sqrt{2}} 
\left(
\begin{array}{c}
\mp 1 \\
-i \\
0 \\
\end{array}
\right), \ 
\epsilon^i (\vec p,0) = 
\left(
\begin{array}{c}
0 \\
0 \\
1 \\
\end{array}
\right)
\end{equation} 
and $u (\vec p, \lambda_2)$ {reduces} to a two-component spinor $\chi(\lambda_2)$. 

Then the matrix element of the operator $\mbox{\boldmath{$\Sigma$}} \cdot \vec q$ 
representing the spin structure of the $DD\phi$ vertex required for the
$DD\to DD$, $ND\to ND$ and $ND\to DD$ transition potentials 
is given in {the} helicity basis by
\begin{eqnarray}
\nonumber
\langle \Lambda' | \ \mbox{\boldmath{$\Sigma$}} \cdot {\vec q} \ | \Lambda \rangle 
&=& 3\sum_{\lambda_{1'},\lambda_{2'},\lambda_1,\lambda_2}
\langle 1\lambda_{1'} \frac{1}{2} \lambda_{2'} | \frac{3}{2} \Lambda'\rangle 
\langle 1\lambda_1 \frac{1}{2} \lambda_2 | \frac{3}{2} \Lambda \rangle \\
\nonumber
&\times&\epsilon^* (\vec p ', \lambda_{1'}) \epsilon (\vec p, \lambda_1) \,
\langle \lambda_{2'} | \ {\vec \sigma}\cdot {\vec q} \ | \lambda_{2} \rangle .
\label{helicity}
\end{eqnarray}
The helicity matrix elements for $\mbox{\boldmath $\sigma$} \cdot \vec q$ can be found, 
e.g., in ref.~\cite{Erkelenz:1971}. 
Analytic expressions {of} helicity amplitudes for the transitions $BB\to BD$ and $BB\to DD$ 
that involve the operators $\mbox{\boldmath{$S$}} \cdot \vec q$ and/or
$\mbox{\boldmath{$S^\dagger$}} \cdot \vec q$
can {easily be} deduced from those in refs.~\cite{Holzenkamp,Holinde:1977}.
Note that an alternative method to perform
the partial-wave projections, which exploits rotational invariance  by
averaging over the total angular momentum projection $M$, has been
formulated in ref. \cite{Golak:2009}.

\subsection{Lagrangians {involving} contact terms} \label{sec:ctLagr}

The construction of the {minimal Lagrangian collecting the contact} terms at leading order for the various combinations of octet 
and decuplet baryons is achieved by writing down all chirally invariant flavor structures combined with all possible spin 
structures.
{This} minimal set is obtained by eliminating redundant terms until the rank of the matrix formed by all transitions matches the number of terms in the Lagrangian.
Redundant terms are deleted in such a way that one obtains a maximal number of {contact} terms with different spin structures and a minimal number of terms with different flavor structures%
\footnote{
In the case of \(DB\to DB\) we eliminated the ``exchange'' spin structures \(\vec S\otimes \vec S^{\dagger}\) 
and \(S^{\alpha\beta}\otimes S^{\alpha\beta\dagger}\) by including a larger number of flavor structures.
}.

In the following we present the minimal leading-order Lagrangians in the non-relativistic limit {and rewritten in terms of particle fields, using} conventional matrix notation.
For the low-energy constants \(c^f_{ij}\) of the contact terms, the subscript $ij$ denotes the number of decuplet baryons 
in the initial and final state.
Note that Lagrangian terms for \(\mathcal{L}_\mathrm{BBBB}\) can be found in \cite{Polinder:2006zh,Petschauer:2013} 
and the terms for \(\mathcal{L}_\mathrm{DBBB}\) have been constructed in \cite{Petschauer:2016pbn}.
We have:

\begin{align*}
\mathcal{L}_\mathrm{BBBB} = &\quad\,
\Czz^1\sum_{a,b,c,d=1}^3 \left(\bar B_{ab} B_{bc}\right)\left(\bar B_{cd} B_{da}\right) \\
&+ \Czz^2\sum_{a,b,c,d=1}^3 \left(\bar B_{ab} \vec\sigma B_{bc}\right)\cdot\left(\bar B_{cd} \vec\sigma B_{da}\right) \\
&+ \Czz^3\sum_{a,b,c,d=1}^3 \left(\bar B_{ab} B_{cd}\right)\left(\bar B_{bc} B_{da}\right) \displaybreak[0] \\
&+ \Czz^4\sum_{a,b,c,d=1}^3 \left(\bar B_{ab} \vec\sigma B_{cd}\right)\cdot\left(\bar B_{bc} \vec\sigma B_{da}\right) \\
&+ \Czz^5\sum_{a,b,c,d=1}^3 \left(\bar B_{ab} B_{ba}\right)\left(\bar B_{cd} B_{dc}\right) \\
&+ \Czz^6\sum_{a,b,c,d=1}^3 \left(\bar B_{ab} \vec\sigma B_{ba}\right)\cdot\left(\bar B_{cd} \vec\sigma B_{dc}\right)
\, \ .  \numberthis
\end{align*}
\begin{align*}
\mathcal{L}_\mathrm{DBBB} = &\quad\,
\Czo^1\sum_{a,b,c,d,e,f=1}^3 \epsilon_{abc}
\big[
\left(\bar T_{ade}\vec S^\dagger B_{db}\right)\cdot\left(\bar B_{fc}\vec\sigma B_{ef}\right)+ \\
&\qquad\qquad\qquad\qquad\qquad
\left(\bar B_{bd}\vec S\, T_{ade}\right)\cdot\left(\bar B_{fe}\vec\sigma B_{cf}\right)
\big]\\
&+\Czo^2
\sum_{a,b,c,d,e,f=1}^3 \epsilon_{abc}
\big[
\left(\bar T_{ade}\vec S^\dagger B_{fb}\right)\cdot\left(\bar B_{dc}\vec\sigma B_{ef}\right)+ \\
&\qquad\qquad\qquad\qquad\qquad
\left(\bar B_{bf}\vec S\, T_{ade}\right)\cdot\left(\bar B_{fe}\vec\sigma B_{cd}\right)
\big] \,. \numberthis
\end{align*}
\begin{align*}
\mathcal{L}_\mathrm{DBDB} = &\quad\,
\Coo^1\sum_{a,b,c,d,e=1}^3 \left(\bar T_{abc} T_{abc}\right)\left(\bar B_{de} B_{ed}\right)\\
&+ \Coo^2\sum_{a,b,c,d,e=1}^3 \left(\bar T_{abc} \vec\Sigma\, T_{abc}\right)\cdot\left(\bar B_{de} \vec\sigma\, B_{ed}\right) \\
&+\Coo^3\sum_{a,b,c,d,e=1}^3 \left(\bar T_{abc} T_{abd}\right)\left(\bar B_{ce} B_{ed}\right)\\
&+ \Coo^4\sum_{a,b,c,d,e=1}^3 \left(\bar T_{abc} \vec\Sigma\, T_{abd}\right)\cdot\left(\bar B_{ce} \vec\sigma\, B_{ed}\right) \displaybreak[0] \\
&+\Coo^5\sum_{a,b,c,d,e=1}^3 \left(\bar T_{abc} T_{abd}\right)\left(\bar B_{ed} B_{ce}\right)\\
&+ \Coo^6\sum_{a,b,c,d,e=1}^3 \left(\bar T_{abc} \vec\Sigma\, T_{abd}\right)\cdot\left(\bar B_{ed} \vec\sigma\, B_{ce}\right) \\
&+\Coo^7\sum_{a,b,c,d,e=1}^3 \left(\bar T_{abc} T_{ade}\right)\left(\bar B_{bd} B_{ce}\right)\\
&+ \Coo^8\sum_{a,b,c,d,e=1}^3 \left(\bar T_{abc} \vec\Sigma\, T_{ade}\right)\cdot\left(\bar B_{bd} \vec\sigma\, B_{ce}\right)
\,. \numberthis
\end{align*}
\begin{align*}
&\mathcal{L}_\mathrm{DDBB} = \\
&=
\Czt^1\sum_{a,b,c,d,e,f,g,h=1}^3 \epsilon_{abc}\epsilon_{def}
\big[
\left(\bar T_{adg}\vec S^\dagger B_{gc}\right)\cdot\left(\bar T_{beh}\vec S^\dagger B_{hf}\right) \\
&\qquad\qquad\qquad\qquad+
\left(\bar B_{cg}\vec S\, T_{adg}\right)\cdot\left(\bar B_{fh}\vec S\, T_{beh}\right)
\big]\\
&+\Czt^2
\sum_{a,b,c,d,e,f,g,h=1}^3 \epsilon_{abc}\epsilon_{def}
\big[
\left(\bar T_{adg}S^{\alpha\beta\dagger} B_{gc}\right)\left(\bar T_{beh}S^{\alpha\beta\dagger} B_{hf}\right) \\
&\qquad\qquad\qquad+
\left(\bar B_{cg}S^{\alpha\beta} T_{adg}\right)\left(\bar B_{fh}S^{\alpha\beta} T_{beh}\right)
\big] \,. \numberthis
\end{align*}
\begin{align*}
\mathcal{L}_\mathrm{DDDB} = &\quad\,
\Cot^1\sum_{a,b,c,d,e,f,g=1}^3 \epsilon_{abc}
\big[
\left(\bar T_{ade}\vec\Sigma T_{def}\right)\cdot\left(\bar T_{bfg}\vec S^\dagger B_{gc}\right) \\
&\qquad\qquad\qquad\qquad+
\left(\bar T_{def}\vec\Sigma T_{ade}\right)\cdot\left(\bar B_{cg}\vec S\, T_{bfg}\right)
\big]\\
&+\Cot^2\sum_{a,b,c,d,e,f,g=1}^3 \epsilon_{abc}
\big[
\left(\bar T_{ade}\Sigma^{\alpha\beta} T_{def}\right)\left(\bar T_{bfg}S^{\alpha\beta\dagger} B_{gc}\right) \\
&\qquad\qquad\qquad\qquad+
\left(\bar T_{def}\Sigma^{\alpha\beta} T_{ade}\right)\left(\bar B_{cg}S^{\alpha\beta} T_{bfg}\right)
\big] \,. \numberthis
\end{align*}
\begin{align*}
\mathcal{L}_\mathrm{DDDD} = &\phantom{ + }
\Ctt^1\sum_{a,b,c,d,f=1}^3 \left(\bar T_{abc} T_{abc}\right)\left(\bar T_{def} T_{def}\right) \\
&+ \Ctt^2\sum_{a,b,c,d,f=1}^3 \left(\bar T_{abc} \vec\Sigma T_{abc}\right)\cdot\left(\bar T_{def} \vec\Sigma T_{def}\right) \\
\ &+ \Ctt^3\sum_{a,b,c,d,f=1}^3 \left(\bar T_{abc} \Sigma^{\alpha\beta} T_{abc}\right)\left(\bar T_{def} \Sigma^{\alpha\beta} T_{def}\right) \\
&+ \Ctt^4\sum_{a,b,c,d,f=1}^3 \left(\bar T_{abc} \Sigma^{\alpha\beta\gamma} T_{abc}\right)\left(\bar T_{def} \Sigma^{\alpha\beta\gamma} T_{def}\right) \displaybreak[0] \\
\ &+ \Ctt^5\sum_{a,b,c,d,f=1}^3 \left(\bar T_{abc} T_{abd}\right)\left(\bar T_{def} T_{cef}\right) \\
&+ \Ctt^6\sum_{a,b,c,d,f=1}^3 \left(\bar T_{abc} \vec\Sigma T_{abd}\right)\cdot\left(\bar T_{def} \vec\Sigma T_{cef}\right) \\
\ &+ \Ctt^7\sum_{a,b,c,d,f=1}^3 \left(\bar T_{abc} \Sigma^{\alpha\beta} T_{abd}\right)\left(\bar T_{def} \Sigma^{\alpha\beta} T_{cef}\right) \\
&+ \Ctt^8\sum_{a,b,c,d,f=1}^3 \left(\bar T_{abc} \Sigma^{\alpha\beta\gamma} T_{abd}\right)\left(\bar T_{def} \Sigma^{\alpha\beta\gamma} T_{cef}\right)
\,. \numberthis
\end{align*}

\subsection{Group theoretical considerations} \label{sec:group}

Group {theory} can be used to deduce the number of {terms} in the chiral contact Lagrangian.
First of all we need to express the two-baryon states in terms of irreducible representations of 
SU(3) (flavor) and SU(2) (spin).
For octet-octet baryon states \(BB\) the decompositions in flavor and spin space {are, respectively}:
\begin{equation}
\mathbf8\otimes\mathbf8 =
{\mathbf{27}}\oplus{\mathbf{8_s}}\oplus{\mathbf1}\oplus
\mathbf{10}\oplus\mathbf{\overline{10}}\oplus\mathbf{8_a}\,,\
\mathbf{2} \otimes \mathbf{2} = \mathbf{1} \oplus \mathbf{3} \,,
\end{equation}
where in flavor space the representations \(\mathbf{27},{\mathbf{8_s}},{\mathbf1}\) are symmetric and 
the representations \(\mathbf{10},\mathbf{\overline{10}},\mathbf{8_a}\) are antisymmetric. 
In spin space \(\mathbf{1}\) is antisymmetric and \(\mathbf{3}\) is symmetric.
For decuplet-octet baryon states \(DB\) one obtains
\begin{equation}
 \mathbf{10}\otimes\mathbf8 = {\mathbf{35}}\oplus{\mathbf{27}}\oplus\mathbf{10}\oplus\mathbf{8}\,,\
 \mathbf{4} \otimes \mathbf{2} = \mathbf{3} \oplus \mathbf{5} \,.
\end{equation}
Decuplet-decuplet baryon states \(DD\) take the form
\begin{equation}
\mathbf{10}\otimes\mathbf{10} = {\mathbf{35}}\oplus\mathbf{\overline{10}}\oplus{\mathbf{28}}\oplus\mathbf{27}\,,\
\mathbf{4} \otimes \mathbf{4} =\mathbf{1} \oplus \mathbf{5} \oplus \mathbf{3} \oplus \mathbf{7} \,.
\end{equation}

Combining this information with the Pauli principle 
(for $\mathbf8\otimes\mathbf8$ and $\mathbf{10}\otimes\mathbf{10}$) 
and the fact that, at leading order, only transitions between the same flavor and spin irreducible representations can occur, we can write down the possible combinations of flavor and spin representations (flavor,spin) in which the transition can occur:
\begin{align*}
BB\to BB\colon&\ (\ir{27},\ir{1}),(\ir{8_s},\ir{1}),(\ir{1},\ir{1}),(\ir{10},\ir{3}),(\ir{10^*},\ir{3}),(\ir{8_a},\ir{3}) \\
BB\to DB\colon&\ (\ir{10},\ir{3}),(\ir{8},\ir{3}) \\
BB\to DD\colon&\ (\ir{27},\ir{1}),(\ir{\overline{10}},\ir{3}) \\
DB\to DB\colon&\ (\ir{35},\ir{3}),(\ir{27},\ir{3}),(\ir{10},\ir{3}),(\ir{8},\ir{3}),(\ir{35},\ir{5}),(\ir{27},\ir{5}),\\ &\ (\ir{10},\ir{5}),(\ir{8},\ir{5}) \\
DB\to DD\colon&\ (\ir{35},\ir{3}),(\ir{27},\ir{5}) \\
DD\to DD\colon&\ (\ir{35},\ir{3}),(\ir{35},\ir{7}),(\ir{\overline{10}},\ir{3}),(\ir{\overline{10}},\ir{7}),(\ir{28},\ir{1}),(\ir{28},\ir{5}), \\ &\ (\ir{27},\ir{1}),(\ir{27},\ir{5})
\end{align*}
The {resulting} number of different (flavor,spin) representations 
is 6 for $BB\to BB$, 8 for $DB\to DB$, 8 for $DD\to DD$, 
and 2 each for the transitions  $BB\to DB$,  $BB\to DD$, and  $DB\to DD$. 
This fits well to the number of low-energy constants of the minimal Lagrangians 
in sec.~\ref{sec:ctLagr}. It serves as a non-trivial check of our calculation.

{The SU(3) relations for the various two-body \(S\)-wave contact 
interactions} among octet and decuplet baryons {are summarized in Appendix C}.
Furthermore, the relations between the LECs of the Lagrangians and the irreducible representations are given.
The SU(3) relations for the decuplet-decuplet interaction can be found already in {Table}~\ref{tab:PWDDDDD} as they are of {prime} interest for {applications discussed in} the following sections. 
All {those} relations are relevant for checking the {consistent} construction of the {LO Lagrangian 
and potentials}.
{By an approach analogous to the one} in ref.~\cite{Petschauer:2015elq}, we can establish which 
irreducible representations contribute to a specific strangeness-isospin channel.
{These} relations have to conform with corresponding results that follow directly from the isoscalar factors of the 
$\mathbf8 \otimes \mathbf8$, $\mathbf{10} \otimes \mathbf8$, and $\mathbf{10} \otimes \mathbf{10}$
representations given in ref.~\cite{deSwart:1963gc}\footnote{Note in particular the symmetry relations 
between $\mathbf{8} \otimes \mathbf{10}$ and $\mathbf{10} \otimes \mathbf{8}$ as given in 
Table~I of ref.~\cite{deSwart:1963gc}.}. 
{They are also important} for actual calculations as {they specify}
how different strangeness and isospin channels are connected with each other 
via SU(3) symmetry. This aspect will be {further} exploited in sec.~\ref{sec:Results}.  

\begin{table*}
\centering
\begin{tabular}{ttttt}
\toprule
S & I & \text{transition} & V_i\,,\ i\in\{{}^1S_{0},{}^5S_{2}\} & V_i\,,\ i\in\{{}^3S_{1},{}^7S_{3}\} \\
\cmidrule(lr){1-3}\cmidrule(lr){4-5}
0 & 0 & \Delta \Delta \leftrightarrow \Delta \Delta & 0 & \GTtt^{\overline{10},i}  \\ 
0 & 1 & \Delta \Delta \leftrightarrow \Delta \Delta & \GTtt^{27,i} & 0 \\ 
0 & 2 & \Delta \Delta \leftrightarrow \Delta \Delta & 0 & \GTtt^{35,i} \\ 
0 & 3 & \Delta \Delta \leftrightarrow \Delta \Delta & \GTtt^{28,i} & 0 \\ 

\cmidrule(lr){1-3}\cmidrule(lr){4-5}
-1 & \frac{1}{2} & \Sigma^*  \Delta \leftrightarrow \Sigma^*  \Delta & \GTtt^{27,i} & \GTtt^{\overline{10},i}  \\ 
-1 & \frac{3}{2} & \Sigma^*  \Delta \leftrightarrow \Sigma^*  \Delta & \GTtt^{27,i} & \GTtt^{35,i} \\ 
-1 & \frac{5}{2} & \Sigma^*  \Delta \leftrightarrow \Sigma^*  \Delta & \GTtt^{28,i} & \GTtt^{35,i} \\ 

\cmidrule(lr){1-3}\cmidrule(lr){4-5}
-2 & 0 & \Sigma^*  \Sigma^*  \leftrightarrow \Sigma^*  \Sigma^*  & \GTtt^{27,i} & 0 \\ 
-2 & 1 & \Xi^*  \Delta \leftrightarrow \Xi^*  \Delta & \GTtt^{27,i} & \frac{1}{3} (2 \GTtt^{\overline{10},i} +\GTtt^{35,i}) \\ 
-2 & 1 & \Sigma^*  \Sigma^*  \leftrightarrow \Xi^*  \Delta & 0 & \frac{1}{3} \sqrt{2} (\GTtt^{35,i}-\GTtt^{\overline{10},i} ) \\ 
-2 & 1 & \Sigma^*  \Sigma^*  \leftrightarrow \Sigma^*  \Sigma^*  & 0 & \frac{1}{3} (\GTtt^{\overline{10},i} +2 \GTtt^{35,i}) \\ 
-2 & 2 & \Xi^*  \Delta \leftrightarrow \Xi^*  \Delta & \frac{1}{5} (3 \GTtt^{27,i}+2 \GTtt^{28,i}) & \GTtt^{35,i} \\ 
-2 & 2 & \Sigma^*  \Sigma^*  \leftrightarrow \Xi^*  \Delta & \frac{1}{5} \sqrt{6} (\GTtt^{28,i}-\GTtt^{27,i}) & 0 \\ 
-2 & 2 & \Sigma^*  \Sigma^*  \leftrightarrow \Sigma^*  \Sigma^*  & \frac{1}{5} (2 \GTtt^{27,i}+3 \GTtt^{28,i}) & 0 \\ 

\cmidrule(lr){1-3}\cmidrule(lr){4-5}
-3 & \frac{1}{2} & \Xi^*  \Sigma^*  \leftrightarrow \Xi^*  \Sigma^*  & \GTtt^{27,i} & \GTtt^{35,i} \\ 
-3 & \frac{3}{2} & \Omega \Delta \leftrightarrow \Omega \Delta & \frac{1}{10} (9 \GTtt^{27,i}+\GTtt^{28,i}) & \frac{\GTtt^{\overline{10},i} +\GTtt^{35,i}}{2} \\ 
-3 & \frac{3}{2} & \Xi^*  \Sigma^*  \leftrightarrow \Omega \Delta & -\frac{3}{10} (\GTtt^{27,i}-\GTtt^{28,i}) & \frac{\GTtt^{35,i}-\GTtt^{\overline{10},i} }{2} \\ 
-3 & \frac{3}{2} & \Xi^*  \Sigma^*  \leftrightarrow \Xi^*  \Sigma^*  & \frac{1}{10} (\GTtt^{27,i}+9 \GTtt^{28,i}) & \frac{\GTtt^{\overline{10},i} +\GTtt^{35,i}}{2} \\ 

\cmidrule(lr){1-3}\cmidrule(lr){4-5}
-4 & 0 & \Xi^*  \Xi^*  \leftrightarrow \Xi^*  \Xi^*  & 0 & \GTtt^{35,i} \\ 
-4 & 1 & \Omega \Sigma^*  \leftrightarrow \Omega \Sigma^*  & \frac{1}{5} (3 \GTtt^{27,i}+2 \GTtt^{28,i}) & \GTtt^{35,i} \\ 
-4 & 1 & \Xi^*  \Xi^*  \leftrightarrow \Omega \Sigma^*  & \frac{1}{5} \sqrt{6} (\GTtt^{28,i}-\GTtt^{27,i}) & 0 \\ 
-4 & 1 & \Xi^*  \Xi^*  \leftrightarrow \Xi^*  \Xi^*  & \frac{1}{5} (2 \GTtt^{27,i}+3 \GTtt^{28,i}) & 0 \\ 

\cmidrule(lr){1-3}\cmidrule(lr){4-5}
-5 & \frac{1}{2} & \Omega \Xi^*  \leftrightarrow \Omega \Xi^*  & \GTtt^{28,i} & \GTtt^{35,i} \\ 

\cmidrule(lr){1-3}\cmidrule(lr){4-5}
-6 & 0 & \Omega \Omega \leftrightarrow \Omega \Omega & \GTtt^{28,i} & 0 \\ 

\bottomrule
\end{tabular}
\caption{SU(3) relations of \(DD\to DD\) in non-vanishing partial waves. 
The subscript $\{22\}$ of the constants \(\Gtt^r\) that denotes the
$\mathcal{B}\mathcal{B}$ channel is omitted in the table.
} \label{tab:PWDDDDD}
\end{table*}

\subsection{Partial wave decomposition of contact spin structures} 
\label{sec:pwd}

In the following we present the explicit forms of the contact potentials \(V_{cont}\)
involving the most general two-body spin operators for the non-vanishing transitions 
between partial waves \({}^{2S+1}L_J\) at leading order.
Following the method described in ref.~\cite{Skibinski:2011db} one {finds} (with constants \(a_i\)):

\vskip 0.2cm 
\(BB\to BB: \ V_{cont} = a_1\ONE + a_2\vec\sigma_1\cdot\vec\sigma_2\)
\begin{align*}
V_{^1S_0} = \langle{}^1S_0|V_{cont}|{}^1S_0\rangle &= a_1 - 3a_2 \\
V_{^3S_1} = \langle{}^3S_1|V_{cont}|{}^3S_1\rangle &= a_1 + a_2
\end{align*}

\(BB\leftarrow DB: \ V_{cont} = a_1 {\vec S}^\dagger_1 \cdot \vec\sigma_2 \\
 ~~~~~~~ BB\rightarrow DB: \ V_{cont} = a_1 {\vec S}_1\cdot\vec\sigma_2\)
\begin{align*}
V_{^3S_1} = \langle{}^3S_1|V_{cont}|{}^3S_1\rangle &= -2\sqrt{\frac23}a_1
\end{align*}

\(BB\leftarrow DD: \ V_{cont} = a_1{\vec S}^\dagger_1\cdot{\vec S}^\dagger_2 + a_2S^{ij\dagger}_1S^{ij\dagger}_2 \\
 ~~~~~~~ BB\rightarrow DD: \ V_{cont} = a_1{\vec S}_1\cdot{\vec S}_2 + a_2S^{ij}_1S^{ij}_2\)
\begin{align*}
V_{^1S_0} = \langle{}^1S_0|V_{cont}|{}^1S_0\rangle &= -\sqrt2 a_1 + \frac{5\sqrt2}3a_2 \\
V_{^3S_1} = \langle{}^3S_1|V_{cont}|{}^3S_1\rangle &= -\frac{\sqrt{10}}3 ( a_1 + a_2 )
\end{align*}

\(DB\to DB: \ V_{cont} = a_1\ONE + a_2\vec\Sigma_1\cdot\vec\sigma_2\)
\begin{align*}
V_{^3S_1} =\langle{}^3S_1|V_{cont}|{}^3S_1\rangle &= a_1 - 5 a_2 \\
V_{^5S_2} =\langle{}^5S_2|V_{cont}|{}^5S_2\rangle &= a_1 + 3 a_2
\end{align*}

\(DB\to BD: \ V_{cont} = a_1{\vec S}_1\cdot{\vec S}^\dagger_2 + a_2S^{ij}_1S^{ij\dagger}_2\)
\begin{align*}
V_{^3S_1}= \langle{}^3S_1|V_{cont}|{}^3S_1\rangle &= \frac13a_1 - \frac53 a_2 \\
V_{^5S_2}= \langle{}^5S_2|V_{cont}|{}^5S_2\rangle &= a_1 + \frac13 a_2
\end{align*}

\(DB\leftarrow DD: \ V_{cont}= a_1{\vec \Sigma}_1\cdot{\vec S}^\dagger_2 + a_2\Sigma^{ij}_1S^{ij\dagger}_2\\ 
 ~~~~~~~  DB\rightarrow DD: \ V_{cont}=a_1{\vec \Sigma}_1\cdot{\vec S}_2 + a_2\Sigma^{ij}_1S^{ij}_2\)
\begin{align*}
V_{^3S_1}=\langle{}^3S_1|V_{cont}|{}^3S_1\rangle &= 2\sqrt{\frac53} a_1 + \sqrt{10} a_2 \\
V_{^5S_2}=\langle{}^5S_2|V_{cont}|{}^5S_2\rangle &= 2\sqrt3 a_1 - \sqrt{2} a_2
\end{align*}

\(DD\to DD: \ V_{cont} = a_1\ONE + a_2\vec\Sigma_1\cdot\vec\Sigma_2 + 
a_3\Sigma^{ij}_1\Sigma^{ij}_2 + a_4\Sigma^{ijk}_1\Sigma^{ijk}_2\)
\begin{align*}
V_{^1S_0}=\langle{}^1S_0|V_{cont}|{}^1S_0\rangle &= a_1 - 15a_2 + \frac{15}{2}a_3 -\frac{350}{3}a_4\\
V_{^3S_1}=\langle{}^3S_1|V_{cont}|{}^3S_1\rangle &= a_1 - 11a_2 + \frac{3}{2}a_3 +70a_4\\
V_{^5S_2}=\langle{}^5S_2|V_{cont}|{}^5S_2\rangle &= a_1 - 3a_2 - \frac{9}{2}a_3 -\frac{70}3a_4\\
V_{^7S_3}=\langle{}^7S_3|V_{cont}|{}^7S_3\rangle &= a_1 + 9a_2 + \frac{3}{2}a_3 +\frac{10}3a_4\\
\end{align*}

The actual potential for a specific channel and partial wave is a combination of low-energy constants 
according to Table~\ref{tab:PWDDDDD} and Tables~\ref{tab:PWDBBBB} - \ref{tab:PWDDBDD} in Appendix \ref{app:SU3}.


\section{Application to lattice QCD results} 
\label{sec:Results}

In this section we exemplify how the formalism developed above 
can be used to analyze results from lattice QCD {computations}. 
{It should be understood that} this study has primarily illustrative character.
{Lattice results for interactions involving decuplet baryons are so far restricted to unphysically large pion masses. It is obvious that LO chiral EFT can give only a qualitative (but nonetheless instructive) picture. More elaborate treatments, at next-to-leading order and beyond, will be feasible as lattice simulations proceed (close) to physical masses and become increasingly accurate.}

The reaction amplitude for a specific potential 
$V=V_{OBE}+V_{cont}$ is obtained by solving a corresponding LS 
equation in partial-wave projected form \cite{Polinder:2006zh},
\begin{eqnarray}
&&T^{L'' L'}(p'',p';E)=V^{L''L'}(p'',p')+
\nonumber\\&&
\sum_{L}\int_0^\infty \frac{dpp^2}{(2\pi)^3} \, V^{L''L}(p'',p)
\frac{2\mu}{k^2-p^2+i\eta}T^{LL'}(p,p';E)\ ,
\nonumber\\&&
\label{LS} 
\end{eqnarray}
where the tensor coupling between different orbital angular momenta 
$L''$, $L'$ is taken into account. 
$\mu$ is the reduced mass, i.e. $\mu = M_1 M_2/(M_1+ M_2)$, with $M_1$, $M_2$ being the 
masses of the baryons in the intermediate state. The on-shell momentum in the intermediate 
state, $k$, and the kinetic energy $E$ in the center-of-mass frame  
are given by the relation $\sqrt{s} = E + M_1 + M_2 = \sqrt{M^2_1+k^2}+\sqrt{M^2_2+k^2}$. 
The scattering amplitude (\ref{LS}) generally includes couplings between different $\mathcal{B}\mathcal{B}$
channels \cite{Polinder:2006zh}. 
{In} the applications described in this section we consider systems without such
{coupled channels, hence} corresponding indices and summations in eq.~(\ref{LS}) {have been omitted}.
For the simulation of the lattice QCD results the respective baryon masses as given
in the corresponding publications are employed. These will be given {in each case considered as we proceed}. 
In the calculations at the physical point we use the following (isospin averaged) 
baryon masses: $M_N=938.92$~MeV, $M_\D=1232$~MeV, $M_\SiS=1385$~MeV, $M_\XiS=1530$~MeV, 
and $M_\Om=1672.45$~MeV. The LS equation is solved in the isospin basis and the 
Coulomb interaction is ignored (as in the lattice QCD calculations).
 
{The} integral in the LS equation (\ref{LS}) is divergent for the chiral potentials
specified above. {A regularization needs to introduced}. We utilize here the same prescription 
as in our $YN$ studies, where the potentials in the LS equation are cut off {in momentum space by multiplication} with a regulator function, 
$f(p',p) = \exp\left[-\left(p'^4+p^4\right)/\Lambda^4\right]$, 
so that the {high-momentum} components of the baryon and pseudoscalar meson fields are removed. 
In the present study we employ the cut-off {scales} $500$ and $700$ MeV, in line with the 
range considered in our LO study of the $\La N$ and $\Si N$ interactions \cite{Polinder:2006zh}. 
The variation of the results with the cutoff {reflect uncertainties that} will be indicated by bands. 
Clearly, a better error analysis based e.g. on the approach advocated in ref.~\cite{Epelbaum:2014efa}
should be done once more lattice data and/or higher order calculations are available.

At LO the only dependence of the potential $V$ on the pion {mass or on other meson masses comes from} 
the meson propagators in eq.~(\ref{OPE}). The SU(3) breaking manifested in 
the masses of the octet and/or decuplet baryons does not affect the potential itself 
at this order \cite{Polinder:2006zh}.  
However, those masses enter the LS equation (\ref{LS}) and, therefore, influence
the actual result for the reaction amplitude $T$.
It is important to take this effect into account, as argued in 
refs.~\cite{Haidenbauer:2011ah,Haidenbauer:2011za}. After all, the binding energy of
a possible bound state results from a delicate interplay between the potential energy 
(that depends on the pion mass) and the kinetic energy in the baryon-baryon Green's
function (that is affected by the baryon masses). 
 
{At} next-to-leading order the contact terms as well as the coupling 
constants depend on the quark masses or, equivalently, on 
the pion mass. For details, we refer to 
refs.~\cite{Beane:2002vs,Beane03,Epe02,Epe02a,Baru:2015,Baru:2016}, 
where the quark mass dependence of the $NN$ interaction has been investigated;
see also ref.~\cite{Petschauer:2013}. {However, as mentioned, for the present limited applications it is sufficient to stay at the LO level of single pseudoscalar meson exchange plus contact term with a much restricted set of low-energy constants.} {For} $NN$ 
scattering~\cite{Beane:2002vs,Beane03,Epe02,Epe02a,Baru:2015,Baru:2016}
the pion-exchange contribution plays an essential role.
In contrast, for some of the systems considered here {such as} $\Om\Om$ or $N\Om$, the 
only {contributor to} pseudoscalar-meson exchange is the $\eta$ meson. 
{Variations} of the pion mass (or the SU(3) breaking due to the 
small pion mass as compared to {the $K$ and $\eta$ masses}) are expected 
to be less important for such systems. 

In the present work we follow closely the strategy utilized in our study/analysis 
of lattice QCD simulations for the $H$-di\-baryon \cite{Haidenbauer:2011ah}
and other possible bound states of two-body systems involving octet baryons in 
the {sectors with} strangeness $S=-2$, $-3$ and $-4$ \cite{Haidenbauer:2011za}.  
{We aim at a reproduction of} the phase shifts and/or scattering lengths from the
{LQCD} calculation within LO chiral EFT. Thereby we {employ} the masses of the 
pseudoscalar mesons and
the baryons corresponding to the lattice simulation in our calculation and 
{fix} the low-energy constant associated with the strength of the contact term
by a fit to the lattice data. 
Results at the physical point are then obtained replacing the masses
of the mesons in the potential by their physical values and correspondingly
those of the baryons in the LS equation. 

\subsection{$\Delta\Delta\pi$ coupling constant}

An essential ingredient of the calculation is the $\Delta\Delta\pi$ coupling constant.
Its value, together with the imposed SU(3) symmetry relations, fixes the 
strengths of all contributions from pseudo\-scalar-meson exchange to the $DD$ interaction. 
Unfortunately, {unlike} the $NN\pi$ and $N\Delta\pi$ coupling constants, the value 
for $f_{\Delta\Delta\pi}$ is not constrained by experimental information. 
{A wide range of} values for the $\Delta\Delta\pi$ coupling constant can be found 
in the literature~\cite{Fettes:2000,Yao:2016,Brown,Slaughter,Dashen:1993,Zhu,Buchmann}. 
In a variety of calculations the {quark model value} \cite{Brown}
is used which amounts to $f_{\D\D\pi} = f_{NN\pi}/5$ \cite{Wiringa}
for the normalization of the spin- and isospin operators {as defined in} Appendix \ref{app:Spin}. 
Large $N_c$ arguments lead to $g_1 = 9g_A/5$ \cite{Fettes:2000,Zhu}, 
where $g_1$ is the $\D\D\pi$ coupling constant commonly used in chiral 
perturbation theory. 
Taking into account the different normalization of the
spin- and isospin operators, this corresponds likewise to $f_{\D\D\pi} = f_{NN\pi}/5$. 
The result in \cite{Zhu} using QCD sum rules is $g_1 = 0.885\pm 0.15$
based on eqs.~(4) and (10) of that work, i.e. roughly half of the large $N_c$ 
prediction. Recent lattice QCD calculations indicate values around $g_1 \approx 0.6$ 
($g_1 \simeq g_A/2$) \cite{Alexandrou:2013} for $m_\pi \approx 300$ MeV.  
In a new study of pion-nucleon scattering within chiral perturbation theory
up to third order that includes the $\Delta$ resonance explicitly \cite{Yao:2016} 
a value $g_1 =1.21\pm 0.46 \pm 0.39$ was deduced. The large uncertainty 
indicates the difficulty to pin down the ${\D\D\pi}$ coupling constant reliably 
from such a calculation.  See, however, ref.~\cite{Gegelia:2016pjm}.

\subsection{Partial waves and isospin factors}

An overview of partial waves up to total angular momentum 
$J=3$ is provided in Table~\ref{tab:pws}.
In general, denoting the states as $|JMLS\rangle$, 
$L$ being the orbital angular momentum and $S$ the total spin,
the $BD$ system (spin $1/2 \otimes 3/2$) can be in the following two sets 
of four states
\begin{eqnarray}
\nonumber
&&|JMJ1\rangle, \ |JMJ2\rangle, \ |JM(J\pm 2)2\rangle; \\
&&|JM(J\pm 1)1\rangle, \ |JM(J\pm 1)2\rangle, 
\end{eqnarray}
which differ by parity and, therefore, do not couple. 
For $DD$ ($3/2 \otimes 3/2$) there are four such sets, 
\begin{eqnarray}
\nonumber
&&|JMJ0\rangle, \ |JMJ2\rangle, \ |JM(J\pm 2)2\rangle; \\
\nonumber
&&|JMJ1\rangle, \ |JMJ3\rangle, \ |JM(J\pm 2)3\rangle; \\
\nonumber
&&|JM(J\pm 1)1\rangle, \ |JM(J \pm 3)3\rangle, \ |JM(J \pm 1)3\rangle; \\
&&|JM(J \pm 1)2\rangle,
\end{eqnarray}
where again there is no coupling between states with different parity. 
The other sets decouple too, as long as the interaction is given only
by pseudoscalar-meson exchange, eq.~(\ref{OPE}). 

\begin{table}[t]
\renewcommand{\arraystretch}{1.2}
\caption{Partial waves for $BB$, $BD$, and $DD$ scattering for angular momenta
$J \le 3$.}
\label{tab:pws}
\centering
\begin{tabular}{|l|l|l|l|}
\hline
$J$  & $BB$ &$BD$ &$DD$ \\
\hline
$0$ & $^1S_0$ & $^5D_0$ & $^1S_0$,$^5D_0$ \\
    & $^3P_0$ & $^3P_0$ & $^3P_0$,$^7F_0$ \\
\hline
$1$ & $^1P_1$ &                          & $^1P_1$,$^5P_1$,$^5F_1$ \\
    & $^3P_1$ & $^3P_1$,$^5P_1$,$^3F_1$ & $^3P_1$,$^7F_1$ \\
    & $^3S_1$,$^3D_1$ & $^3S_1$,$^3D_1$,$^5D_1$ & $^3S_1$,$^3D_1$,$^7D_1$,$^7G_1$ \\
    &                  &                                    & $^5D_1$                            \\
\hline
$2$ & $^1D_2$ & $^5S_2$,$^3D_2$,$^5D_2$,$^5G_2$ & $^5S_2$,$^1D_2$,$^5D_2$,$^5G_2$ \\
    & $^3D_2$ &                                    & $^3D_2$,$^7D_2$,$^7G_2$ \\
    & $^3P_2$,$^3F_2$ & $^3P_2$,$^5P_2$,$^3F_2$,$^5F_2$ & $^3P_2$,$^7P_2$,$^3F_2$,$^7F_2$,$^7H_2$ \\
    &                  &                                    & $^5P_2$,$^5F_2$                            \\
\hline
$3$ & $^1F_3$ &                                    & $^5P_3$,$^1F_3$,$^5F_3$,$^5H_3$ \\
    & $^3F_3$ & $^5P_3$,$^3F_3$,$^5F_3$,$^5H_3$ & $^7P_3$,$^3F_3$,$^7F_3$,$^7H_3$ \\
    & $^3D_3$,$^3G_3$ & $^3D_3$,$^5G_3$,$^5D_3$,$^5G_3$ & $^7S_3$,$^3D_3$,$^7D_3$,$^3G_3$,$^7G_3$,$^7I_3$ \\
    &                  &                                    & $^5D_3$,$^5G_3$                            \\
\hline
\end{tabular}
\renewcommand{\arraystretch}{1.0}
\end{table}

We focus here on (coupled) partial waves that involve 
$S$-wave states, i.e., where contributions from contact terms 
at leading order arise. The $DD$ system can be in the following $S$-wave 
states: $^1S_0$, $^3S_1$, $^5S_2$, $^7S_3$. In channels with identical
particles the Pauli principle reduces the number of possible states.
For example, in the $\Om\Om$ system only the $S$ waves $^1S_0$ and $^5S_2$ 
are allowed. In case of $\Delta\Delta$ the spin-space odd states ($^1S_0$, $^5S_2$) 
can have total isospin $I=1$ and $3$ while the spin-space even states 
($^3S_1$, $^7S_3$) can have total isospin $I=0$ and $2$. 
There is no restriction from the Pauli principle for the $BD$ system 
so that one has the $S$-wave states $^3S_1$ and $^5S_2$  
in all channels. 

\begin{table}[h]
\renewcommand{\arraystretch}{1.4}
\caption{Decuplet-decuplet scattering: isospin factors $I_P$ for the various 
meson exchanges. 
}
\label{tab:2DD}
\centering
\begin{tabular}{|c|r|r|r|r|}
\hline
Channel &Isospin &$\pi$ &$K$ &$\eta$\\
\hline
$\D\D\rightarrow \D\D$ &$0$ &$-15$ &$0$ &$1$ \\
                       &$1$ &$-11$ &$0$ &$1$ \\
                       &$2$ &$-3$  &$0$ &$1$ \\
                       &$3$  &$9$  &$0$ &$1$ \\
\hline
$\SiS\D\rightarrow \SiS\D$ &$\frac{1}{2}$ &$-5$ &$1/3$ &$1$ \\
                           &$\frac{3}{2}$ &$-2$ &$-2/3$ &$1$ \\
                           &$\frac{5}{2}$  &$3$ &$1$ &$1$ \\
\hline
$\XiS\D\rightarrow \XiS\D$     &$1$ &$-5$ &$0$ &$1$ \\
                               &$2$ &$ 3$ &$0$ &$1$ \\
$\SiS\SiS\rightarrow \SiS\SiS$ &$0$ &$-2$ &$0$ &$1$ \\
                               &$1$ &$-1$ &$0$ &$1$ \\
                               &$2$ &$1$ &$0$ &$1$ \\
$\XiS\D  \rightarrow \SiS\SiS$ &$1$ &$ 0$ &$-2/\sqrt{3}$ &$0$ \\
                               &$2$ &$ 0$ &$-2$ &$0$ \\
\hline
$\Om \D\rightarrow \Om \D$ &$\frac{3}{2}$ &$0$ &$0$ &$1$ \\
$\XiS\SiS \rightarrow \XiS\SiS$ &$\frac{1}{2}$ &$-2$ &$-1$ &$1$ \\
                                &$\frac{3}{2}$  &$1$ &$2$ &$1$ \\
$\Om \D\rightarrow \XiS\SiS$ &$\frac{3}{2}$ &$0$ &$1$ &$0$ \\
\hline
$\Om \SiS\rightarrow \Om \SiS$ &$1$ &$ 0$ &$0$ &$1$ \\
$\XiS\XiS\rightarrow \XiS\XiS$ &$0$ &$-3$ &$0$ &$1$ \\
                               &$1$ &$1$  &$0$ &$1$ \\
$\Om \SiS\rightarrow \XiS\XiS$ &$1$ &$0$ &$ 2$ &$0$ \\
\hline
$\Om \XiS\rightarrow \Om \XiS$ &$\frac{1}{2}$ &$0$ &$1$ &$1$ \\
\hline
$\Om \Om \rightarrow \Om \Om $ &$0$ &$0$ &$0$ &$1$ \\
\hline
\end{tabular}
\renewcommand{\arraystretch}{1.0}
\end{table}

In Table~\ref{tab:2DD} the isospin factors that enter into the evaluation
of the meson-exchange contribution to the $DD$ potential are summarized. 
For strangeness $S=-2$ and $-4$ there are channels with non-identical 
and with identical particles which require special treatment. 
Specifically, a proper symmetrization is required which can be
achieved by introducing the flavor-exchange operator $P_f$, cf. 
the procedure applied in the analogous $BB$ case with $S=-2$
($\Si\Si$, $\Xi N$, etc.) described in ref.~\cite{Polinder:2007mp}. 

\begin{figure}
\centering
\includegraphics[width=\columnwidth]{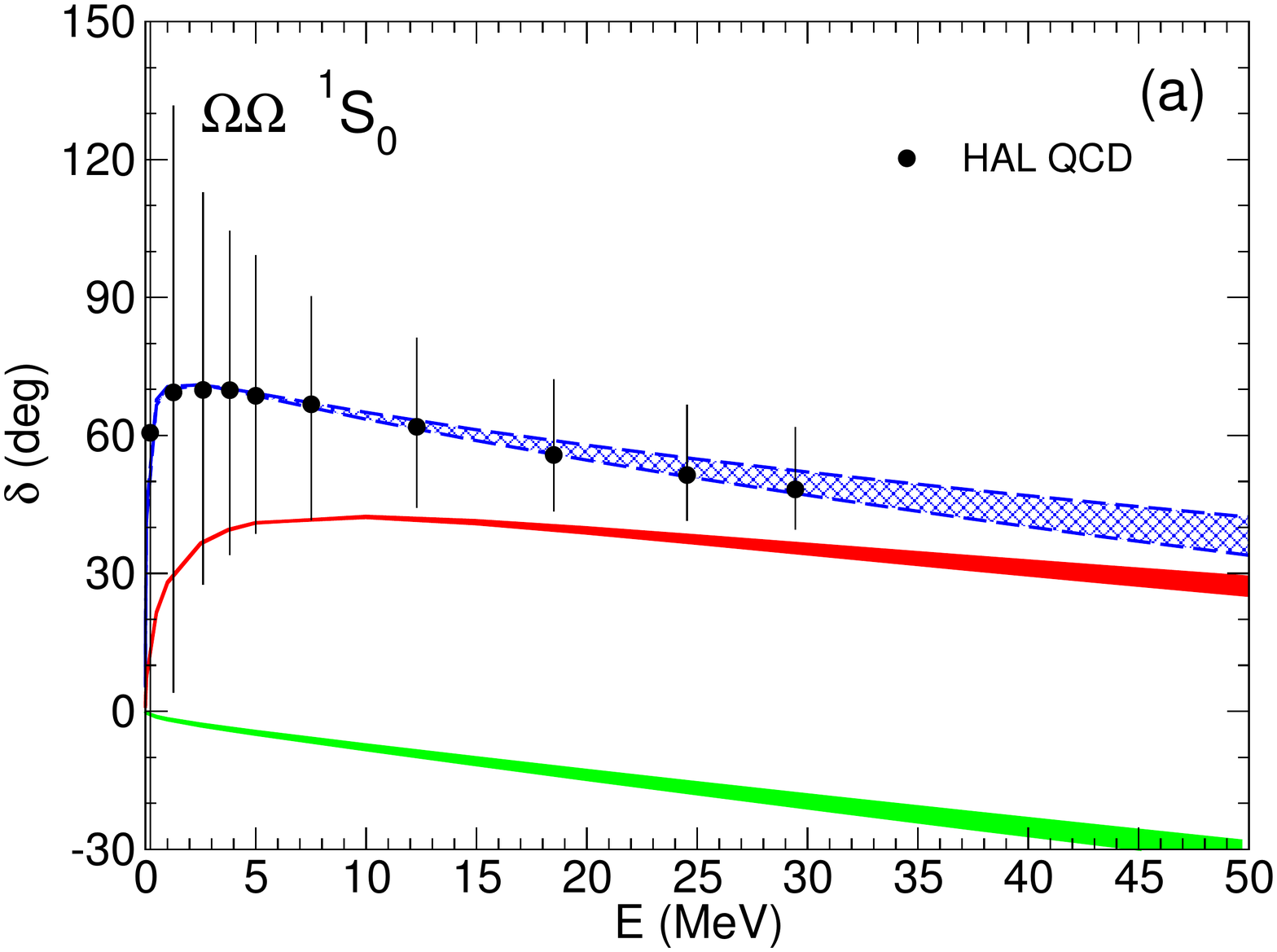}

\includegraphics[width=\columnwidth]{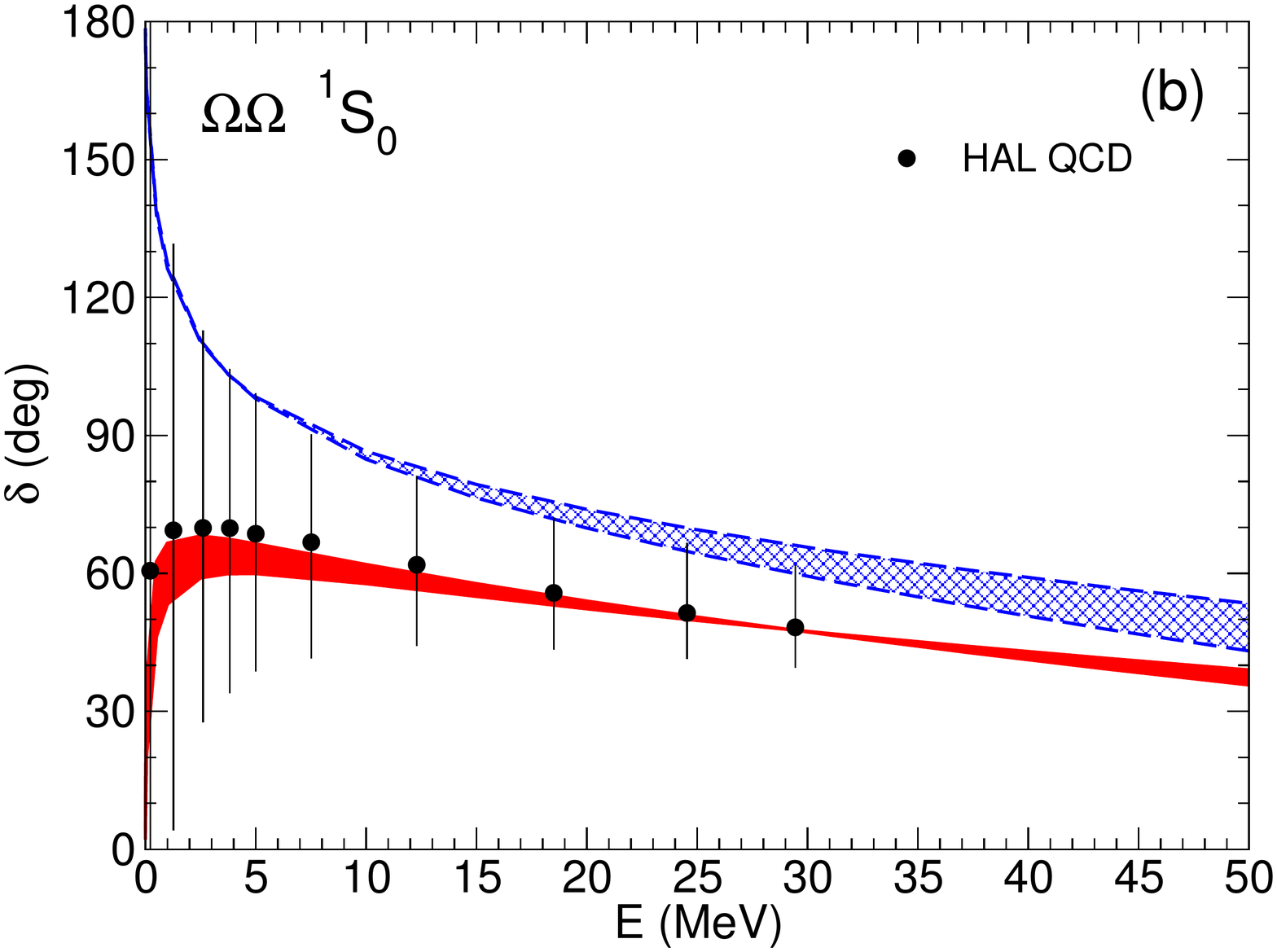}
\caption{$\Om\Om$ $^1S_0$ phase shift as a function 
of the kinetic energy in the center-of-mass system. 
The result of the HAL QCD collaboration is taken from ref.~\cite{Yamada:2015}.
Fits to the central value (a) and maximal value (b) 
of the lattice simulation are shown by hatched (blue) bands, corresponding
to cutoff variations between $500$ and $700$ MeV.
Corresponding results at the physical point are indicated by dark (red) bands. 
The  gray (green) band is the result based on the lattice simulation by 
Buchoff et al. \cite{Buchoff:2012}.
 }
\label{fig:ph1s0}
\end{figure}

\subsection{$\Om\Om$ and SU(3) related channels}

{The Paul principle implies that} the only allowed $S$-wave states for $\Om\Om$ are
$^1S_0$ and $^5S_2$. Moreover, the associated potentials depend only on a single 
LEC for each partial wave, namely the one corresponding to the SU(3) irreducible 
representation {\bf 28}, see Table~\ref{tab:PWDDDDD}.
There are lattice QCD simulations for $\Om\Om$ scattering by two groups. 
The earlier one by Buchoff et al.~\cite{Buchoff:2012} corresponds to a pion 
mass of $m_\pi\approx 390$~MeV and an $\Om$ mass close to the physical value.  
It suggests a fairly weak and repulsive interaction in the $^1S_0$ state
with a scattering length of $a = (0.16\pm 0.22)$~fm in the $^1S_0$ partial wave. 
The results of the HAL QCD collaboration, published soon after, turned
out to be {qualitatively} different \cite{Yamada:2015}. 
That calculation suggests a strongly attractive interaction for the $^1S_0$
channel, with phase shifts comparable to those of the $^1S_0$ partial wave
in neutron-proton scattering. Indeed, taking into account the large
statistical errors, even an $\Om\Om$ bound state might be supported.
The lattice set-up in ref.~\cite{Yamada:2015} corresponds to the masses:
$m_\pi = 701$~MeV, $m_K = 789$~MeV, $M_\Om = 1966$~MeV, quite far from the physical point.

Results of our analysis of the HAL QCD results are presented in Fig.~\ref{fig:ph1s0} 
as a function of the kinetic energy in the center-of-mass frame. These are achieved 
by appropriately adjusting the LEC $C^{28,\,^1S_0}_{22}$ (denoted simply by $C^{28}$
in the {following} discussion) to the lattice data for each cutoff. The variation of the 
results with the cutoffs $\Lambda=500-700$~MeV is indicated by {uncertainty} bands.
Fig.~\ref{fig:ph1s0}(a) {shows} a fit to the central HAL QCD prediction 
(hatched/blue band) while 
(b) is based on a fit to the upper limit given for the $^1S_0$ phase shift. 
{In this case} an $\Om\Om$ bound state with binding energy around $1$ MeV is produced.  
Since the $\eta$ mass is not given in ref.~\cite{Yamada:2015}, we assume
that $m_\eta \approx m_K$. 
In principle, the GMO mass formula could have been used to fix $m_\eta$ based on the 
values of $m_K$ and $m_\pi$. However, we refrain from doing so in this illustrative study. 
Anyway, it should be said that variations of $m_\eta$
{by} $20$ or $30$ MeV have very little influence on the actual
results and can be accommodated by a slight readjustment of $C^{28}$.   

The extrapolation to the physical point is indicated by dark (red)  
bands in
Fig.~\ref{fig:ph1s0}. Obviously, the interaction becomes less attractive 
and for neither of the two cases considered there is a bound state.
It should be noted that preliminary results for $\Om\Om$ corresponding to
a pion mass close to the physical point ($m_\pi\approx 145$ MeV) reported 
{recently} by the HAL QCD collaboration \cite{Sasaki:2016} 
suggest that there could be indeed a bound state, with a binding energy 
{roughly comparable to} that of the deuteron.  However, the corresponding 
mass of the $\Om$ baryon is not specified so that it remains unclear how 
close the latter is to its physical value. 
In any case, it will be interesting to see the final result.

The results in Fig.~\ref{fig:ph1s0} were obtained with a $\D\D\pi$ coupling
constant corresponding to $g_1 =1.5$. In the course of fitting to the lattice 
predictions we varied $g_1$ and it turned out that values around $1.3-1.5$ 
allowed for the best reproduction of the energy dependence suggested by
the lattice calculation. For larger values, specifically for values
close to the large $N_c$ prediction, $g_1 =9 g_A/5 \approx 2.3$, we
observed a very strong energy dependence of the $^1S_0$ phase shift
close to threshold that is not in line with the lattice data. 
It is caused by a dramatic increase in the coupling between the 
$^1S_0$ and $^5D_0$ partial waves. The larger coupling constant
increases the strength of the tensor force due to $\eta$ exchange 
so that the mixing angle becomes larger than the $^1S_0$ phase shift
itself already at small energies. We consider such a {scenario} as not
realistic. But it should be noted at the same time that the conclusions of ref.~\cite{Yamada:2015}
are based on an effective purely central $\Om\Om$ potential, used for calculating the $^1S_0$ phase shift. 
Accordingly, the role of the coupling to the $^5D_0$ channel in the evaluation of the lattice
data remains unclear. 

Results based on the lattice calculation in ref.~\cite{Buchoff:2012} are 
indicated by the gray (green) band in Fig.~\ref{fig:ph1s0}, where we
fixed the LEC by a fit to the scattering length $a = 0.16$ fm.
Since the relevant masses in that study are already fairly close to the 
physical point ($M_\Om = 1632$ MeV \cite{Luu}, $m_\eta = 587$ MeV \cite{Beane:2011}) 
we show only the {evaluation} for physical masses. 
{The obvious} discrepancy between the 
predictions in refs.~\cite{Buchoff:2012} and \cite{Yamada:2015} cannot be 
resolved by the present study. 

\begin{figure}[h]
\centering
\includegraphics[width=\columnwidth]{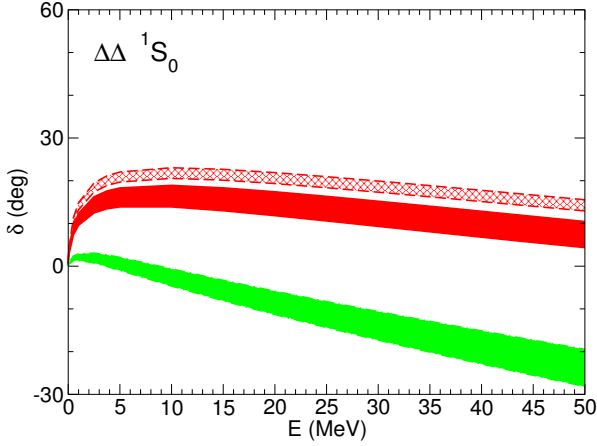}
\caption{Prediction for the $\D\D$ $^1S_0$ phase shift with isospin $I=3$  
based on the $\Om\Om$ result and SU(3) symmetry. 
The dark (red) band is based on the $C^{28}$ value fixed by a fit to 
the central HAL QCD result for $\Om\Om$, while the hatched (red) band
corresponds to the maximum HAL QCD result, cf. Fig.~\ref{fig:ph1s0}. 
The gray (green) band corresponds to the lattice result of ref.~\cite{Buchoff:2012}.
All results are for physical {meson and baryon} masses. 
}
\label{fig:ph1s0D}
\end{figure}

\begin{figure}[h]
\centering
\includegraphics[width=\columnwidth]{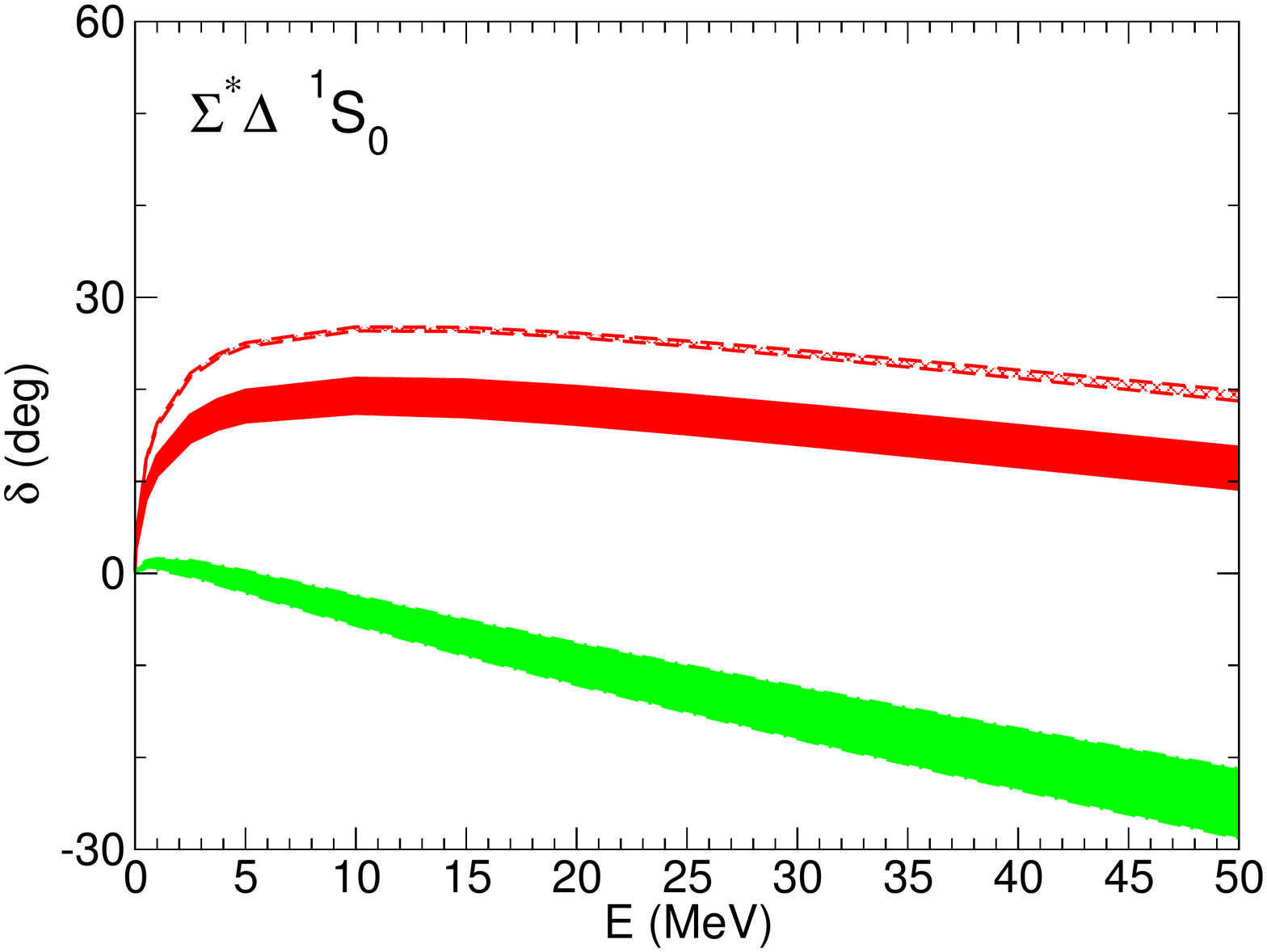}
\caption{Prediction for the $\SiS\D$ $^1S_0$ phase shift with isospin $I=5/2$  
based on the $\Om\Om$ result and SU(3) symmetry. 
Same description of curves as in Fig.~\ref{fig:ph1s0D}. 
}
\label{fig:ph1s0S}
\end{figure}

\begin{figure}[h]
\centering
\includegraphics[width=\columnwidth]{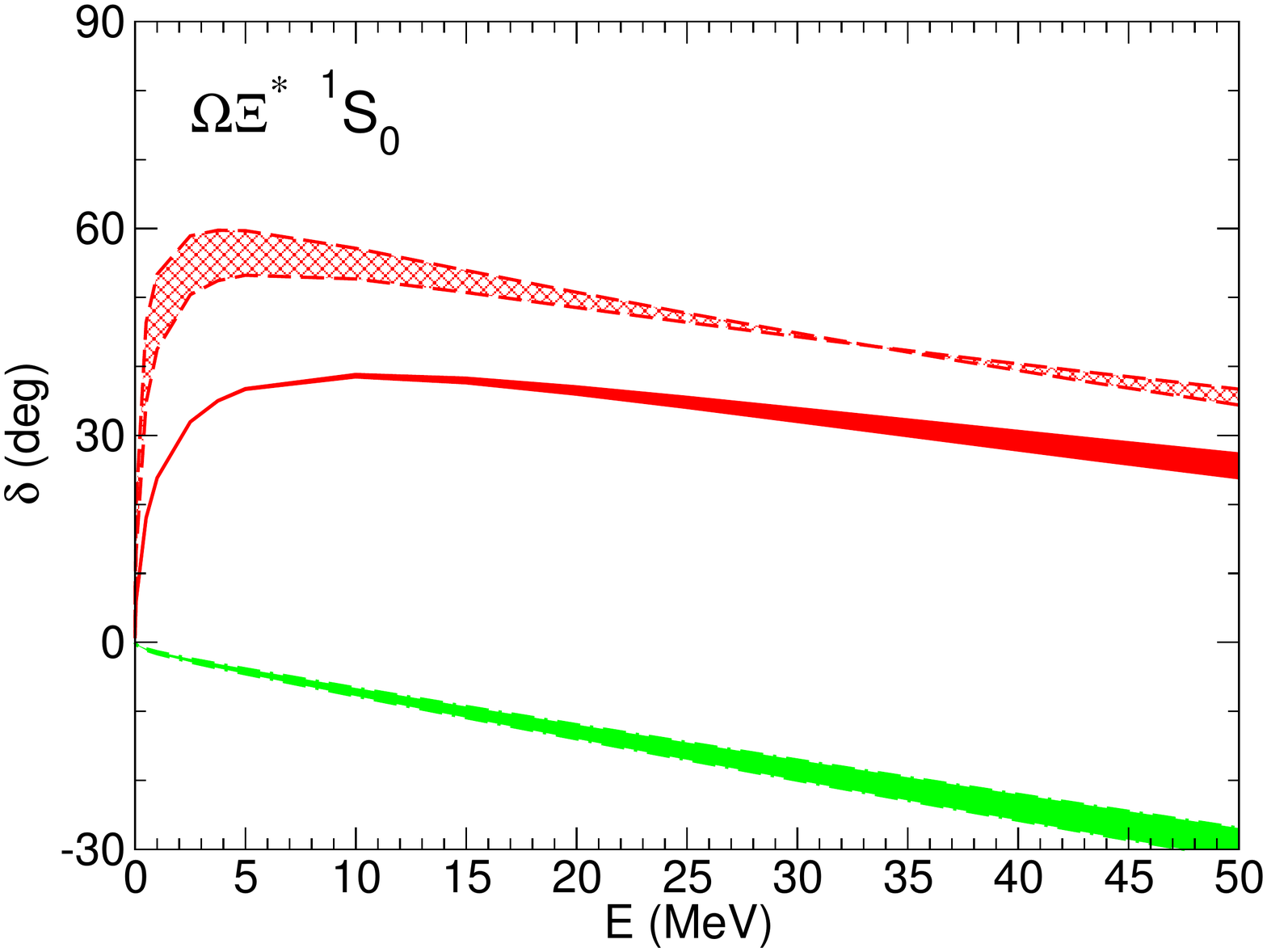}
\caption{Prediction for the $\XiS\Om$ $^1S_0$ phase shift with isospin $I=1/2$  
based on the $\Om\Om$ result and SU(3) symmetry. 
Same description of curves as in Fig.~\ref{fig:ph1s0D}. 
}
\label{fig:ph1s0X}
\end{figure}

Once the LEC $C^{28}$ is fixed by a fit to the $\Om\Om$ channel,
SU(3) symmetry can be exploited to obtain corresponding $^1S_0$ results/predictions 
for $\D\D$ ($I=3$), $\SiS\D$ ($I=5/2$), and $\Om\XiS$ ($I=1/2$). 
The interaction in all those cases is determined by the same LEC,
cf. Table~\ref{tab:PWDDDDD}, together with contributions from meson exchange.
And in all those cases there is no coupling to $BB$ or $BD$ channels. 

The corresponding phase shifts are presented in Figs.~\ref{fig:ph1s0D}-\ref{fig:ph1s0X}.
They illustrate the amount of SU(3) breaking that arises in our calculation from the
mass differences between $\pi$, $\eta$ and $K$, and between the decuplet baryons. 
Clearly, given that the $\D$ and $\SiS$ resonances are fairly broad, the $\D\D$ and $\SiS\D$ 
phase shifts are interesting {just for purely} academic reasons. This is different for the $\XiS$ 
because its width is only around $10$ MeV, i.e. comparable to the one of the $\omega$ meson, 
so that one {can} view it practically as a stable particle.  

While in case of $\Om\Om$ scattering only the $\eta$-meson can contribute, 
$\eta$ as well as $K$ exchange is possible for $\Om\XiS$, see Table~\ref{tab:2DD}.
However, given that $m_K\approx m_\eta$ there is not much difference in the actual 
potentials. Since the reduced mass $\mu$ of the latter system is smaller we 
still expect that the attraction should be slightly {weaker}, cf. Fig.~\ref{fig:ph1s0X}. 
Concerning $\D\D$ and $\SiS\D$ there is a sizable contribution from pion exchange
so that the corresponding potentials are more strongly influenced by the SU(3) breaking
due to the meson masses. At the same time the reduced masses for these systems
are noticeably smaller. Both effects together lead to interactions that are
appreciably less attractive, as can be seen from the phase shifts
in Figs.~\ref{fig:ph1s0D} and \ref{fig:ph1s0S}.
The predictions based on the lattice result of ref.~\cite{Buchoff:2012} 
(cf. green/gray bands) suggest a basically repulsive interaction for all
considered systems. Only in the $\D\D$ and $\SiS\D$ channels and for energies
close to the threshold {does one observe} slightly positive values for the $^1S_0$ phase shift 
signalling a weakly attractive tail of the potential. 

In this context, let us mention that quark-model studies \cite{Huang:2014}, but also 
Faddeev-type calculations \cite{Gal:2014}, often suggest the existence of a $\D\D$ dibaryon 
with $I=3$. Recently, an experimental search for {such a dibaryon state} has been performed by the 
WASA-at-COSY {collaboration} in the reaction $pp\to pp\pi^+\pi^+\pi^-\pi^-$~\cite{Adlarson:2016}. 
However, no clear-cut evidence for such a state was found.

\begin{figure}[h]
\centering
\includegraphics[width=\columnwidth]{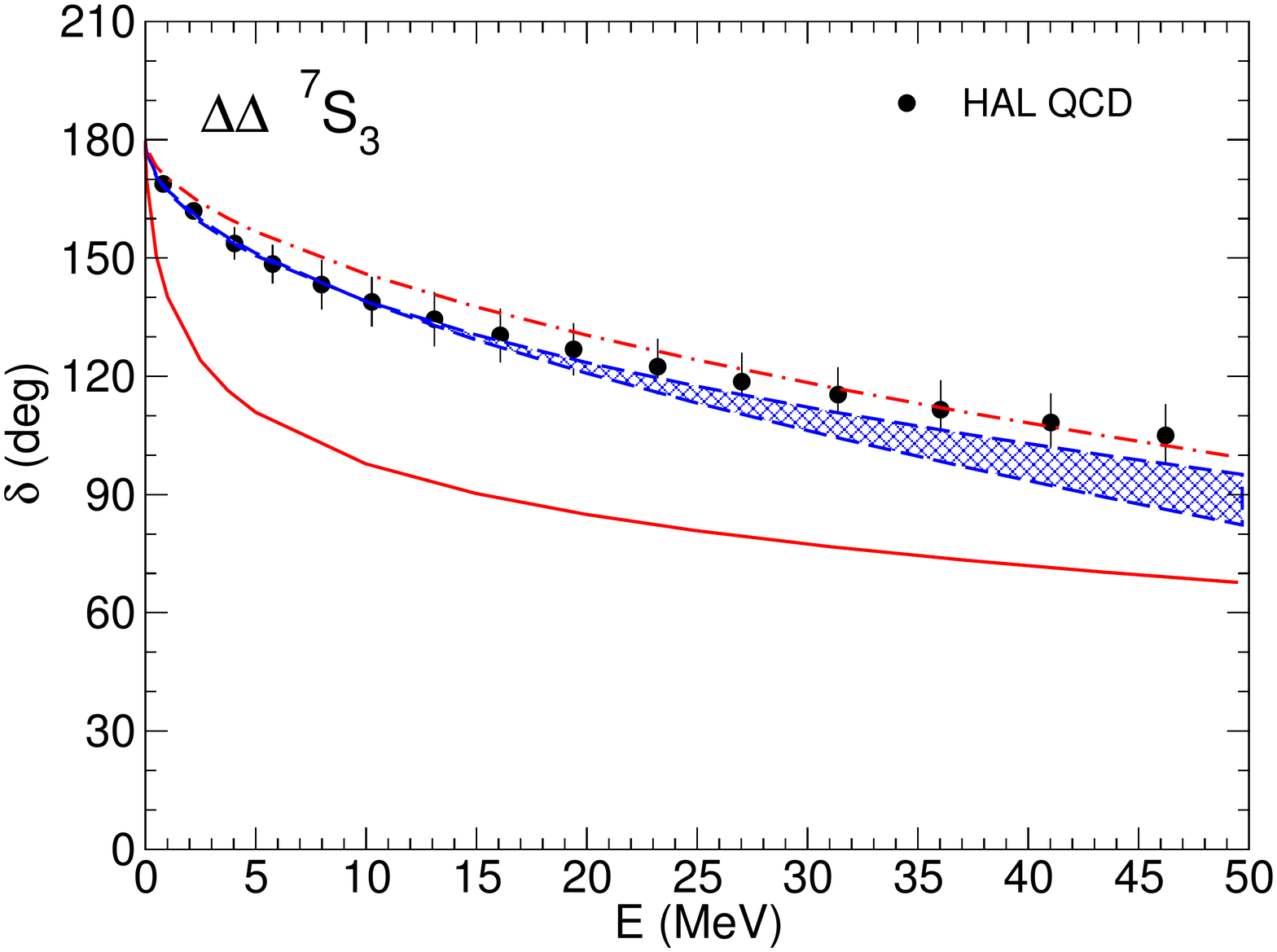}
\caption{Prediction for the $\D\D$ $^7S_3$ phase shift with isospin $I=0$. 
The result of the HAL QCD collaboration is taken from ref.~\cite{Sasaki:2016}.
Results of a fit to the lattice simulation is shown by a hatched band, 
corresponding to cutoff variations between $500$ and $700$ MeV.
The corresponding results for physical masses are shown by 
solid ($700$ MeV) and dash-dotted ($500$ MeV) lines. 
}
\label{fig:ph7s3}
\end{figure}

\subsection{$\D\D$ with $J=3$ and $I=0$}

Dibaryon candidates for the $\D\D$ state with $J=3$, $I=0$ are often 
mentioned together with the ones with $J=0$, $I=3$ discussed above. 
In particular, model studies {point out a} close connection between the $\D\D$ 
states with $I(J^P)=0(3^+)$ and {its $I(J^P)=3(0^+)$ mirror} state   
\cite{Huang:2014,Gal:2014}.
Indeed, at least with regard to the $S$-wave interactions the product of the 
expectation values for the spin and isospin operators is identical for the two 
channels so that the corresponding potential due to pion-exchange is the same. 
However, from Table~\ref{tab:PWDDDDD} one can see that the states actually belong 
to different SU(3) irreducible representations, so that the interactions
involve different LECs. 
Furthermore, for the state with $J=3$ there are many more coupled
partial waves, see Table~\ref{tab:pws}. And finally, the $\D\D$ state with 
$J=3$, $I=0$ couples to the $NN$ system. 

Nonetheless, the $\D\D$ state with $J=3$, $I=0$ is {of special interest} 
because, as discussed in the Introduction, the $d^*(2380)$ dibaryon candidate 
seen by the WASA-at-COSY collaboration \cite{Adlarson:2014} could be a {quasibound state}
produced by this system. Indeed, the analysis of data on quasifree polarized 
$np$ scattering presented in that paper suggests the $d^*(2380)$ dibaryon to be 
a resonance in the $^3D_3$-$^3G_3$ $NN$ partial wave. This state 
can couple to the $\D\D$ system with the partial waves 
$^7S_3$-$^3D_3$-$^7D_3$-$^3G_3$-$^7G_3$-$^7I_3$, see Table~\ref{tab:pws}.
The {position} of the resonance pole given by the WASA-at-COSY collaboration is 
$2380\pm 10 - i (40\pm 5)$ MeV. Since the nominal $\D\D$ threshold is at $2464$ MeV
this would imply a binding energy in the order of $84\pm 10$ MeV. 
{Of course this naive binding energy assignment ignores the width associated with the $\Delta\rightarrow\pi N$ decay channels.}
For an overview of the rapidly growing literature on the $d^*(2380)$ dibaryons
see for example ref.~\cite{Clement:2016}.
 
Addressing the experimental results directly is beyond the scope of the
present study. Fortunately, there are also lattice QCD calculations for 
the $\D\D$ system in question by the HAL QCD collaboration to which we can 
connect. It should be said, however, that so far only preliminary results are 
available that have been presented at conferences \cite{Sasaki:2016}.
Accordingly, we emphasize that our considerations here have likewise
preliminary character.
The $\D\D$ lattice QCD calculation of the $^7S_3$ partial wave with $I=0$
corresponds to a pion mass of $m_\pi = 1015$~MeV \cite{Sasaki:2016}. 
Unfortunately, the pertinent value for $M_\D$ is not given. {Supposedly} the 
lattice set-up with $m_\pi = 1015$~MeV corresponds to an SU(3) symmetric calculation 
where the octet baryon mass is $M_B=2030$~MeV \cite{Inoue:2010es}, and so we assume 
that using this mass is a {meaningful} option. 
The results of the HAL QCD collaboration for the $^7S_3$ $\D\D$ partial wave is shown in 
Fig.~\ref{fig:ph7s3} together with our fits. The LEC is fixed in such a way that 
the results at low energies are well reproduced. The hypothesis of a $\D\D$ bound 
state \cite{Sasaki:2016} is taken over and imposed in the fitting procedure. 
Again the cutoff values $\Lambda=500$ and $700$~MeV have been adopted. 

Given that the lattice calculation is for a rather large pion mass, the phase shifts
are basically determined by the contact term in $^7S_3$ alone. The binding 
energy implied by the fit is around $50-100$~MeV. 
When physical masses are used the results for the two cutoffs show different trends, see 
Fig.~\ref{fig:ph7s3}. In one case, $\Lambda=500$~MeV (dash-dotted 
line), there is very little change and the binding energy is still around $75$~MeV. 
{This happens to be close to the $d^*(2380)$ but this coincidence should of course not be overinterpreted.} 
In the other case (solid line), 
the attraction is strongly reduced and only a fairly shallow bound state 
with a binding energy of {about} $2.2$~MeV survives. 
{Lowering} the $\D$ mass to its physical value {weakens} the attraction because
the reduced mass becomes smaller and {along} with it the contribution of the loop
integral in the LS equation~(\ref{LS}). 
At the same time, reducing the pion mass increases the tensor coupling to the
other partial waves, in particular, to the two $D$ waves. Thereby, the 
attraction is increased so that there is a compensating effect. 
Obviously, the details of this compensation depends strongly on the cutoff 
so that no clear trend emerges. One has to keep in mind that the $^7S_3$ state 
can couple to many other partial waves, see Table~\ref{tab:pws}, which 
increases the sensitivity to the employed cutoff mass. 
A further complication is certainly the circumstance that the {so far} available lattice 
QCD calculation is for rather large masses. An extrapolation over a large mass region 
and, moreover, based on an LO calculation cannot be very reliable. Nevertheless, even for masses closer to the
physical point it will be a challenge to perform {an accurate} extrapolation
in view of the complicated angular-momentum structure of that state. 

Anyway, based on the present results one might still conclude that the existence 
of a {quasibound $\D\D$ state} with $J=3$, $I=0$ is at least not {totally} implausible. 
We should add that, given the very preliminary character of the $\D\D$ LQCD calculation,
we ignored here the coupling to $NN$ (in the $^3D_3-{}^3G_3$ partial wave, 
cf. Table~\ref{tab:pws}) and the fact that the $\Delta$ has a sizable width, i.e. 
there is a coupling to the $NN\pi$ and $NN\pi\pi$ continuum \cite{Gal:2016}. 
Both should have a significant influence on the location of the state and 
certainly on its width. 

\subsection{$N \Om$ with $J=2$}
Finally, let us consider the $N\Om$ interaction in the $^5S_2$
partial wave. Also for this case results from a lattice QCD
calculation by the HAL QCD collaboration are available~\cite{Etminan:2014}.
The lattice set-up corresponds to the masses:
$M_\Om = 2105$ MeV, $M_N = 1806$ MeV $m_\pi = 875$~MeV, $m_K = 916$~MeV 
(again we assume that $m_\eta \approx m_K$). 

The $N\Om$ system is interesting because it involves only stable particles
(stable against hadronic decay) so that, in principle, even scattering
experiments are feasible. However, unlike $\Om\Om$ discussed above, 
$N\Om$ can couple to various other $BB$ and $BD$ channels.
Specifically, the $N\Om$ $^5S_2$ system (with threshold at $\sqrt{s}=2611$ MeV) 
can couple to $\Lambda\Xi$ ($^1D_2$, threshold at $2434$ MeV), 
$\Sigma\Xi$ ($^1D_2$, threshold at $2511$ MeV) 
and $\Lambda\Xi\pi$ ($^3P_2s$, threshold at $2574$ MeV). 
In addition, and more unfavorable for a concrete calculation, it couples
to $\Lambda\Xi^*$ (threshold at $2647$ MeV), $\Sigma\Xi^*$ (threshold at $2725$ MeV), 
and $\Xi\Sigma^*$ (threshold at $2703$ MeV) in all partial waves, 
see Table~\ref{tab:pws}. 
{Thus all four LECs} that contribute to 
$BD$ scattering (cf. Table~\ref{tab:PWDDBDB}) are needed. There is no way to 
determine the individual values of those four LECs from the $N\Om$ results in 
ref.~\cite{Etminan:2014}. 
Indeed, the only ``selective'' LEC in $BD$ scattering is $C^{35}$ which, once 
fixed, could be used to relate the interactions for $N\D$ ($I=2$), $\Si\D$ ($I=5/2$), 
and $\Xi\Om$ ($I=1/2$). 
Unfortunately, lattice results are not available for any of
these channels.

\begin{figure}[t]
\centering
\includegraphics[width=\columnwidth]{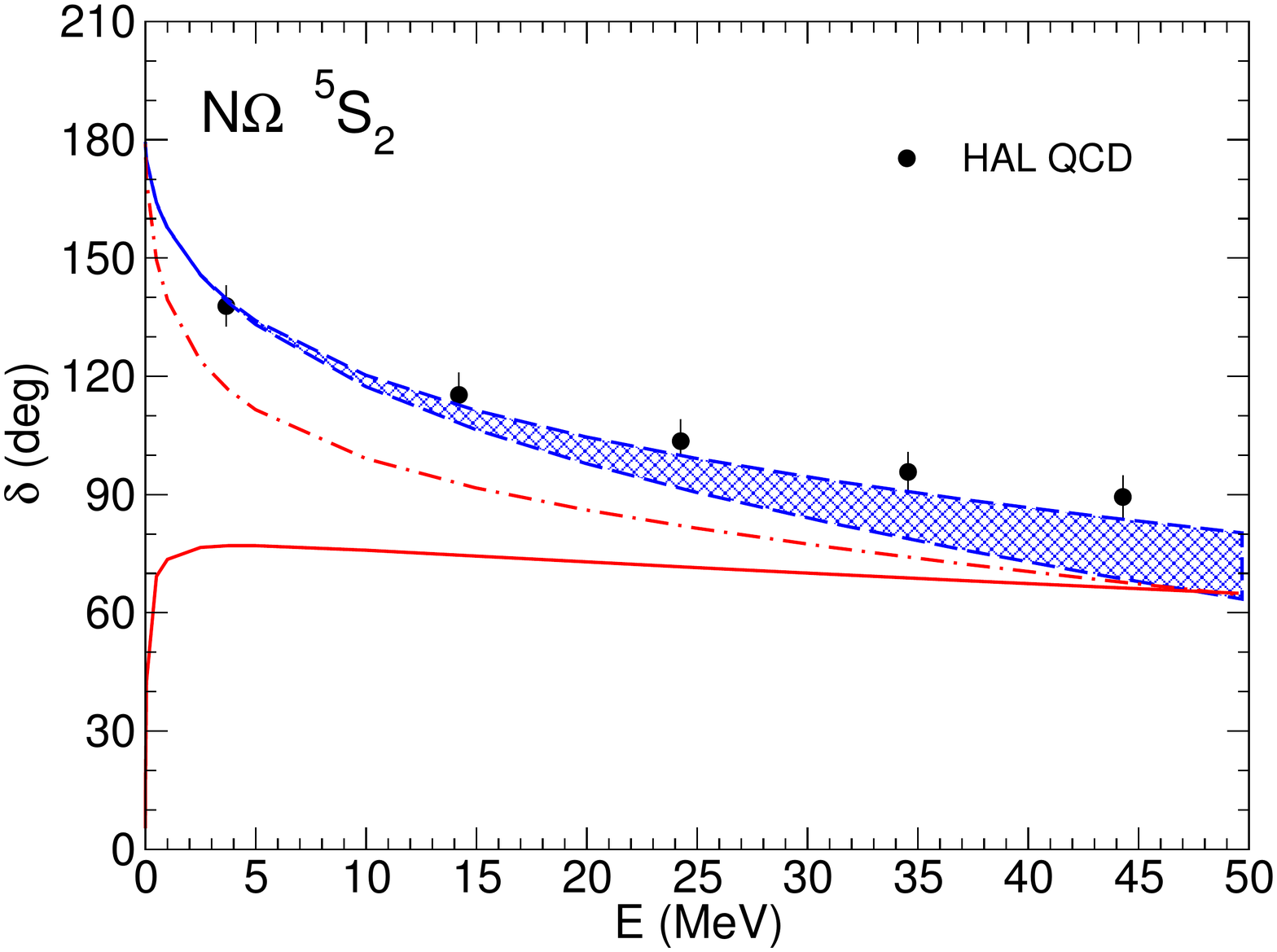}
\caption{$N\Om$ $^5S_2$ phase shift.
The result of the HAL QCD collaboration is taken from ref.~\cite{Etminan:2014}.
Results of a fit to the lattice simulation is shown by a hatched band, 
corresponding to cutoff variations between $500$ and $700$ MeV.
The corresponding results for physical masses are shown by 
solid ($700$ MeV) and dash-dotted ($500$ MeV) lines. 
}
\label{fig:ph5s2}
\end{figure}

{We focus in the following} on the only $BD$ lattice results available
and ignore all couplings of $N\Om$ to other channels. Then only one specific
combination of LECs enter, see Table~\ref{tab:PWDDBDB}, which can be fixed by 
a fit to the lattice predictions. Note that the 
interaction in the $N\Om$ system is particularly simple because, {apart from} the
contact interaction, only $\eta$-meson exchange can contribute. For other $BD$
reactions there are contributions from direct diagrams, involving $BB\phi$ 
and $DD\phi$ vertices, and from exchange diagrams that involve $BD\phi$ and 
$DB\phi$ vertices. 
 
Fits to the lattice calculation are presented in Fig.~\ref{fig:ph5s2}.
Again the standard {cutoffs} $\Lambda=500$ MeV and $700$ MeV are
adopted. The lattice results of ref.~\cite{Etminan:2014} support the
existence of a bound state in the $N\Om$ system. The binding energy 
estimated from our fits is in the {range} $9-11$ MeV. When extrapolating
to the physical point we observe the same difficulty as already in the
$\D\D$ case above. There is no clear trend because the results depend
significantly on the cutoff. In one case ($\Lambda=500$ MeV, dash-dotted line) 
the bound state would survive, with a binding energy of around $2$ MeV, while 
for the other cutoff (solid line) it disappears. Thus conclusions - even 
qualitative ones - are difficult to draw. 
This is even more the case if we recall the various other channels that
can couple to $N\Om$ and that are open at the physical point. In view of these
additional channels, it is to be expected that the dynamics in the $N\Om$ system
is highly complex. {It will therefore}
be challenging to establish the existence of a bound
state in this system from a lattice calculation. Actually, the 
situation is even more involved than the one for the $H$-dibaryon, where there is 
a delicate interplay between the interactions in the $\La\La$, $\Xi N$,
and $\Si\Si$ channels 
\cite{Haidenbauer:2011ah,Haidenbauer:2011za,Sasaki:20162,Inoue:2011,Yamaguchi:2016}. 

\section{Summary and outlook} 
\label{sec:Summary}

In this paper we have derived the general form of the baryon-baryon interaction 
involving octet and decuplet baryons in a chiral effective field theory approach 
based on the Weinberg power counting. The present work is an extension of earlier
studies on the nucleon-nucleon \cite{Epelbaum:2008}, 
hyperon-nucleon \cite{Polinder:2006zh,Haidenbauer:2013}
and hyperon-hyperon \cite{Polinder:2007mp,Haidenbauer:2015} systems
within the same framework. Specific attention 
is paid to the {connections between} the interactions in the various baryon-baryon
channels that follow from the underlying (approximate) SU(3) flavor symmetry.
The leading-order potential presented in this paper consists of two components: 
{i) long-ranged} one-pseudo{\-}scalar-meson exchanges ($\pi$, $K$, $\eta$)
{with} coupling constants at the various baryon-baryon-meson vertices
related via ${\rm SU(3)}$ symmetry; {ii) short-ranged} four-baryon contact terms without derivatives. 
For the latter the most general, minimal and ${\rm SU(3)}$ invariant Lagrangian {has been derived}.  
The number of independent contact terms at LO amounts to: six for the
scattering of two octet baryons (as {already established in} \cite{Polinder:2006zh,Polinder:2007mp}),
eight for the scattering of two decuplet baryons, eight for 
the scattering of an octet baryon on a decuplet baryon, and  
two (each) for the transitions between these systems. 

The low-energy constants associated with those contact terms need to
be determined from scattering data. Given the lack of empirical information
on the scattering of decuplet baryons we have illustrated how
lattice QCD simulations \cite{Buchoff:2012,Yamada:2015,Etminan:2014,Sasaki:2016} 
can be used to constrain or even fix some of the LECs. Admittedly, since the {presently available} 
lattice QCD calculations for scattering of decuplet baryons still involve large pion masses 
(in general $m_\pi \approx 700 - 1000$ MeV) and large baryon masses,
the considered extrapolations to the physical point have to be taken with a grain of salt.
Nonetheless, it is clear from the presented applications that chiral EFT 
provides a useful tool to analyze lattice results once calculations 
corresponding to masses closer to the physical point will become available.

Finally, since the $\Om$ can decay only {through weak interactions}, in principle, actual experiments 
where $\Om$ baryons are scattered on nucleons are feasible, analogous to 
those performed for the $\La N$ and $\Si N$ systems. 
Indeed, there are plans for studying the $N\Om$ interaction experimentally 
at J-PARC \cite{Takahashi:2013}, where it is intended to produce
the $\Om$ in the reaction $K^-p \to \Om K^+ K^0$. 
There is also a proposal to establish a secondary $K^0_L$ beam at JLab 
that can then be used for producing $\Om$s  \cite{Amaryan:2015}.
$\Om$ baryons can also be produced in the reaction $\bar pp\to \bar \Om\Om$, {a case for}
the $\bar {\rm P}$ANDA project at the FAIR facility 
in Darmstadt \cite{PANDA,PANDA2}. 
Furthermore, heavy-ion collisions could allow one to access information on the
$N\Om$ \cite{Han:2015,Morita:2016} but also the $\Om\Om$ interaction. 

Another and completely different field of application for our formalism might
be in studies of pion production in $NN$ collisions. Here, an explicit inclusion 
of the direct $N\D$ and $\D\D$ interactions could be a sensible next step for 
refining the treatment of the reaction $NN \to NN\pi$ within 
chiral perturbation theory~\cite{Baru:2013}.  
Furthermore, a better knowledge of the $\Delta N$ and $\Delta\Delta$
interaction could shed light on the so-called $\Delta$ puzzle in neutron
star matter, see e.g. ref. \cite{Drago:2014}.
 
\bigskip

{\bf Acknowledgments:}
We acknowledge stimulating discussions with Tom Luu. 
This work is supported in part by the DFG and the NSFC through
funds provided to the Sino-German CRC 110 ``Symmetries and
the Emergence of Structure in QCD''. The work of UGM was also
supported by the Chinese Academy of Sciences (CAS) President's
International Fellowship Initiative (PIFI) (Grant No. 2017VMA025).

\appendix
\section{Spin matrices}
\label{app:Spin}

In this appendix we summarize the spin transition operators involving spin 1/2 and spin 3/2 states.
In order to construct these matrices (cf.\ ref.~\cite{Ericson:1988gk}), we use the Wigner-Eckart 
theorem, which relates matrix elements of a spherical tensor operator to a reduced matrix element 
and Clebsch-Gordon coefficients. For writing a Cartesian tensor (up to rank 3) in terms of spherical 
tensors, we express the usual scalar \(\ONE\), vector \(x^i\), irreducible tensor of rank two 
\(x^ix^j - \vec r^{\,2}\delta^{ij}/3\) and irreducible tensor of rank 
three \(5x^ix^jx^k - (x^i\delta^{jk}+x^j\delta^{ik}+x^k\delta^{ij})\vec r^{\,2}\) into spherical 
harmonics of the same rank.

Following this approach, we obtain for the vector spin matrices (see also refs.~\cite{Wiringa,Ericson:1988gk})%
\footnote{Note that in Wiringa et al. \cite{Wiringa} ${\bf S}$ is defined as $4\times 2$
matrix, which is equivalent to ${\bf S}^\dagger$ in the present paper.}:
\begin{equation}
 \sigma_1 =\begin{pmatrix}0 & 1\\1 & 0 \end{pmatrix}\,, \quad
 \sigma_2 =\begin{pmatrix}0 & -\mathrm i\\\mathrm i & 0 \end{pmatrix}\,, \quad
 \sigma_3 =\begin{pmatrix}1 & 0\\0 & -1 \end{pmatrix}\,,
\end{equation}%
\begin{align*}
&S_1 = \begin{pmatrix}
 -\frac{1}{\sqrt{2}} & 0 & \frac{1}{\sqrt{6}} & 0 \\
 0 & -\frac{1}{\sqrt{6}} & 0 & \frac{1}{\sqrt{2}}
\end{pmatrix},\
S_2 = \begin{pmatrix}
 -\frac{i}{\sqrt{2}} & 0 & -\frac{i}{\sqrt{6}} & 0 \\
 0 & -\frac{i}{\sqrt{6}} & 0 & -\frac{i}{\sqrt{2}}
\end{pmatrix},\\
&S_3 = \begin{pmatrix}
 0 & \sqrt{\frac{2}{3}} & 0 & 0 \\
 0 & 0 & \sqrt{\frac{2}{3}} & 0
\end{pmatrix}, \numberthis
\label{eq:Spin1} 
\end{align*}%
\begin{align*}
&\Sigma_1 = \begin{pmatrix}
 0 & \sqrt{3} & 0 & 0 \\
 \sqrt{3} & 0 & 2 & 0 \\
 0 & 2 & 0 & \sqrt{3} \\
 0 & 0 & \sqrt{3} & 0
\end{pmatrix},\
\Sigma_2 = \begin{pmatrix}
 0 & -i \sqrt{3} & 0 & 0 \\
 i \sqrt{3} & 0 & -2 i & 0 \\
 0 & 2 i & 0 & -i \sqrt{3} \\
 0 & 0 & i \sqrt{3} & 0
\end{pmatrix},\\
&\Sigma_3 = \begin{pmatrix}
 3 & 0 & 0 & 0 \\
 0 & 1 & 0 & 0 \\
 0 & 0 & -1 & 0 \\
 0 & 0 & 0 & -3
\end{pmatrix}. \numberthis
\end{align*}
The analogous isospin matrices corresponding to the spin matrices \(\vec\sigma\), \(\vec S\) 
and \(\vec\Sigma\) are denoted by \(\vec\tau\), \(\vec T\) and \(\vec\theta\).

The spin matrices with rank two and three can be expressed through vector and scalar spin matrices:
\begin{align*}
S^{ij} ={}& -\frac1{\sqrt6}\Big( \sigma^i S^{j} + \sigma^j S^{i} \Big)\,,\
S^{ij\dagger} = -\frac1{\sqrt6}\Big( S^{i\dagger}\sigma^j + S^{j\dagger}\sigma^i \Big)\,, \\
\Sigma^{ij} ={}& \frac18\Big( \Sigma^i \Sigma^j + \Sigma^j \Sigma^i - 10\delta^{ij}\ONE \Big)
= \delta^{ij} \ONE - \frac32\Big( S^{i\dagger}S^j + S^{j\dagger}S^i \Big)\,, \\
\Sigma^{ijk} ={}& \frac1{36\sqrt3}\Big( 5( \Sigma^i \Sigma^j \Sigma^k + \Sigma^k \Sigma^i \Sigma^j + \Sigma^j \Sigma^k \Sigma^i +\Sigma^i \Sigma^k \Sigma^j \\
&+ \Sigma^j \Sigma^i \Sigma^k + \Sigma^k \Sigma^j \Sigma^i)
-82( \Sigma^i \delta^{jk} + \Sigma^j \delta^{ik} + \Sigma^k \delta^{ij} ) \Big)\,. \numberthis
\end{align*}

\section{Two-body potentials in particle representation} \label{app:BBpart}

In this section, we show the two-body interaction potentials in particle basis for the transitions
\(BB\to BB\), \(BB\leftrightarrow DB\), \(DB\to DB\), \(BB\leftrightarrow DD\),
\(DB\leftrightarrow DD\), and \(DD\to DD\).

First, we write the Lagrangians presented in sections \ref{subsec:LagMes} and \ref{sec:ctLagr} in terms of their particle fields, as defined in equations \eqref{eq:mesonmat}, \eqref{eq:baryonmat} and \eqref{eq:baryonDecTensor}.
The physical fields comprise the sets:
\begin{align*}
 \phi_i &\in \left\{\pi^0,\pi^+,\pi^-,K^+,K^-,K^0,\bar K^0,\eta\right\}\,,\\
 B_i &\in \left\{n,p,\Sigma^0,\Sigma^+,\Sigma^-,\Lambda,\Xi^0,\Xi^-\right\}\,,\\
 B^*_i &\in \left\{\Delta^-,\Delta^0,\Delta^+,\Delta^{++},\Sigma^{*0},\Sigma^{*+},\Sigma^{*-},\Xi^{*0},\Xi^{*-},\Omega^-\right\}\,.
\numberthis \label{eq:particleFields}
\end{align*}

For the octet-baryon meson Lagrangian one obtains
\begin{align*}
 \mathcal{L}_\mathrm{BB\phi} = -\frac{1}{2f_0}\sum_{i,j,k} N_{B_iB_j\phi_k} (\bar B_i \vec\sigma B_j) \cdot (\vec\nabla \phi_k) \,, \numberthis 
\label{eq:BBMP}
\end{align*}
where the sum over \(i,j,k\) runs over all particles fields, specified in eq.~\eqref{eq:particleFields}.
Furthermore, we have introduced the SU(3) factors \(N\). 
These factors can be easily obtained by multiplying the flavor matrices and taking the traces.
For example \(N_{\Lambda\Lambda\eta}\) can be calculated as (see eq.~\eqref{eq:BBM})
\begin{equation}
 N_{\Lambda\Lambda\eta} = D\, {\rm tr}(Y^\top_\Lambda \lbrace Y_\eta,Y_\Lambda\rbrace)
+ F\, {\rm tr}( Y^\top_\Lambda \left[Y_\eta,Y_\Lambda\right])  = -{2D \over \sqrt{3}}\,,
\end{equation}
with the flavor matrices
\begin{equation}
 Y_\Lambda :=
 \begin{pmatrix}
  \frac{1}{\sqrt 6} & 0 & 0 \\
  0 & \frac{1}{\sqrt 6} & 0 \\
  0 & 0 & -\frac{2}{\sqrt 6}
 \end{pmatrix}\,,\quad
Y_\eta :=
 \begin{pmatrix}
  {1\over\sqrt{3}}  & 0 & 0 \\
  0 & {1\over\sqrt{3}} & 0 \\
  0 & 0 & -{2\over\sqrt{3}}
 \end{pmatrix}\,.
\end{equation}
We will use the representation in terms of SU(3) factors \(N\) in the following, as it allows for a simple 
presentation of the two-baryon potentials.

For the corresponding Lagrangian terms involving decuplet-baryons one obtains in the same way
\begin{align*}
\mathcal{L}_\mathrm{DB\phi} &= \frac C{f_0} \sum_{i,j,k} N_{B_i^*B_j\phi_k}
\Big[
(\bar B_i^* \vec S^{\,\dagger} B_j) \cdot \left(\vec\nabla \phi_k\right) \\
&\qquad\qquad\qquad\qquad+
(\bar B_j \vec S B_i^*) \cdot \left(\vec\nabla \phi_k^\dagger\right)
\Big] \,, \numberthis
\label{eq:BDP} 
\end{align*}
\begin{align*}
\mathcal{L}_\mathrm{DD\phi} = \frac{H}{f_0} \sum_{i,j,k} N_{B^*_iB^*_j\phi_k} (\bar B^*_i \vec\Sigma B^*_j) \cdot (\vec\nabla \phi_k) \,. \numberthis
\label{eq:DDP}
\end{align*}

The minimal set of contact Lagrangian terms of sec.~\ref{sec:ctLagr} can also be rewritten in terms of particle fields, with SU(3) factors \(N\):
\begin{align*}
\mathcal{L}_\mathrm{BBBB} =&\quad\,\sumCzza \sum_{i,j,k,l} N^{f}_{B_iB_jB_kB_l}
\left(\bar B_iB_j\right)\left(\bar B_kB_l\right) \\
&+\sumCzzb \sum_{i,j,k,l} N^{f}_{B_iB_jB_kB_l}
\left(\bar B_i\vec \sigma B_j\right)\cdot\left(\bar B_k\vec\sigma B_l\right) \,. \numberthis
\end{align*}
\begin{align*}
\mathcal{L}_\mathrm{DBBB} = \sumCzoa \sum_{i,j,k,l} N^f_{B_i^*B_jB_kB_l}
\big[
&\left(\bar B_i^*\vec S^\dagger B_j\right)\cdot\left(\bar B_k\vec\sigma B_l\right) \\
&+\left(\bar B_j\vec S\, B_i^*\right)\cdot\left(\bar B_l\vec\sigma B_k\right)
\big] \,. \numberthis
\end{align*}
\begin{align*}
\mathcal{L}_\mathrm{DBDB} =\phantom{ + }{}& \sumCooa \sum_{i,j,k,l} N^{f}_{B^*_iB^*_jB_kB_l}
\left(\bar B^*_i B^*_j\right)\left(\bar B_k B_l\right) \\
+&\sumCoob \sum_{i,j,k,l} N^{f}_{B^*_iB^*_jB_kB_l}
\left(\bar B^*_i\vec\Sigma B^*_j\right)\cdot\left(\bar B_k\vec\sigma B_l\right)
\,. \numberthis
\end{align*}
\begin{align*}
\mathcal{L}_\mathrm{DDBB} =\phantom{ + }{}& \sumCzta \sum_{i,j,k,l} N^{1}_{B^*_iB_jB^*_kB_l}
\big[
\left(\bar B^*_i\vec S^\dagger B_j\right)\cdot\left(\bar B^*_k\vec S^\dagger B_l\right) \\
&\qquad\qquad\qquad\quad+
\left(\bar B_j\vec S B^*_i\right)\cdot\left(\bar B_l\vec S B^*_k\right)
\big] \\
+ \,& \sumCztb \sum_{i,j,k,l} N^{2}_{B^*_iB_jB^*_kB_l}
\big[
\left(\bar B^*_iS^{\alpha\beta\dagger} B_j\right)\left(\bar B^*_kS^{\alpha\beta\dagger} B_l\right) \\
&\qquad\qquad\qquad\quad+
\left(\bar B_jS^{\alpha\beta} B^*_i\right)\left(\bar B_lS^{\alpha\beta} B^*_k\right)
\big] \,. \numberthis
\end{align*}
\begin{align*}
\mathcal{L}_\mathrm{DDDB} =\phantom{ + }{}&\sumCota \sum_{i,j,k,l} N^{1}_{B^*_iB^*_jB^*_kB_l}
\big[
\left(\bar B^*_i\vec\Sigma B^*_j\right)\cdot\left(\bar B^*_k\vec S^\dagger B_l\right)\\
&\qquad\qquad\qquad\quad+
\left(\bar B^*_j\vec\Sigma B^*_i\right)\cdot\left(\bar B_l\vec S B^*_k\right)
\big] \\
+\,&\sumCotb \sum_{i,j,k,l} N^{2}_{B^*_iB^*_jB^*_kB_l}
\big[
\left(\bar B^*_i\Sigma^{\alpha\beta} B^*_j\right)\left(\bar B^*_k S^{\alpha\beta\dagger} B_l\right)\\
&\qquad\qquad\qquad\quad+
\left(\bar B^*_j\Sigma^{\alpha\beta} B^*_i\right)\left(\bar B_l S^{\alpha\beta} B^*_k\right)
\big] \,. \numberthis
\end{align*}
\begin{align*}
\mathcal{L}_\mathrm{DDDD} =\phantom{ + }{}& \sumCtta \sum_{i,j,k,l} N^{f}_{B^*_iB^*_jB^*_kB^*_l}
\left(\bar B^*_iB^*_j\right)\left(\bar B^*_kB^*_l\right) \\
+&\sumCttb \sum_{i,j,k,l} N^{f}_{B^*_iB^*_jB^*_kB^*_l}
\left(\bar B^*_i\vec \Sigma B^*_j\right)\cdot\left(\bar B^*_k\vec\Sigma B^*_l\right) \\
+&\sumCttc \sum_{i,j,k,l} N^{f}_{B^*_iB^*_jB^*_kB^*_l}
\left(\bar B^*_i \Sigma^{\alpha\beta} B^*_j\right)\left(\bar B^*_k\Sigma^{\alpha\beta} B^*_l\right) \\
+&\sumCttd \sum_{i,j,k,l} N^{f}_{B^*_iB^*_jB^*_kB^*_l}
\left(\bar B^*_i\Sigma^{\alpha\beta\gamma} B^*_j\right)\left(\bar B^*_k\Sigma^{\alpha\beta\gamma} B^*_l\right)
\,. \numberthis
\end{align*}

Let us now come to the calculation of the two-body interaction potentials
based on the Lagrangian in particle basis as shown above.
The calculation is done in the center-of-mass frame with momentum assignments \(A(\vec p\,)B(-\vec p\,)\to C(\vec p^{\,\prime})D(-\vec p^{\,\prime})\).
We use the common definitions \(\vec q = \vec p^{\,\prime} - \vec p\) and \(\vec k = \vec p^{\,\prime} + \vec p\), where \(\vec q\) appears in the direct one-meson exchange and \(\vec k\) appears in the exchanged one-meson exchange\footnote{
We choose the conventions \([B_i,B^*_j]_+ = B_iB^*_j + B^*_jB_i = 0\) (and \([B_i,B_j]_+ = [B^*_i,B^*_j]_+ = 0\)), i.e., all (octet and decuplet) baryons anti-commute.}.

For all considered transitions we show in the first line, how the potential is calculated from Feynman diagrams.
In the Feynman diagrams themselves, particles that are vertically above each other are defined to be in the same baryon bilinear.
Octet baryons are denoted by single lines, decuplet baryons by double lines, and mesons by dashed lines.

Below the Feynman diagrams, we provide the full potentials in particle basis, for a general assignment of baryons \(B_i\), decuplet baryons \(B_i^*\) and mesons \(\phi\).
For a clearer presentation, we introduce prefactors \(X_i\), which are linear combinations of SU(3) coefficients \(N\) and low-energy constants.
The factor \(X_1\) is always related to the direct one-meson exchange, \(X_2\) is related to the exchanged one-meson exchange and the remaining \(X_i\) concern contact interaction with various spin-structures.
The \(X_i\) for the considered interactions are summarized in tables~\ref{tab:xi} and \ref{tab:xi2}.

For the construction of the potentials we also need various spin exchange operators, defined by 
\(P^{(\sigma)} \vert\chi_1,\chi_2\rangle = \vert\chi_2,\chi_1\rangle\),
where the first position in \(\vert.,.\rangle\) is spin space 1 (or baryon bilinear 1) and the second position in \(\vert.,.\rangle\) is in spin space 2 (or baryon bilinear 2).
For two spin-1/2 states the well-known spin exchange operator is given by
\begin{equation}
\psd = {\psd}^\dagger = \frac12 ( \ONE + \vec\sigma_1\cdot\vec\sigma_2 ) \,.
\end{equation}
For two spin-3/2 states, the spin exchange operator can be expressed through
\begin{equation}
\psq = {\psq}^\dagger =
\frac{1}{4} \ONE +
\frac{1}{20} \Sigma^i_1\Sigma^i_2 +
\frac{1}{6} \Sigma_1^{ij}\Sigma_2^{ij} +
\frac{3}{200} \Sigma_1^{ijk}\Sigma_2^{ijk}~.
\end{equation}
The spin exchange operator exchanging a spin-1/2 state and a spin-3/2 state is given by\footnote{
The spin exchange operator fulfills \(\psdq\vert\frac12,\frac32\rangle=\vert\frac32,\frac12\rangle\) and \(\psqd\vert\frac32,\frac12\rangle=\vert\frac12,\frac32\rangle\).
}
\begin{equation}
\psdq = \frac34\left( \vec S_1^{\,\dagger}\cdot\vec S_2 + S_1^{\dagger ij} S_2^{ij} \right) \,.
\end{equation}

Let us now list all considered transitions and their potentials, starting with the well-known potential 
involving only octet baryons, and with an increasing number of involved decuplet baryons.
\begin{itemize}
\item \(B_1B_2\to B_3B_4\) (with bilinears in spin space \(B_1\)-\(B_3\) and \(B_2\)-\(B_4\)):
\begin{align*}
V ={}& 
\Fpic{B_1}{B_2}{B_3}{B_4}{BBBBome} - \psd\cdot\Fpic{B_1}{B_2}{B_4}{B_3}{BBBBome} +
\Fpic{B_1}{B_2}{B_3}{B_4}{BBBBct} - \psd\cdot\Fpic{B_1}{B_2}{B_4}{B_3}{BBBBct} \\
={}& 
-\frac{X_1}{4f^2_0} \frac{\vec\sigma_1\cdot\vec q\ \vec\sigma_2\cdot\vec q}{\vec q^{\,2}+m^2_\phi}
+\frac{X_2}{4f^2_0} \psd\frac{\vec\sigma_1\cdot\vec k\ \vec\sigma_2\cdot\vec k}{\vec k^{\,2}+m^2_\phi} \\
&+ X_3 \ONE + X_4 \vec\sigma_1\cdot\vec\sigma_2 \,, \numberthis
\end{align*} 

\item \(B_1B_2\to B_3^*B_4\): 
\begin{align*}
V ={}&
\Fpic{B_1}{B_2}{B_3^*}{B_4}{BBDBome} - \psdq\cdot\Fpic{B_1}{B_2}{B_4}{B_3^*}{BBBDome} +
\Fpic{B_1}{B_2}{B_3^*}{B_4}{BBDBct} - \psdq\cdot\Fpic{B_1}{B_2}{B_4}{B_3^*}{BBBDct} \\
={}&
\frac{C X_1}{2f^2_0} \frac{\vec S^{\,\dagger}_1\cdot\vec q\ \vec\sigma_2\cdot\vec q}{\vec q^{\,2}+m^2_\phi}
-\frac{C X_2}{2f^2_0} \psdq \frac{\vec\sigma_1\cdot\vec k\ \vec S^{\,\dagger}_2\cdot\vec k}{\vec k^{\,2}+m^2_\phi} \\
&+ X_3 \vec S^{\,\dagger}_1 \cdot \vec\sigma_2 \,, \numberthis
\end{align*}

\item \(B_1^*B_2\to B_3B_4\): 
\begin{align*}
V ={}&
\Fpic{B^*_1}{B_2}{B_3}{B_4}{DBBBome} - \psd\cdot\Fpic{B^*_1}{B_2}{B_4}{B_3}{DBBBome} +
\Fpic{B^*_1}{B_2}{B_3}{B_4}{DBBBct} - \psd\cdot\Fpic{B^*_1}{B_2}{B_4}{B_3}{DBBBct} \\
={}&
\frac{C X_1}{2f^2_0} \frac{\vec S_1\cdot\vec q\ \vec\sigma_2\cdot\vec q}{\vec q^{\,2}+m^2_\phi}
-\frac{C X_2}{2f^2_0} \psd \frac{\vec S_1\cdot\vec k\ \vec \sigma_2\cdot\vec k}{\vec k^{\,2}+m^2_\phi}
+ X_3 \vec S_1 \cdot \vec\sigma_2 \,, \numberthis
\end{align*}

\item \(B_1^*B_2\to B^*_3B_4\): 
\begin{align*}
V &= 
\Fpic{B^*_1}{B_2}{B_3^*}{B_4}{DBDBome} - \psdq\cdot\Fpic{B^*_1}{B_2}{B_4}{B_3^*}{DBBDome} +
\Fpic{B^*_1}{B_2}{B_3^*}{B_4}{DBDBct} - \psdq\cdot\Fpic{B^*_1}{B_2}{B_4}{B_3^*}{DBBDct} \\
={}& 
\frac{X_1 H}{2f^2_0} \frac{\vec\Sigma_1\cdot\vec q\ \vec\sigma_2\cdot\vec q}{\vec q^{\,2}+m^2_\phi}
+\frac{X_2 C^2}{f^2_0} \psdq\frac{\vec S_1\cdot\vec k\ \vec S^{\,\dagger}_2\cdot\vec k}{\vec k^{\,2}+m^2_\phi} \\
&+ X_3 \ONE + X_4 \vec\Sigma_1\cdot\vec\sigma_2 \,, \numberthis
\end{align*}

\item \(B_1B_2\to B^*_3B^*_4\): 
\begin{align*}
V ={}&
\Fpic{B_1}{B_2}{B_3^*}{B^*_4}{BBDDome} - \psq\cdot\Fpic{B_1}{B_2}{B^*_4}{B_3^*}{BBDDome} +
\Fpic{B_1}{B_2}{B_3^*}{B^*_4}{BBDDct} - \psq\cdot\Fpic{B_1}{B_2}{B^*_4}{B_3^*}{BBDDct} \\
={}&
-\frac{X_1 C^2}{f^2_0} \frac{\vec S^{\,\dagger}_1\cdot\vec q\ \vec S^{\,\dagger}_2\cdot\vec q}{\vec q^{\,2}+m^2_\phi}
+\frac{X_2 C^2}{f^2_0} \psq\frac{\vec S^{\,\dagger}_1\cdot\vec k\ \vec S^{\,\dagger}_2\cdot\vec k}{\vec k^{\,2}+m^2_\phi} \\
&+ X_3 \vec S^\dagger_1 \cdot \vec S^\dagger_2 + X_4 S^{\alpha\beta\dagger}_1 S^{\alpha\beta\dagger}_2 \,, \numberthis
\end{align*}

\item \(B^*_1B^*_2\to B_3B_4\): 
\begin{align*}
V ={}& 
\Fpic{B^*_1}{B^*_2}{B_3}{B_4}{DDBBome} - \psd\cdot\Fpic{B^*_1}{B^*_2}{B_4}{B_3}{DDBBome} +
\Fpic{B^*_1}{B^*_2}{B_3}{B_4}{DDBBct} - \psd\cdot\Fpic{B^*_1}{B^*_2}{B_4}{B_3}{DDBBct} \\
={}& 
-\frac{X_1C^2}{f^2_0} \frac{\vec S_1\cdot\vec q\ \vec S_2\cdot\vec q}{\vec q^{\,2}+m^2_\phi}
+\frac{X_2C^2}{f^2_0} \psd\frac{\vec S_1\cdot\vec k\ \vec S_2\cdot\vec k}{\vec k^{\,2}+m^2_\phi} \\
&+ X_3 \vec S_1 \cdot \vec S_2 + X_4 S^{\alpha\beta}_1 S^{\alpha\beta}_2 \,, \numberthis
\end{align*} 

\item \(B^*_1B_2\to B^*_3B^*_4\): 
\begin{align*}
V ={}& 
\Fpic{B^*_1}{B_2}{B_3^*}{B^*_4}{DBDDome} - \psq\cdot\Fpic{B^*_1}{B_2}{B^*_4}{B_3^*}{DBDDome} +
\Fpic{B^*_1}{B_2}{B_3^*}{B^*_4}{DBDDct} - \psq\cdot\Fpic{B^*_1}{B_2}{B^*_4}{B_3^*}{DBDDct} \\
={}& 
-\frac{X_1C H}{f^2_0} \frac{\vec\Sigma_1\cdot\vec q\ \vec S^{\,\dagger}_2\cdot\vec q}{\vec q^{\,2}+m^2_\phi}
+\frac{X_2C H}{f^2_0} \psq\frac{\vec\Sigma_1\cdot\vec k\ \vec S^{\,\dagger}_2\cdot\vec k}{\vec k^{\,2}+m^2_\phi} \\
&+ X_3 \vec \Sigma_1\cdot\vec S_2^\dagger + X_4 \Sigma_1^{\alpha\beta}S_2^{\alpha\beta\dagger} \,, \numberthis
\end{align*}

\item \(B^*_1B^*_2\to B^*_3B_4\): 
\begin{align*}
V ={}& 
\Fpic{B^*_1}{B^*_2}{B_3^*}{B_4}{DDDBome} - \psdq\cdot\Fpic{B^*_1}{B^*_2}{B_4}{B_3^*}{DDBDome} +
\Fpic{B^*_1}{B^*_2}{B_3^*}{B_4}{DDDBct} - \psdq\cdot\Fpic{B^*_1}{B^*_2}{B_4}{B_3^*}{DDBDct} \\
={}& 
-\frac{X_1C H}{f^2_0} \frac{\vec\Sigma_1\cdot\vec q\ \vec S_2\cdot\vec q}{\vec q^{\,2}+m^2_\phi}
+\frac{X_2C H}{f^2_0} \psdq\frac{\vec S_1\cdot\vec k\ \vec\Sigma_2\cdot\vec k}{\vec k^{\,2}+m^2_\phi} \\
&+ X_3 \vec \Sigma_1\cdot\vec S_2 + X_4 \Sigma_1^{\alpha\beta}S_2^{\alpha\beta} \,, \numberthis
\end{align*}

\item \(B^*_1B^*_2\to B^*_3B^*_4\): 
\begin{align*}
V ={}&
\Fpic{B^*_1}{B^*_2}{B_3^*}{B^*_4}{DDDDome} - \psq\cdot\Fpic{B^*_1}{B^*_2}{B^*_4}{B_3^*}{DDDDome} +
\Fpic{B^*_1}{B^*_2}{B_3^*}{B^*_4}{DDDDct} - \psq\cdot\Fpic{B^*_1}{B^*_2}{B^*_4}{B_3^*}{DDDDct} \\
={}&
-\frac{X_1H^2}{f^2_0} \frac{\vec\Sigma_1\cdot\vec q\ \vec\Sigma_2\cdot\vec q}{\vec q^{\,2}+m^2_\phi}
+\frac{X_2H^2}{f^2_0} \psq\frac{\vec\Sigma_1\cdot\vec k\ \vec\Sigma_2\cdot\vec k}{\vec k^{\,2}+m^2_\phi} \\
&+ X_3 \ONE + X_4 \vec\Sigma_1\cdot\vec\Sigma_2
+ X_5 \Sigma_1^{\alpha\beta}\Sigma_2^{\alpha\beta} + X_6 \Sigma_1^{\alpha\beta\gamma}\Sigma_2^{\alpha\beta\gamma} \,, \numberthis
\end{align*}

\end{itemize}

\newcolumntype{f}{>{\normalsize$}c<{$}}
\newcolumntype{F}{>{\centering\normalsize\arraybackslash$}p{8.cm}<{$}}

\begin{table}[p]
\centering
\rotatebox{90}{ %
\begin{tabular}{fffFF}
\toprule
\text{process} & X_1 & X_2 & X_3 & X_4 \\
\midrule
B_1B_2\to B_3B_4 &
N_{B_3B_1\bar\phi} N_{B_4B_2\phi} &
N_{B_4B_1\bar\phi} N_{B_3B_2\phi} &
- \sumCzza (N^f_{B_{3}B_{1}B_{4}B_{2}}+N^f_{B_{4}B_{2}B_{3}B_{1}}) 
+ \frac12 \sumCzza (N^f_{B_{4}B_{1}B_{3}B_{2}}+N^f_{B_{3}B_{2}B_{4}B_{1}})
+ \frac32 \sumCzzb (N^f_{B_{4}B_{1}B_{3}B_{2}}+N^f_{B_{3}B_{2}B_{4}B_{1}}) &
- \sumCzzb (N^f_{B_{3}B_{1}B_{4}B_{2}}+N^f_{B_{4}B_{2}B_{3}B_{1}})
+ \frac12 \sumCzza (N^f_{B_{4}B_{1}B_{3}B_{2}}+N^f_{B_{3}B_{2}B_{4}B_{1}})
- \frac12 \sumCzzb (N^f_{B_{4}B_{1}B_{3}B_{2}}+N^f_{B_{3}B_{2}B_{4}B_{1}}) \\
B_1B_2\to B^*_3B_4 &
N_{B^*_3B_1\bar\phi} N_{B_4B_2\phi} &
N_{B_4B_1\bar\phi} N_{B^*_3B_2\phi} &
-\sumCzoa N^f_{B^*_3B_{1}B_{4}B_{2}}
+ \sumCzoa N^f_{B^*_3B_{2}B_{4}B_{1}} \\
B^*_1B_2\to B_3B_4 &
N_{B^*_1B_3\phi} N_{B_4B_2\phi} &
N_{B^*_1B_4\phi} N_{B_3B_2\phi} &
-\sumCzoa N^f_{B^*_{1}B_3B_{2}B_{4}}
+ \sumCzoa N^f_{B^*_{1}B_4B_{2}B_{3}} \\
B^*_1B_2\to B^*_3B_4 &
N_{B^*_3B^*_1\bar\phi} N_{B_4B_2\phi} &
N_{B^*_1B_4\phi} N_{B^*_3B_2\phi} &
- \sumCooa N^f_{B^*_{3}B^*_{1}B_{4}B_{2}} &
- \sumCoob N^f_{B^*_{3}B^*_{1}B_{4}B_{2}} \\
B_1B_2\to B^*_3B^*_4 &
N_{B^*_3B_1\bar\phi} N_{B^*_4B_2\phi} &
N_{B^*_4B_1\bar\phi} N_{B^*_3B_2\phi} &
-\sumCzta (N^1_{B^*_{3}B_{1}B^*_{4}B_{2}}+N^1_{B^*_{4}B_{2}B^*_{3}B_{1}})
+ \frac14 \sumCzta (N^1_{B^*_{4}B_{1}B^*_{3}B_{2}}+N^1_{B^*_{3}B_{2}B^*_{4}B_{1}})
+ \frac54 \sumCztb (N^2_{B^*_{4}B_{1}B^*_{3}B_{2}}+N^2_{B^*_{3}B_{2}B^*_{4}B_{1}}) &
-\sumCztb (N^2_{B^*_{3}B_{1}B^*_{4}B_{2}}+N^2_{B^*_{4}B_{2}B^*_{3}B_{1}})
+ \frac34 \sumCzta (N^1_{B^*_{4}B_{1}B^*_{3}B_{2}}+N^1_{B^*_{3}B_{2}B^*_{4}B_{1}})
- \frac14 \sumCztb (N^2_{B^*_{4}B_{1}B^*_{3}B_{2}}+N^2_{B^*_{3}B_{2}B^*_{4}B_{1}}) \\
B^*_1B^*_2\to B_3B_4 &
N_{B^*_1B_3\phi} N_{B^*_2B_4\bar\phi} &
N_{B^*_1B_4\phi} N_{B^*_2B_3\bar\phi} &
- \sumCzta (N^1_{B^*_{1}B_{3}B^*_{2}B_{4}}+N^1_{B^*_{2}B_{4}B^*_{1}B_{3}})
+ \frac14 \sumCzta (N^1_{B^*_{1}B_{4}B^*_{2}B_{3}}+N^1_{B^*_{2}B_{3}B^*_{1}B_{4}})
+ \frac54 \sumCztb (N^2_{B^*_{1}B_{4}B^*_{2}B_{3}}+N^2_{B^*_{2}B_{3}B^*_{1}B_{4}}) &
- \sumCztb (N^2_{B^*_{1}B_{3}B^*_{2}B_{4}}+N^2_{B^*_{2}B_{4}B^*_{1}B_{3}})
+ \frac34 \sumCzta (N^1_{B^*_{1}B_{4}B^*_{2}B_{3}}+N^1_{B^*_{2}B_{3}B^*_{1}B_{4}})
- \frac14 \sumCztb (N^2_{B^*_{1}B_{4}B^*_{2}B_{3}}+N^2_{B^*_{2}B_{3}B^*_{1}B_{4}}) \\
B^*_1B_2\to B^*_3B^*_4 &
N_{B^*_3B^*_1\bar\phi} N_{B^*_4B_2\phi} &
N_{B^*_4B^*_1\bar\phi} N_{B^*_3B_2\phi} &
-\sumCota N^1_{B^*_{3}B^*_{1}B^*_{4}B_{2}}
- \frac12 \sumCota N^1_{B^*_{4}B^*_{1}B^*_{3}B_{2}}
+ \frac12\sqrt{\frac32} \sumCotb N^2_{B^*_{4}B^*_{1}B^*_{3}B_{2}} &
- \sumCotb N^2_{B^*_{3}B^*_{1}B^*_{4}B_{2}}
+ \sqrt{\frac32}\sumCota N^1_{B^*_{4}B^*_{1}B^*_{3}B_{2}}
+ \frac12 \sumCotb N^2_{B^*_{4}B^*_{1}B^*_{3}B_{2}} \\
B^*_1B^*_2\to B^*_3B_4 &
N_{B^*_3B^*_1\bar\phi} N_{B^*_2B_4\bar\phi} &
N_{B^*_1B_4\phi} N_{B^*_3B^*_2\phi} &
- \sumCota N^1_{B^*_{1}B^*_{3}B^*_{2}B_{4}}
- \frac12 \sumCota N^1_{B^*_{2}B^*_{3}B^*_{1}B_{4}}
+ \frac12\sqrt{\frac32} \sumCotb N^2_{B^*_{2}B^*_{3}B^*_{1}B_{4}} &
- \sumCotb N^2_{B^*_{1}B^*_{3}B^*_{2}B_{4}}
+ \sqrt{\frac32} \sumCota N^1_{B^*_{2}B^*_{3}B^*_{1}B_{4}}
+ \frac12 \sumCotb N^2_{B^*_{2}B^*_{3}B^*_{1}B_{4}} \\
\bottomrule
\end{tabular}
}
\caption{Coefficients \(X_i\) for various interactions.} \label{tab:xi}
\end{table}

\begin{table}[p]
\centering
\rotatebox{90}{ %
\begin{tabular}{f>{\centering\normalsize\arraybackslash$}p{22.5cm}<{$}}
\toprule
X_1 &
N_{B^*_3B^*_1\bar\phi} N_{B^*_4B^*_2\phi} \\
X_2 &
N_{B^*_4B^*_1\bar\phi} N_{B^*_3B^*_2\phi} \\
X_3 &
- \sumCtta (N^f_{B^*_{3}B^*_{1}B^*_{4}B^*_{2}}+N^f_{B^*_{4}B^*_{2}B^*_{3}B^*_{1}})
+ \frac14 \sumCtta (N^f_{B^*_{4}B^*_{1}B^*_{3}B^*_{2}}+N^f_{B^*_{3}B^*_{2}B^*_{4}B^*_{1}})
+ \frac{15}4 \sumCttb (N^f_{B^*_{4}B^*_{1}B^*_{3}B^*_{2}}+N^f_{B^*_{3}B^*_{2}B^*_{4}B^*_{1}})
+ \frac{15}8 \sumCttc (N^f_{B^*_{4}B^*_{1}B^*_{3}B^*_{2}}+N^f_{B^*_{3}B^*_{2}B^*_{4}B^*_{1}})
+ \frac{175}6 \sumCttd (N^f_{B^*_{4}B^*_{1}B^*_{3}B^*_{2}}+N^f_{B^*_{3}B^*_{2}B^*_{4}B^*_{1}}) \\
X_4 &
- \sumCttb (N^f_{B^*_{3}B^*_{1}B^*_{4}B^*_{2}}+N^f_{B^*_{4}B^*_{2}B^*_{3}B^*_{1}})
+ \frac1{20} \sumCtta (N^f_{B^*_{4}B^*_{1}B^*_{3}B^*_{2}}+N^f_{B^*_{3}B^*_{2}B^*_{4}B^*_{1}})
+ \frac{11}{20} \sumCttb (N^f_{B^*_{4}B^*_{1}B^*_{3}B^*_{2}}+N^f_{B^*_{3}B^*_{2}B^*_{4}B^*_{1}})
+ \frac3{40} \sumCttc (N^f_{B^*_{4}B^*_{1}B^*_{3}B^*_{2}}+N^f_{B^*_{3}B^*_{2}B^*_{4}B^*_{1}})
- \frac72 \sumCttd (N^f_{B^*_{4}B^*_{1}B^*_{3}B^*_{2}}+N^f_{B^*_{3}B^*_{2}B^*_{4}B^*_{1}}) \\
X_5 &
- \sumCttc (N^f_{B^*_{3}B^*_{1}B^*_{4}B^*_{2}}+N^f_{B^*_{4}B^*_{2}B^*_{3}B^*_{1}})
+ \frac16 \sumCtta (N^f_{B^*_{4}B^*_{1}B^*_{3}B^*_{2}}+N^f_{B^*_{3}B^*_{2}B^*_{4}B^*_{1}})
+ \frac12 \sumCttb (N^f_{B^*_{4}B^*_{1}B^*_{3}B^*_{2}}+N^f_{B^*_{3}B^*_{2}B^*_{4}B^*_{1}})
- \frac34 \sumCttc (N^f_{B^*_{4}B^*_{1}B^*_{3}B^*_{2}}+N^f_{B^*_{3}B^*_{2}B^*_{4}B^*_{1}})
+ \frac{35}9 \sumCttd (N^f_{B^*_{4}B^*_{1}B^*_{3}B^*_{2}}+N^f_{B^*_{3}B^*_{2}B^*_{4}B^*_{1}}) \\
X_6 &
- \sumCttd (N^f_{B^*_{3}B^*_{1}B^*_{4}B^*_{2}}+N^f_{B^*_{4}B^*_{2}B^*_{3}B^*_{1}})
+ \frac3{200} \sumCtta (N^f_{B^*_{4}B^*_{1}B^*_{3}B^*_{2}}+N^f_{B^*_{3}B^*_{2}B^*_{4}B^*_{1}})
- \frac{27}{200} \sumCttb (N^f_{B^*_{4}B^*_{1}B^*_{3}B^*_{2}}+N^f_{B^*_{3}B^*_{2}B^*_{4}B^*_{1}})
+ \frac9{400} \sumCttc (N^f_{B^*_{4}B^*_{1}B^*_{3}B^*_{2}}+N^f_{B^*_{3}B^*_{2}B^*_{4}B^*_{1}})
- \frac1{20} \sumCttd (N^f_{B^*_{4}B^*_{1}B^*_{3}B^*_{2}}+N^f_{B^*_{3}B^*_{2}B^*_{4}B^*_{1}}) \\
\bottomrule
\end{tabular}
}
\caption{Coefficients \(X_i\) for the transition \(B^*_1B^*_2\to B^*_3B^*_4\).} \label{tab:xi2}
\end{table}

Analogous to eq.~(19) of ref.~\cite{Petschauer:2015elq}, we obtain a representation of the potentials in isospin
basis, by applying the relation\footnote{
In order to conform with the convention used in our previous works \cite{Polinder:2007mp,Haidenbauer:2015} 
the potential is multiplied with a factor \(1/\sqrt2\) each, for identical particles in the initial and/or final state.
For example, the potential of the transition \(\Xi^* N\to \Sigma^*\Sigma^*\) is, therefore, multiplied with a 
factor \(1/\sqrt2\) and the potential of the transition \(\Sigma^*\Sigma^*\to \Sigma^*\Sigma^*\) with a factor \(1/2\).}
\begin{align} \label{eq:iso}
&\langle(i_{3}i_{4})IM|V|(i_1i_2)IM\rangle
=\sum_{\substack{m_1,m_2,\\m_{3 },m_{4}}}
\delta_{M,m_{3}+m_{4}}
\delta_{M,m_1+m_2} \notag \\
&\qquad \times
C^{i_{3}i_{4}I}_{m_{3}m_{4}M}
C^{i_1i_2I}_{m_1m_2M}
\langle i_{3}m_{3};i_{4}m_{4}|V|i_1m_1;i_2m_2\rangle
\end{align}
to the potential \(V\) in particle basis, where \(I\) is the total 
isospin and \(i_j\) is the isospin of the particles.
This approach is used in order to obtain the SU(3) relations shown in appendix~\ref{app:SU3}.

In order to employ the transformation to the isospin basis, eq.~\eqref{eq:iso}, with the conventional Clebsch-Gordan coefficients \(C\), 
the following sign changes in identifying the isospin eigenstates with the particle fields are necessary:
\begin{align*}
&\Sigma^+=-\vert 1,+1\rangle\,,\
\Xi^-=-\vert 1/2,-1/2\rangle\,,\ \numberthis \\
&\pi^+=-\vert 1,+1\rangle\,,\
K^-=-\vert 1/2,-1/2\rangle\,, \\
&\Sigma^{*+}=-\vert 1,+1\rangle\,,\
\Sigma^{*0}=-\vert 1,0\rangle\,,\
\Sigma^{*-}=-\vert 1,-1\rangle\,,\
\Omega^-=-\vert 0,0\rangle\,.
\end{align*}

\clearpage
\onecolumn
\section{SU(3) relations} 
\label{app:SU3}

In this appendix, we display our results for the SU(3) relations of the two-baryon interactions, 
following the approach described in app.~\ref{app:BBpart}.
In order to obtain the various SU(3) relations for the considered two-body interactions, 
we have projected the contact terms onto partial waves (see subsect.~\ref{sec:pwd}) and switched to the isospin basis (see Appendix \ref{app:BBpart}).

The low-energy constants $c^f_{ij}$ of the minimal contact Lagrangian terms have been redefined according 
to the group theoretical considerations in sec.~\ref{sec:group}. One obtains the following relations. \\
\begin{itemize}
\item \(BB\to BB\):
\begin{align*}
\Gzz^{1}  &= \frac{2}{3} (\Czz^1 -3 \Czz^2 -8 \Czz^3 +24 \Czz^4 -3 \Czz^5 +9 \Czz^6 ) \\ 
\Gzz^{8_s}  &= \frac{4 \Czz^1 }{3}-4 \Czz^2 -\frac{5 \Czz^3 }{3}+5 \Czz^4 -2 \Czz^5 +6 \Czz^6  \\ 
\Gzz^{8_a}  &= 3 \Czz^3 +3 \Czz^4 -2 (\Czz^5 +\Czz^6 ) \\ 
\Gzz^{10}  &= 2 (\Czz^1 +\Czz^2 -\Czz^5 -\Czz^6 ) \\ 
\Gzz^{\overline{10}}  &= -2 (\Czz^1 +\Czz^2 +\Czz^5 +\Czz^6 ) \\ 
\Gzz^{27} &= -2 (\Czz^1 -3 \Czz^2 +\Czz^5 -3 \Czz^6 )
\numberthis
\end{align*}
\item \(BB\to BD\) and \(BD\to BB\):
\begin{align*}
\Gzo^{8}  &= -\frac{1}{3} \sqrt{2} (\Czo^1 +3 \Czo^2 ) \\ 
\Gzo^{10}  &= -\frac{2 \sqrt{2} \Czo^1 }{3}
\numberthis
\end{align*}
\item \(DB\to DB\):
\begin{align*}
\Goo^{35,3S1} &= -\Coo^1 +5 \Coo^2 -\Coo^5 +5 \Coo^6  \\ 
\Goo^{27,3S1} &= \frac{1}{3} (-3 \Coo^1 +15 \Coo^2 +\Coo^5 -5 \Coo^6 ) \\ 
\Goo^{10,3S1}  &= \frac{1}{3} (-3 \Coo^1 +15 \Coo^2 -4 \Coo^3 +20 \Coo^4 -\Coo^5 +5 \Coo^6 -4 \Coo^7 +20 \Coo^8 ) \\ 
\Goo^{8,3S1}  &= \frac{1}{6} (-6 \Coo^1 +30 \Coo^2 -10 \Coo^3 +50 \Coo^4 +2 \Coo^5 -10 \Coo^6 +5 \Coo^7 -25 \Coo^8 ) \\ 
\Goo^{35,5S2} &= -\Coo^1 -3 \Coo^2 -\Coo^5 -3 \Coo^6  \\ 
\Goo^{27,5S2} &= -\Coo^1 -3 \Coo^2 +\frac{\Coo^5 }{3}+\Coo^6  \\ 
\Goo^{10,5S2}  &= \frac{1}{3} (-3 \Coo^1 -9 \Coo^2 -4 \Coo^3 -12 \Coo^4 -\Coo^5 -3 \Coo^6 -4 \Coo^7 -12 \Coo^8 ) \\ 
\Goo^{8,5S2}  &= -\Coo^1 -3 \Coo^2 -\frac{5 \Coo^3 }{3}-5 \Coo^4 +\frac{\Coo^5 }{3}+\Coo^6 +\frac{5 \Coo^7 }{6}+\frac{5 \Coo^8 }{2}
\numberthis
\end{align*}
\item \(BB\to DD\) and \(DD\to BB\):
\begin{align*}
\Gzt^{27} &= \frac{4}{9} \sqrt{5} (3 \Czt^1 -5 \Czt^2 ) \\ 
\Gzt^{\overline{10}}  &= \frac{4}{3} \sqrt{5} (\Czt^1 +\Czt^2 )
\numberthis
\end{align*}
\item \(DB\to DD\) and \(DD\to DB\):
\begin{align*}
\Got^{27} &= -\frac{2}{9} \sqrt{5} \left(6 \Cot^1 -\sqrt{6} \Cot^2 \right) \\ 
\Got^{35} &= -\frac{2}{3} \sqrt{5} \left(2 \Cot^1 +\sqrt{6} \Cot^2 \right)
\numberthis
\end{align*}
\item \(DD\to DD\):
\begin{align*}
\Gtt^{\overline{10},3S1}  &= \frac{1}{3} (-6 \Ctt^1 +66 \Ctt^2 -9 \Ctt^3 -420 \Ctt^4 +2 \Ctt^5 -22 \Ctt^6 +3 \Ctt^7 +140 \Ctt^8 ) \\ 
\Gtt^{\overline{10},7S3}  &= -2 \Ctt^1 -18 \Ctt^2 -3 \Ctt^3 -\frac{20 \Ctt^4 }{3}+\frac{2 \Ctt^5 }{3}+6 \Ctt^6 +\Ctt^7 +\frac{20 \Ctt^8 }{9} \\ 
\Gtt^{27,1S0} &= \frac{1}{27} (-54 \Ctt^1 +810 \Ctt^2 -405 \Ctt^3 +6300 \Ctt^4 +6 \Ctt^5 -90 \Ctt^6 +45 \Ctt^7 -700 \Ctt^8 ) \\ 
\Gtt^{27,5S2} &= -2 \Ctt^1 +6 \Ctt^2 +9 \Ctt^3 +\frac{140 \Ctt^4 }{3}+\frac{2 \Ctt^5 }{9}-\frac{2 \Ctt^6 }{3}-\Ctt^7 -\frac{140 \Ctt^8 }{27} \\ 
\Gtt^{35,3S1} &= \frac{1}{3} (-6 \Ctt^1 +66 \Ctt^2 -9 \Ctt^3 -420 \Ctt^4 -2 \Ctt^5 +22 \Ctt^6 -3 \Ctt^7 -140 \Ctt^8 ) \\ 
\Gtt^{35,7S3} &= -2 \Ctt^1 -18 \Ctt^2 -3 \Ctt^3 -\frac{20 \Ctt^4 }{3}-\frac{2 \Ctt^5 }{3}-6 \Ctt^6 -\Ctt^7 -\frac{20 \Ctt^8 }{9} \\ 
\Gtt^{28,1S0} &= \frac{1}{3} (-6 \Ctt^1 +90 \Ctt^2 -45 \Ctt^3 +700 \Ctt^4 -6 \Ctt^5 +90 \Ctt^6 -45 \Ctt^7 +700 \Ctt^8 ) \\ 
\Gtt^{28,5S2} &= -2 \Ctt^1 +6 \Ctt^2 +9 \Ctt^3 +\frac{140 \Ctt^4 }{3}-2 \Ctt^5 +6 \Ctt^6 +9 \Ctt^7 +\frac{140 \Ctt^8 }{3}
\numberthis
\end{align*}
\end{itemize}

The SU(3) tables of the two-body transitions (employing the redefined constants above) are given in 
Tables \ref{tab:PWDDDDD}, \ref{tab:PWDBBBB}, \ref{tab:PWDBBDB}, \ref{tab:PWDBBDD}, \ref{tab:PWDDBDB}, \ref{tab:PWDDBDD}.

\LTcapwidth=\textwidth

\renewcommand{\sharedhead}{
\toprule
S & I & \text{transition} & V_{{}^1S_{0}} & V_{{}^3S_{1}} \\
\cmidrule(lr){1-3}\cmidrule(lr){4-5}
}
\begin{longtable}{ttttt}
\caption{SU(3) relations of \(BB\to BB\) in non-vanishing partial waves. 
The subscript $\{00\}$ of the constants \(\Gzz^r\) that denotes the 
$\mathcal{B}\mathcal{B}$ channel is omitted in the table.
} \label{tab:PWDBBBB} \\ \sharedhead
\endfirsthead
\caption{(\dots continued)} \\ \sharedhead
\endhead
\bottomrule \multicolumn{5}{l}{\scriptsize \dots continues on next page}
\endfoot
\bottomrule
\endlastfoot
0 & 0 & N N \leftrightarrow N N & 0 & \GTzz^{\overline{10}}  \\ 
0 & 1 & N N \leftrightarrow N N & \GTzz^{27} & 0 \\ 

\cmidrule(lr){1-3}\cmidrule(lr){4-5}
-1 & \frac{1}{2} & \Lambda N \leftrightarrow \Lambda N & \frac{1}{10} (9 \GTzz^{27}+\GTzz^{8_s} ) & \frac{\GTzz^{\overline{10}} +\GTzz^{8_a} }{2} \\ 
-1 & \frac{1}{2} & \Lambda N \leftrightarrow \Sigma N & -\frac{3}{10} (\GTzz^{27}-\GTzz^{8_s} ) & \frac{\GTzz^{\overline{10}} -\GTzz^{8_a} }{2} \\ 
-1 & \frac{1}{2} & \Sigma N \leftrightarrow \Sigma N & \frac{1}{10} (\GTzz^{27}+9 \GTzz^{8_s} ) & \frac{\GTzz^{\overline{10}} +\GTzz^{8_a} }{2} \\ 
-1 & \frac{3}{2} & \Sigma N \leftrightarrow \Sigma N & \GTzz^{27} & \GTzz^{10}  \\ 

\cmidrule(lr){1-3}\cmidrule(lr){4-5}
-2 & 0 & \Lambda \Lambda \leftrightarrow \Lambda \Lambda & \frac{1}{40} (5 \GTzz^{1} +27 \GTzz^{27}+8 \GTzz^{8_s} ) & 0 \\ 
-2 & 0 & \Lambda \Lambda \leftrightarrow \Sigma \Sigma & -\frac{1}{40} \sqrt{3} (5 \GTzz^{1} +3 \GTzz^{27}-8 \GTzz^{8_s} ) & 0 \\ 
-2 & 0 & \Xi N \leftrightarrow \Lambda \Lambda & \frac{1}{20} (5 \GTzz^{1} -9 \GTzz^{27}+4 \GTzz^{8_s} ) & 0 \\ 
-2 & 0 & \Xi N \leftrightarrow \Xi N & \frac{1}{10} (5 \GTzz^{1} +3 \GTzz^{27}+2 \GTzz^{8_s} ) & \GTzz^{8_a}  \\ 
-2 & 0 & \Xi N \leftrightarrow \Sigma \Sigma & \frac{1}{20} \sqrt{3} (-5 \GTzz^{1} +\GTzz^{27}+4 \GTzz^{8_s} ) & 0 \\ 
-2 & 0 & \Sigma \Sigma \leftrightarrow \Sigma \Sigma & \frac{1}{40} (15 \GTzz^{1} +\GTzz^{27}+24 \GTzz^{8_s} ) & 0 \\ 
-2 & 1 & \Xi N \leftrightarrow \Xi N & \frac{1}{5} (2 \GTzz^{27}+3 \GTzz^{8_s} ) & \frac{1}{3} (\GTzz^{10} +\GTzz^{\overline{10}} +\GTzz^{8_a} ) \\ 
-2 & 1 & \Xi N \leftrightarrow \Sigma \Lambda & \frac{1}{5} \sqrt{6} (\GTzz^{27}-\GTzz^{8_s} ) & \frac{\GTzz^{10} -\GTzz^{\overline{10}} }{\sqrt{6}} \\ 
-2 & 1 & \Xi N \leftrightarrow \Sigma \Sigma & 0 & \frac{\GTzz^{10} +\GTzz^{\overline{10}} -2 \GTzz^{8_a} }{3 \sqrt{2}} \\ 
-2 & 1 & \Sigma \Lambda \leftrightarrow \Sigma \Lambda & \frac{1}{5} (3 \GTzz^{27}+2 \GTzz^{8_s} ) & \frac{\GTzz^{10} +\GTzz^{\overline{10}} }{2} \\ 
-2 & 1 & \Sigma \Lambda \leftrightarrow \Sigma \Sigma & 0 & \frac{\GTzz^{10} -\GTzz^{\overline{10}} }{2 \sqrt{3}} \\ 
-2 & 1 & \Sigma \Sigma \leftrightarrow \Sigma \Sigma & 0 & \frac{1}{6} (\GTzz^{10} +\GTzz^{\overline{10}} +4 \GTzz^{8_a} ) \\ 
-2 & 2 & \Sigma \Sigma \leftrightarrow \Sigma \Sigma & \GTzz^{27} & 0 \\ 

\cmidrule(lr){1-3}\cmidrule(lr){4-5}
-3 & \frac{1}{2} & \Xi \Lambda \leftrightarrow \Xi \Lambda & \frac{1}{10} (9 \GTzz^{27}+\GTzz^{8_s} ) & \frac{\GTzz^{10} +\GTzz^{8_a} }{2} \\ 
-3 & \frac{1}{2} & \Xi \Lambda \leftrightarrow \Xi \Sigma & -\frac{3}{10} (\GTzz^{27}-\GTzz^{8_s} ) & \frac{\GTzz^{10} -\GTzz^{8_a} }{2} \\ 
-3 & \frac{1}{2} & \Xi \Sigma \leftrightarrow \Xi \Sigma & \frac{1}{10} (\GTzz^{27}+9 \GTzz^{8_s} ) & \frac{\GTzz^{10} +\GTzz^{8_a} }{2} \\ 
-3 & \frac{3}{2} & \Xi \Sigma \leftrightarrow \Xi \Sigma & \GTzz^{27} & \GTzz^{\overline{10}}  \\ 

\cmidrule(lr){1-3}\cmidrule(lr){4-5}
-4 & 0 & \Xi \Xi \leftrightarrow \Xi \Xi & 0 & \GTzz^{10}  \\ 
-4 & 1 & \Xi \Xi \leftrightarrow \Xi \Xi & \GTzz^{27} & 0 \\ 

\end{longtable}

\renewcommand{\sharedhead}{
\toprule
S & I & \text{transition} & V_{{}^3S_{1}} \\
\cmidrule(lr){1-3}\cmidrule(lr){4-4}
}
\begin{longtable}{tttt}
\caption{SU(3) relations of \(BB\to DB\) (and  \(DB\to BB\)) in non-vanishing partial waves. 
The subscript $\{01\}$ of the constants \(\Gzo^r\) that denotes the
$\mathcal{B}\mathcal{B}$ channel is omitted in the table.
} \label{tab:PWDBBDB} \\ \sharedhead
\endfirsthead
\caption{(\dots continued)} \\ \sharedhead
\endhead
\bottomrule \multicolumn{4}{l}{\scriptsize \dots continues on next page}
\endfoot
\bottomrule
\endlastfoot
-1 & \frac{1}{2} & \Lambda N \leftrightarrow \Sigma^*  N & \GTzo^{8}  \\ 
-1 & \frac{1}{2} & \Lambda N \leftrightarrow \Delta \Sigma & -2 \GTzo^{8}  \\ 
-1 & \frac{1}{2} & \Sigma N \leftrightarrow \Sigma^*  N & -\GTzo^{8}  \\ 
-1 & \frac{1}{2} & \Sigma N \leftrightarrow \Delta \Sigma & 2 \GTzo^{8}  \\ 
-1 & \frac{3}{2} & \Sigma N \leftrightarrow \Sigma^*  N & \GTzo^{10}  \\ 
-1 & \frac{3}{2} & \Sigma N \leftrightarrow \Delta \Lambda & -\frac{\GTzo^{10} }{\sqrt{2}} \\ 
-1 & \frac{3}{2} & \Sigma N \leftrightarrow \Delta \Sigma & -\sqrt{\frac{5}{2}} \GTzo^{10}  \\ 

\cmidrule(lr){1-3}\cmidrule(lr){4-4}
-2 & 0 & \Xi N \leftrightarrow \Xi^*  N & -2 \GTzo^{8}  \\ 
-2 & 0 & \Xi N \leftrightarrow \Sigma^*  \Sigma & \sqrt{6} \GTzo^{8}  \\ 
-2 & 1 & \Xi N \leftrightarrow \Xi^*  N & \frac{2 (\GTzo^{10} +\GTzo^{8} )}{3} \\ 
-2 & 1 & \Xi N \leftrightarrow \Sigma^*  \Lambda & -\sqrt{\frac{2}{3}} \GTzo^{8}  \\ 
-2 & 1 & \Xi N \leftrightarrow \Sigma^*  \Sigma & -\frac{2}{3} (\GTzo^{10} +\GTzo^{8} ) \\ 
-2 & 1 & \Xi N \leftrightarrow \Delta \Xi & -\frac{2}{3} (\GTzo^{10} -2 \GTzo^{8} ) \\ 
-2 & 1 & \Sigma \Lambda \leftrightarrow \Xi^*  N & \sqrt{\frac{2}{3}} \GTzo^{10}  \\ 
-2 & 1 & \Sigma \Lambda \leftrightarrow \Sigma^*  \Sigma & -\sqrt{\frac{2}{3}} \GTzo^{10}  \\ 
-2 & 1 & \Sigma \Lambda \leftrightarrow \Delta \Xi & -\sqrt{\frac{2}{3}} \GTzo^{10}  \\ 
-2 & 1 & \Sigma \Sigma \leftrightarrow \Xi^*  N & \frac{1}{3} \sqrt{2} (\GTzo^{10} -2 \GTzo^{8} ) \\ 
-2 & 1 & \Sigma \Sigma \leftrightarrow \Sigma^*  \Lambda & \frac{2 \GTzo^{8} }{\sqrt{3}} \\ 
-2 & 1 & \Sigma \Sigma \leftrightarrow \Sigma^*  \Sigma & -\frac{1}{3} \sqrt{2} (\GTzo^{10} -2 \GTzo^{8} ) \\ 
-2 & 1 & \Sigma \Sigma \leftrightarrow \Delta \Xi & -\frac{1}{3} \sqrt{2} (\GTzo^{10} +4 \GTzo^{8} ) \\ 

\cmidrule(lr){1-3}\cmidrule(lr){4-4}
-3 & \frac{1}{2} & \Xi \Lambda \leftrightarrow \Omega N & \frac{\GTzo^{10} +2 \GTzo^{8} }{\sqrt{2}} \\ 
-3 & \frac{1}{2} & \Xi \Lambda \leftrightarrow \Xi^*  \Lambda & \frac{1}{2} (\GTzo^{10} -2 \GTzo^{8} ) \\ 
-3 & \frac{1}{2} & \Xi \Lambda \leftrightarrow \Xi^*  \Sigma & -\frac{\GTzo^{10} }{2}-\GTzo^{8}  \\ 
-3 & \frac{1}{2} & \Xi \Lambda \leftrightarrow \Sigma^*  \Xi & \GTzo^{8} -\GTzo^{10}  \\ 
-3 & \frac{1}{2} & \Xi \Sigma \leftrightarrow \Omega N & \frac{\GTzo^{10} -2 \GTzo^{8} }{\sqrt{2}} \\ 
-3 & \frac{1}{2} & \Xi \Sigma \leftrightarrow \Xi^*  \Lambda & \frac{\GTzo^{10} }{2}+\GTzo^{8}  \\ 
-3 & \frac{1}{2} & \Xi \Sigma \leftrightarrow \Xi^*  \Sigma & \GTzo^{8} -\frac{\GTzo^{10} }{2} \\ 
-3 & \frac{1}{2} & \Xi \Sigma \leftrightarrow \Sigma^*  \Xi & -\GTzo^{10} -\GTzo^{8}  \\ 

\cmidrule(lr){1-3}\cmidrule(lr){4-4}
-4 & 0 & \Xi \Xi \leftrightarrow \Omega \Lambda & \sqrt{2} \GTzo^{10}  \\ 
-4 & 0 & \Xi \Xi \leftrightarrow \Xi^*  \Xi & -\sqrt{2} \GTzo^{10}  \\ 

\end{longtable}

\renewcommand{\sharedhead}{
\toprule
S & I & \text{transition} & V_{{}^1S_{0}} & V_{{}^3S_{1}} \\
\cmidrule(lr){1-3}\cmidrule(lr){4-5}
}
\begin{longtable}{ttttt}
\caption{SU(3) relations of \(BB\to DD\) (and  \(DD\to BB\)) in non-vanishing partial waves. 
The subscript $\{02\}$ of the constants \(\Gzt^r\) that denotes the
$\mathcal{B}\mathcal{B}$ channel is omitted in the table.
} \label{tab:PWDBBDD} \\ \sharedhead
\endfirsthead
\caption{(\dots continued)} \\ \sharedhead
\endhead
\bottomrule \multicolumn{5}{l}{\scriptsize \dots continues on next page}
\endfoot
\bottomrule
\endlastfoot
0 & 0 & N N \leftrightarrow \Delta \Delta & 0 & \GTzt^{\overline{10}}  \\ 
0 & 1 & N N \leftrightarrow \Delta \Delta & \GTzt^{27} & 0 \\ 

\cmidrule(lr){1-3}\cmidrule(lr){4-5}
-1 & \frac{1}{2} & \Lambda N \leftrightarrow \Sigma^*  \Delta & \frac{3 \GTzt^{27}}{\sqrt{10}} & \frac{\GTzt^{\overline{10}} }{\sqrt{2}} \\ 
-1 & \frac{1}{2} & \Sigma N \leftrightarrow \Sigma^*  \Delta & -\frac{\GTzt^{27}}{\sqrt{10}} & \frac{\GTzt^{\overline{10}} }{\sqrt{2}} \\ 
-1 & \frac{3}{2} & \Sigma N \leftrightarrow \Sigma^*  \Delta & \GTzt^{27} & 0 \\ 

\cmidrule(lr){1-3}\cmidrule(lr){4-5}
-2 & 0 & \Lambda \Lambda \leftrightarrow \Sigma^*  \Sigma^*  & \frac{3}{2} \sqrt{\frac{3}{10}} \GTzt^{27} & 0 \\ 
-2 & 0 & \Xi N \leftrightarrow \Sigma^*  \Sigma^*  & -\sqrt{\frac{3}{10}} \GTzt^{27} & 0 \\ 
-2 & 0 & \Sigma \Sigma \leftrightarrow \Sigma^*  \Sigma^*  & -\frac{\GTzt^{27}}{2 \sqrt{10}} & 0 \\ 
-2 & 1 & \Xi N \leftrightarrow \Xi^*  \Delta & \sqrt{\frac{2}{5}} \GTzt^{27} & \frac{\sqrt{2} \GTzt^{\overline{10}} }{3} \\ 
-2 & 1 & \Xi N \leftrightarrow \Sigma^*  \Sigma^*  & 0 & -\frac{\GTzt^{\overline{10}} }{3} \\ 
-2 & 1 & \Sigma \Lambda \leftrightarrow \Xi^*  \Delta & \sqrt{\frac{3}{5}} \GTzt^{27} & -\frac{\GTzt^{\overline{10}} }{\sqrt{3}} \\ 
-2 & 1 & \Sigma \Lambda \leftrightarrow \Sigma^*  \Sigma^*  & 0 & \frac{\GTzt^{\overline{10}} }{\sqrt{6}} \\ 
-2 & 1 & \Sigma \Sigma \leftrightarrow \Xi^*  \Delta & 0 & \frac{\GTzt^{\overline{10}} }{3} \\ 
-2 & 1 & \Sigma \Sigma \leftrightarrow \Sigma^*  \Sigma^*  & 0 & -\frac{\GTzt^{\overline{10}} }{3 \sqrt{2}} \\ 
-2 & 2 & \Sigma \Sigma \leftrightarrow \Xi^*  \Delta & \sqrt{\frac{3}{5}} \GTzt^{27} & 0 \\ 
-2 & 2 & \Sigma \Sigma \leftrightarrow \Sigma^*  \Sigma^*  & -\sqrt{\frac{2}{5}} \GTzt^{27} & 0 \\ 

\cmidrule(lr){1-3}\cmidrule(lr){4-5}
-3 & \frac{1}{2} & \Xi \Lambda \leftrightarrow \Xi^*  \Sigma^*  & \frac{3 \GTzt^{27}}{\sqrt{10}} & 0 \\ 
-3 & \frac{1}{2} & \Xi \Sigma \leftrightarrow \Xi^*  \Sigma^*  & -\frac{\GTzt^{27}}{\sqrt{10}} & 0 \\ 
-3 & \frac{3}{2} & \Xi \Sigma \leftrightarrow \Omega \Delta & \frac{3 \GTzt^{27}}{\sqrt{10}} & \frac{\GTzt^{\overline{10}} }{\sqrt{2}} \\ 
-3 & \frac{3}{2} & \Xi \Sigma \leftrightarrow \Xi^*  \Sigma^*  & -\frac{\GTzt^{27}}{\sqrt{10}} & -\frac{\GTzt^{\overline{10}} }{\sqrt{2}} \\ 

\cmidrule(lr){1-3}\cmidrule(lr){4-5}
-4 & 1 & \Xi \Xi \leftrightarrow \Omega \Sigma^*  & \sqrt{\frac{3}{5}} \GTzt^{27} & 0 \\ 
-4 & 1 & \Xi \Xi \leftrightarrow \Xi^*  \Xi^*  & -\sqrt{\frac{2}{5}} \GTzt^{27} & 0 \\ 

\end{longtable}

\renewcommand{\sharedhead}{
\toprule
S & I & \text{transition} & V_i\,,\ i\in\{{}^3S_{1},{}^5S_{2}\}\\
\cmidrule(lr){1-3}\cmidrule(lr){4-4}
}
\begin{longtable}{tttt}
\caption{SU(3) relations of \(DB\to DB\) in non-vanishing partial waves. 
The subscript $\{11\}$ of the constants \(\Goo^r\) that denotes the
$\mathcal{B}\mathcal{B}$ channel is omitted in the table.
} \label{tab:PWDDBDB} \\ \sharedhead
\endfirsthead
\caption{(\dots continued)} \\ \sharedhead
\endhead
\bottomrule \multicolumn{4}{l}{\scriptsize \dots continues on next page}
\endfoot
\bottomrule
\endlastfoot
0 & 1 & \Delta N \leftrightarrow \Delta N & \GToo^{27,i} \\ 
0 & 2 & \Delta N \leftrightarrow \Delta N & \GToo^{35,i} \\ 

\cmidrule(lr){1-3}\cmidrule(lr){4-4}
-1 & \frac{1}{2} & \Sigma^*  N \leftrightarrow \Sigma^*  N & \frac{1}{5} (4 \GToo^{27,i}+\GToo^{8,i} ) \\ 
-1 & \frac{1}{2} & \Delta \Sigma \leftrightarrow \Sigma^*  N & \frac{2 (\GToo^{27,i}-\GToo^{8,i} )}{5} \\ 
-1 & \frac{1}{2} & \Delta \Sigma \leftrightarrow \Delta \Sigma & \frac{1}{5} (\GToo^{27,i}+4 \GToo^{8,i} ) \\ 
-1 & \frac{3}{2} & \Sigma^*  N \leftrightarrow \Sigma^*  N & \frac{1}{8} (2 \GToo^{10,i} +\GToo^{27,i}+5 \GToo^{35,i}) \\ 
-1 & \frac{3}{2} & \Delta \Lambda \leftrightarrow \Sigma^*  N & -\frac{2 \GToo^{10,i} +3 \GToo^{27,i}-5 \GToo^{35,i}}{8 \sqrt{2}} \\ 
-1 & \frac{3}{2} & \Delta \Lambda \leftrightarrow \Delta \Lambda & \frac{1}{16} (2 \GToo^{10,i} +9 \GToo^{27,i}+5 \GToo^{35,i}) \\ 
-1 & \frac{3}{2} & \Delta \Sigma \leftrightarrow \Sigma^*  N & \frac{1}{8} \sqrt{\frac{5}{2}} (-2 \GToo^{10,i} +\GToo^{27,i}+\GToo^{35,i}) \\ 
-1 & \frac{3}{2} & \Delta \Sigma \leftrightarrow \Delta \Lambda & \frac{1}{16} \sqrt{5} (2 \GToo^{10,i} -3 \GToo^{27,i}+\GToo^{35,i}) \\ 
-1 & \frac{3}{2} & \Delta \Sigma \leftrightarrow \Delta \Sigma & \frac{1}{16} (10 \GToo^{10,i} +5 \GToo^{27,i}+\GToo^{35,i}) \\ 
-1 & \frac{5}{2} & \Delta \Sigma \leftrightarrow \Delta \Sigma & \GToo^{35,i} \\ 

\cmidrule(lr){1-3}\cmidrule(lr){4-4}
-2 & 0 & \Xi^*  N \leftrightarrow \Xi^*  N & \frac{1}{5} (3 \GToo^{27,i}+2 \GToo^{8,i} ) \\ 
-2 & 0 & \Sigma^*  \Sigma \leftrightarrow \Xi^*  N & \frac{1}{5} \sqrt{6} (\GToo^{27,i}-\GToo^{8,i} ) \\ 
-2 & 0 & \Sigma^*  \Sigma \leftrightarrow \Sigma^*  \Sigma & \frac{1}{5} (2 \GToo^{27,i}+3 \GToo^{8,i} ) \\ 
-2 & 1 & \Xi^*  N \leftrightarrow \Xi^*  N & \frac{1}{15} (5 \GToo^{10,i} +3 \GToo^{27,i}+5 \GToo^{35,i}+2 \GToo^{8,i} ) \\ 
-2 & 1 & \Sigma^*  \Lambda \leftrightarrow \Xi^*  N & -\frac{3 \GToo^{27,i}-5 \GToo^{35,i}+2 \GToo^{8,i} }{5 \sqrt{6}} \\ 
-2 & 1 & \Sigma^*  \Lambda \leftrightarrow \Sigma^*  \Lambda & \frac{1}{10} (3 \GToo^{27,i}+5 \GToo^{35,i}+2 \GToo^{8,i} ) \\ 
-2 & 1 & \Sigma^*  \Sigma \leftrightarrow \Xi^*  N & \frac{1}{30} (-10 \GToo^{10,i} +9 \GToo^{27,i}+5 \GToo^{35,i}-4 \GToo^{8,i} ) \\ 
-2 & 1 & \Sigma^*  \Sigma \leftrightarrow \Sigma^*  \Lambda & \frac{-9 \GToo^{27,i}+5 \GToo^{35,i}+4 \GToo^{8,i} }{10 \sqrt{6}} \\ 
-2 & 1 & \Sigma^*  \Sigma \leftrightarrow \Sigma^*  \Sigma & \frac{1}{60} (20 \GToo^{10,i} +27 \GToo^{27,i}+5 \GToo^{35,i}+8 \GToo^{8,i} ) \\ 
-2 & 1 & \Delta \Xi \leftrightarrow \Xi^*  N & \frac{1}{30} (-10 \GToo^{10,i} -3 \GToo^{27,i}+5 \GToo^{35,i}+8 \GToo^{8,i} ) \\ 
-2 & 1 & \Delta \Xi \leftrightarrow \Sigma^*  \Lambda & \frac{3 \GToo^{27,i}+5 \GToo^{35,i}-8 \GToo^{8,i} }{10 \sqrt{6}} \\ 
-2 & 1 & \Delta \Xi \leftrightarrow \Sigma^*  \Sigma & \frac{1}{60} (20 \GToo^{10,i} -9 \GToo^{27,i}+5 \GToo^{35,i}-16 \GToo^{8,i} ) \\ 
-2 & 1 & \Delta \Xi \leftrightarrow \Delta \Xi & \frac{1}{60} (20 \GToo^{10,i} +3 \GToo^{27,i}+5 \GToo^{35,i}+32 \GToo^{8,i} ) \\ 
-2 & 2 & \Sigma^*  \Sigma \leftrightarrow \Sigma^*  \Sigma & \frac{1}{4} (\GToo^{27,i}+3 \GToo^{35,i}) \\ 
-2 & 2 & \Delta \Xi \leftrightarrow \Sigma^*  \Sigma & \frac{1}{4} \sqrt{3} (\GToo^{35,i}-\GToo^{27,i}) \\ 
-2 & 2 & \Delta \Xi \leftrightarrow \Delta \Xi & \frac{1}{4} (3 \GToo^{27,i}+\GToo^{35,i}) \\ 

\cmidrule(lr){1-3}\cmidrule(lr){4-4}
-3 & \frac{1}{2} & \Omega N \leftrightarrow \Omega N & \frac{1}{40} (10 \GToo^{10,i} +9 \GToo^{27,i}+5 \GToo^{35,i}+16 \GToo^{8,i} ) \\ 
-3 & \frac{1}{2} & \Xi^*  \Lambda \leftrightarrow \Omega N & \frac{10 \GToo^{10,i} -9 \GToo^{27,i}+15 \GToo^{35,i}-16 \GToo^{8,i} }{40 \sqrt{2}} \\ 
-3 & \frac{1}{2} & \Xi^*  \Lambda \leftrightarrow \Xi^*  \Lambda & \frac{1}{80} (10 \GToo^{10,i} +9 \GToo^{27,i}+45 \GToo^{35,i}+16 \GToo^{8,i} ) \\ 
-3 & \frac{1}{2} & \Xi^*  \Sigma \leftrightarrow \Omega N & \frac{-10 \GToo^{10,i} +21 \GToo^{27,i}+5 \GToo^{35,i}-16 \GToo^{8,i} }{40 \sqrt{2}} \\ 
-3 & \frac{1}{2} & \Xi^*  \Sigma \leftrightarrow \Xi^*  \Lambda & \frac{1}{80} (-10 \GToo^{10,i} -21 \GToo^{27,i}+15 \GToo^{35,i}+16 \GToo^{8,i} ) \\ 
-3 & \frac{1}{2} & \Xi^*  \Sigma \leftrightarrow \Xi^*  \Sigma & \frac{1}{80} (10 \GToo^{10,i} +49 \GToo^{27,i}+5 \GToo^{35,i}+16 \GToo^{8,i} ) \\ 
-3 & \frac{1}{2} & \Sigma^*  \Xi \leftrightarrow \Omega N & \frac{-10 \GToo^{10,i} -3 \GToo^{27,i}+5 \GToo^{35,i}+8 \GToo^{8,i} }{20 \sqrt{2}} \\ 
-3 & \frac{1}{2} & \Sigma^*  \Xi \leftrightarrow \Xi^*  \Lambda & \frac{1}{40} (-10 \GToo^{10,i} +3 \GToo^{27,i}+15 \GToo^{35,i}-8 \GToo^{8,i} ) \\ 
-3 & \frac{1}{2} & \Sigma^*  \Xi \leftrightarrow \Xi^*  \Sigma & \frac{1}{40} (10 \GToo^{10,i} -7 \GToo^{27,i}+5 \GToo^{35,i}-8 \GToo^{8,i} ) \\ 
-3 & \frac{1}{2} & \Sigma^*  \Xi \leftrightarrow \Sigma^*  \Xi & \frac{1}{20} (10 \GToo^{10,i} +\GToo^{27,i}+5 \GToo^{35,i}+4 \GToo^{8,i} ) \\ 
-3 & \frac{3}{2} & \Xi^*  \Sigma \leftrightarrow \Xi^*  \Sigma & \frac{\GToo^{27,i}+\GToo^{35,i}}{2} \\ 
-3 & \frac{3}{2} & \Sigma^*  \Xi \leftrightarrow \Xi^*  \Sigma & \frac{\GToo^{35,i}-\GToo^{27,i}}{2} \\ 
-3 & \frac{3}{2} & \Sigma^*  \Xi \leftrightarrow \Sigma^*  \Xi & \frac{\GToo^{27,i}+\GToo^{35,i}}{2} \\ 

\cmidrule(lr){1-3}\cmidrule(lr){4-4}
-4 & 0 & \Omega \Lambda \leftrightarrow \Omega \Lambda & \frac{\GToo^{10,i} +\GToo^{35,i}}{2} \\ 
-4 & 0 & \Xi^*  \Xi \leftrightarrow \Omega \Lambda & \frac{\GToo^{35,i}-\GToo^{10,i} }{2} \\ 
-4 & 0 & \Xi^*  \Xi \leftrightarrow \Xi^*  \Xi & \frac{\GToo^{10,i} +\GToo^{35,i}}{2} \\ 
-4 & 1 & \Omega \Sigma \leftrightarrow \Omega \Sigma & \frac{1}{4} (3 \GToo^{27,i}+\GToo^{35,i}) \\ 
-4 & 1 & \Xi^*  \Xi \leftrightarrow \Omega \Sigma & \frac{1}{4} \sqrt{3} (\GToo^{35,i}-\GToo^{27,i}) \\ 
-4 & 1 & \Xi^*  \Xi \leftrightarrow \Xi^*  \Xi & \frac{1}{4} (\GToo^{27,i}+3 \GToo^{35,i}) \\ 

\cmidrule(lr){1-3}\cmidrule(lr){4-4}
-5 & \frac{1}{2} & \Omega \Xi \leftrightarrow \Omega \Xi & \GToo^{35,i} \\ 

\end{longtable}

\renewcommand{\sharedhead}{
\toprule
S & I & \text{transition} & V_{{}^3S_{1}} & V_{{}^5S_{2}} \\
\cmidrule(lr){1-3}\cmidrule(lr){4-5}
}
\begin{longtable}{ttttt}
\caption{SU(3) relations of \(DB\to DD\) (and  \(DD\to DB\)) in non-vanishing partial waves. 
The subscript $\{12\}$ of the constants \(\Got^r\) that denotes the
$\mathcal{B}\mathcal{B}$ channel is omitted in the table.
} \label{tab:PWDDBDD} \\ \sharedhead
\endfirsthead
\caption{(\dots continued)} \\ \sharedhead
\endhead
\bottomrule \multicolumn{5}{l}{\scriptsize \dots continues on next page}
\endfoot
\bottomrule
\endlastfoot
0 & 1 & \Delta N \leftrightarrow \Delta \Delta & 0 & \GTot^{27} \\ 
0 & 2 & \Delta N \leftrightarrow \Delta \Delta & \GTot^{35} & 0 \\ 

\cmidrule(lr){1-3}\cmidrule(lr){4-5}
-1 & \frac{1}{2} & \Sigma^*  N \leftrightarrow \Sigma^*  \Delta & 0 & \frac{2 \GTot^{27}}{\sqrt{5}} \\ 
-1 & \frac{1}{2} & \Delta \Sigma \leftrightarrow \Sigma^*  \Delta & 0 & \frac{\GTot^{27}}{\sqrt{5}} \\ 
-1 & \frac{3}{2} & \Sigma^*  N \leftrightarrow \Sigma^*  \Delta & \frac{1}{2} \sqrt{\frac{5}{2}} \GTot^{35} & \frac{\GTot^{27}}{2 \sqrt{2}} \\ 
-1 & \frac{3}{2} & \Delta \Lambda \leftrightarrow \Sigma^*  \Delta & \frac{\sqrt{5} \GTot^{35}}{4} & -\frac{3 \GTot^{27}}{4} \\ 
-1 & \frac{3}{2} & \Delta \Sigma \leftrightarrow \Sigma^*  \Delta & \frac{\GTot^{35}}{4} & \frac{\sqrt{5} \GTot^{27}}{4} \\ 
-1 & \frac{5}{2} & \Delta \Sigma \leftrightarrow \Sigma^*  \Delta & \GTot^{35} & 0 \\ 

\cmidrule(lr){1-3}\cmidrule(lr){4-5}
-2 & 0 & \Xi^*  N \leftrightarrow \Sigma^*  \Sigma^*  & 0 & \sqrt{\frac{3}{5}} \GTot^{27} \\ 
-2 & 0 & \Sigma^*  \Sigma \leftrightarrow \Sigma^*  \Sigma^*  & 0 & \sqrt{\frac{2}{5}} \GTot^{27} \\ 
-2 & 1 & \Xi^*  N \leftrightarrow \Xi^*  \Delta & \frac{\GTot^{35}}{3} & \frac{\GTot^{27}}{\sqrt{5}} \\ 
-2 & 1 & \Xi^*  N \leftrightarrow \Sigma^*  \Sigma^*  & \frac{\sqrt{2} \GTot^{35}}{3} & 0 \\ 
-2 & 1 & \Sigma^*  \Lambda \leftrightarrow \Xi^*  \Delta & \frac{\GTot^{35}}{\sqrt{6}} & -\sqrt{\frac{3}{10}} \GTot^{27} \\ 
-2 & 1 & \Sigma^*  \Lambda \leftrightarrow \Sigma^*  \Sigma^*  & \frac{\GTot^{35}}{\sqrt{3}} & 0 \\ 
-2 & 1 & \Sigma^*  \Sigma \leftrightarrow \Xi^*  \Delta & \frac{\GTot^{35}}{6} & \frac{3 \GTot^{27}}{2 \sqrt{5}} \\ 
-2 & 1 & \Sigma^*  \Sigma \leftrightarrow \Sigma^*  \Sigma^*  & \frac{\GTot^{35}}{3 \sqrt{2}} & 0 \\ 
-2 & 1 & \Delta \Xi \leftrightarrow \Xi^*  \Delta & \frac{\GTot^{35}}{6} & -\frac{\GTot^{27}}{2 \sqrt{5}} \\ 
-2 & 1 & \Delta \Xi \leftrightarrow \Sigma^*  \Sigma^*  & \frac{\GTot^{35}}{3 \sqrt{2}} & 0 \\ 
-2 & 2 & \Sigma^*  \Sigma \leftrightarrow \Xi^*  \Delta & \frac{\sqrt{3} \GTot^{35}}{2} & \frac{1}{2} \sqrt{\frac{3}{5}} \GTot^{27} \\ 
-2 & 2 & \Sigma^*  \Sigma \leftrightarrow \Sigma^*  \Sigma^*  & 0 & -\frac{\GTot^{27}}{\sqrt{10}} \\ 
-2 & 2 & \Delta \Xi \leftrightarrow \Xi^*  \Delta & \frac{\GTot^{35}}{2} & -\frac{3 \GTot^{27}}{2 \sqrt{5}} \\ 
-2 & 2 & \Delta \Xi \leftrightarrow \Sigma^*  \Sigma^*  & 0 & \sqrt{\frac{3}{10}} \GTot^{27} \\ 

\cmidrule(lr){1-3}\cmidrule(lr){4-5}
-3 & \frac{1}{2} & \Omega N \leftrightarrow \Xi^*  \Sigma^*  & \frac{\GTot^{35}}{2 \sqrt{2}} & \frac{3 \GTot^{27}}{2 \sqrt{10}} \\ 
-3 & \frac{1}{2} & \Xi^*  \Lambda \leftrightarrow \Xi^*  \Sigma^*  & \frac{3 \GTot^{35}}{4} & -\frac{3 \GTot^{27}}{4 \sqrt{5}} \\ 
-3 & \frac{1}{2} & \Xi^*  \Sigma \leftrightarrow \Xi^*  \Sigma^*  & \frac{\GTot^{35}}{4} & \frac{7 \GTot^{27}}{4 \sqrt{5}} \\ 
-3 & \frac{1}{2} & \Sigma^*  \Xi \leftrightarrow \Xi^*  \Sigma^*  & \frac{\GTot^{35}}{2} & -\frac{\GTot^{27}}{2 \sqrt{5}} \\ 
-3 & \frac{3}{2} & \Xi^*  \Sigma \leftrightarrow \Omega \Delta & \frac{\GTot^{35}}{2} & \frac{3 \GTot^{27}}{2 \sqrt{5}} \\ 
-3 & \frac{3}{2} & \Xi^*  \Sigma \leftrightarrow \Xi^*  \Sigma^*  & \frac{\GTot^{35}}{2} & -\frac{\GTot^{27}}{2 \sqrt{5}} \\ 
-3 & \frac{3}{2} & \Sigma^*  \Xi \leftrightarrow \Omega \Delta & \frac{\GTot^{35}}{2} & -\frac{3 \GTot^{27}}{2 \sqrt{5}} \\ 
-3 & \frac{3}{2} & \Sigma^*  \Xi \leftrightarrow \Xi^*  \Sigma^*  & \frac{\GTot^{35}}{2} & \frac{\GTot^{27}}{2 \sqrt{5}} \\ 

\cmidrule(lr){1-3}\cmidrule(lr){4-5}
-4 & 0 & \Omega \Lambda \leftrightarrow \Xi^*  \Xi^*  & \frac{\GTot^{35}}{\sqrt{2}} & 0 \\ 
-4 & 0 & \Xi^*  \Xi \leftrightarrow \Xi^*  \Xi^*  & \frac{\GTot^{35}}{\sqrt{2}} & 0 \\ 
-4 & 1 & \Omega \Sigma \leftrightarrow \Omega \Sigma^*  & \frac{\GTot^{35}}{2} & \frac{3 \GTot^{27}}{2 \sqrt{5}} \\ 
-4 & 1 & \Omega \Sigma \leftrightarrow \Xi^*  \Xi^*  & 0 & -\sqrt{\frac{3}{10}} \GTot^{27} \\ 
-4 & 1 & \Xi^*  \Xi \leftrightarrow \Omega \Sigma^*  & \frac{\sqrt{3} \GTot^{35}}{2} & -\frac{1}{2} \sqrt{\frac{3}{5}} \GTot^{27} \\ 
-4 & 1 & \Xi^*  \Xi \leftrightarrow \Xi^*  \Xi^*  & 0 & \frac{\GTot^{27}}{\sqrt{10}} \\ 

\cmidrule(lr){1-3}\cmidrule(lr){4-5}
-5 & \frac{1}{2} & \Omega \Xi \leftrightarrow \Omega \Xi^*  & \GTot^{35} & 0 \\ 

\end{longtable}

\twocolumn

\end{document}